\documentclass[journal]{IEEEtran}
\usepackage{cite}
\usepackage{amsmath}
\usepackage{graphicx}
\usepackage{float}
\usepackage{caption}
\usepackage{subfigure}
\usepackage{color}
\usepackage{CJK}
\usepackage{algorithm}
\usepackage{algorithmic}
\usepackage{tikz}
\usepackage{epstopdf}
\usepackage{indentfirst}
\usepackage{subfigure}
\usepackage{epstopdf}
\usepackage[latin1]{inputenc}
\usepackage{rotating}
\usepackage{colortbl}
\usepackage{amssymb,amsthm,enumerate,graphicx,float}
\usepackage{booktabs}
\usepackage{hyperref}

\ifCLASSINFOpdf
\else
\fi
\newtheorem{theorem}{Theorem}

\newtheorem{lemma}{Lemma}
\newtheorem{proposition}{Proposition}
\newtheorem{assumption}{Assumption}

\newtheorem{remark}{Remark}

\begin{document}
\title{Distributed Optimization Involving Non-Lipschitzian, Linear Composite Functions: Unified Algorithm Design and Convergence Analysis}

\author{Liang Ran, Huaqing Li, \IEEEmembership{Senior Member, IEEE}, Lifeng Zheng, Jun Li, Run Tang
\thanks{This work was supported in part by the Fundamental Research Funds for the Central Universities under Grant SWU-XDJH202312, in part by the National Natural Science Foundation of China under Grant 62173278, 62573364, and in part by the Chongqing Science Fund for Distinguished Young Scholars under Grant 2024NSCQ-JQX0103. \emph{(Corresponding author: Huaqing Li.)} }
\thanks{L. Ran is College of Electronic and Information Engineering, Southwest University, Chongqing 400715, China, and the Key Laboratory of Intelligent Analysis and Decision on ComplexSystems, School of Mathematics and Statistics, Chongqing University of Posts and Telecommunications, Chongqing 400065, China (e-mail: ranliang\_rl@163.com).}
\thanks{H. Li and R. Tang are the Chongqing Key Laboratory of Nonlinear Circuits and Intelligent Information Processing, College of Electronic and Information Engineering, Southwest University, Chongqing 400715, China (e-mail: huaqingli@swu.edu.cn, RTangtr@163.com).}
\thanks{J. Li is School of Economics and Management, Chongqing University of Posts and Telecommunications, No. 2, Chongwen Road, Nan'an District, Chongqing 400065, China (e-mail: jun\_li2023@163.com).}
\thanks{L. Zheng is the School of Computer Science and Engineering, Chongqing University of Technology, No. 69 Hongguang Avenue, Banan District, Chongqing 400054, China (e-mail: zlf\_swu@163.com).}
}

\maketitle

\begin{abstract}
This paper formulates a generic distributed nonsmooth optimization problem subject to edge-based equality constraints, local inequality constraints characterized by general convex functions, as well as globally coupling constraints including affine equality and nonlinear inequality. The terminal goal is to minimize the sum of local non-differentiable convex functions and proximable nonsmooth convex composite terms attached to linear operators, over networked multi-agent systems. This formulated framework not only applies in complex practical problems but also reduces to existing distributed optimization structures. To cope with technical issues caused by local linear mappings, potential multiple nonsmooth functions and coupling nonlinear constraints, we first derive a new local first-order condition, and further present a novel fully distributed optimization algorithm based on variable decompositions and proximal operators. Specifically, the proposed algorithm allows each agent to update states via local information exchange and through {locally fixed step-sizes}, and to find inexact solutions of subproblems without loss of convergence speed. In contrast to existing tailor-made algorithms for the bounded gradients of objective functions or affine constraints, our algorithm is capable of solving more practical cases involving general convex functions and nonlinear constraints. {Under mild assumptions, the convergence rate $o(1/k)$ in terms of the optimality and successive-difference metrics is achieved, by constructing the equivalent variational inequalities of the proposed algorithm.} Finally, numerical experiments on practical application problems illustrate the theoretical results.
\end{abstract}


\section{Introduction} \label{sec:introduction}

This paper considers the agent set ${\cal V} = \{ 1, \cdots ,m\} $ over the undirected network ${\cal G} = ({\cal V}, {\cal E})$ with the edge set ${\cal E} \subseteq {\cal V} \times {\cal V}$. All agents exchange information over ${\cal G}$ in order to collectively seek $x = {[x_1^{\rm T}, \cdots ,x_m^{\rm T}]^{\rm T}}$ that minimizes the following general constrained problem:
\begin{subequations}\label{OrigProblem}
  \begin{equation}\label{OrigProblema}
   \mathop {{\text{minimize}}}\limits_{{x_1}, \cdots ,{x_m}} {\text{  }}\sum\limits_{i = 1}^m {\left( {{f_i}\left( {{x_i}} \right) + {g_i}\left( {{R_i}{x_i}} \right)} \right)} \qquad\qquad\qquad\qquad\quad
  \end{equation}
  \begin{equation}\label{OrigProblemb}
   {\text{subject to  }}{B_{ij}}{x_i} + {B_{ji}}{x_j} = {b_{\left( {i,j} \right)}},{\text{ }} \forall \left( {i,j} \right) \in \mathcal{E} \qquad\qquad\,\,
  \end{equation}
  \begin{equation}\label{OrigProblemc}
  {h_i}\left( {{x_i}} \right) \leq 0, i = 1, \cdots ,m \qquad\qquad\quad
  \end{equation}
  \begin{equation}\label{OrigProblemd}
   \sum\limits_{i = 1}^m {\left( {{A_i}{x_i} - {a_i}} \right)}  = 0,{\text{ }}\sum\limits_{i = 1}^m {{C_i}\left( {{x_i}} \right)}  \leq 0
  \end{equation}
\end{subequations}
where ${x_i} \in {\mathbb{R}^{{n_i}}}$ is the decision variable of agent $i$, ${f_i}:{\mathbb{R}^{{n_i}}} \to \mathbb{R}$ and ${g_i}:{\mathbb{R}^{{{\tilde r}_i}}} \to \mathbb{R} \cup \{  + \infty \} $ with the local linear mapping ${R_i} \in {\mathbb{R}^{{{\tilde r}_i} \times {n_i}}}$ are possibly nonsmooth functions, while ${h_i}:{\mathbb{R}^{{n_i}}} \to {\mathbb{R}^{{p_i}}}$ and ${C_i}:{\mathbb{R}^{{n_i}}} \to {\mathbb{R}^c}$ are the potentially nonlinear functions. In \eqref{OrigProblemb}, ${B_{ij}} \in {\mathbb{R}^{{l_{\left( {i,j} \right)}} \times {n_i}}},{B_{ji}} \in {\mathbb{R}^{{l_{\left( {i,j} \right)}} \times {n_j}}}$ are proper known matrices, ${b_{\left( {i,j} \right)}} \in {\mathbb{R}^{{l_{\left( {i,j} \right)}}}}$ is a vector. In \eqref{OrigProblemd}, ${A_i} \in {\mathbb{R}^{q \times {n_i}}}$ and ${a_i} \in {\mathbb{R}^q}$ are matrix and vector, respectively. As a fresh optimization framework, the above setup arises in numerous practical applications, such as the optimal power flow problem \cite{Yang2020DCDOPF}, and vehicle formation problem \cite{Wang2026Globally}.

To date, a variety of efficient distributed algorithms have been explored for different optimization problems. In \cite{Nedic2009Subgradient}, the distributed subgradient method was first developed to solve unconstrained consensus optimization problems. After that, many gradient-based variants have appeared for such unconstrained problems in the literature, including the linearly convergent first-order method \cite{Jakovetic2019Unification}, proximal gradient algorithms for nonsmooth objectives \cite{Li2019NIDS}, asynchronous distributed algorithms \cite{Bastianello2021Asynchronous,Cannelli2016Parallel}, and another insightful approach \cite{Nguyen2023Accelerated} for time-varying networks. With the consideration of local constraints arising in consensus problems, the distributed stochastic algorithms for local feasible constraints \cite{Jiang2025Stochastic,Huang2026Constrained}, the quasi-Newton method for linear (in)equality constraints \cite{Shorinwa2024Newton}, and distributed continuous-time algorithms \cite{Zhu2020Projected,Liu2025Continuous} for potential nonlinear constraints characterized by convex functions were successively proposed. However, the above methods do not cope well with global coupling constraints. Fortunately, these advancements \cite{Sherson2019Separable,Liu2021Region,Carli2020Distributed,Liang2018Modified,Xing2025Randomized} have made it possible to crack this difficulty. To be specific, a distributed method of multipliers was produced in \cite{Sherson2019Separable} via the combination of dual lifting and operator splitting approaches, the discrete-time primal-dual algorithm based on region partitioning in \cite{Liu2021Region} and proximal Jacobian alternating direction method of multipliers in \cite{Carli2020Distributed} were further presented for linear coupling constraints. But \cite{Sherson2019Separable} required two rounds of communication at each agent, and \cite{Carli2020Distributed} used a coordinated step-size that is dependent on the number of agents. For the purpose of coping with coupling inequalities, a projected subgradient dynamics and modified Lagrangian function were integrated in \cite{Liang2018Modified} to produce distributed primal-dual methods in continuous-time setting, while a static penalty method was introduced in \cite{Xing2025Randomized} for seeking approximate solutions to original problems. { However, there still exist several challenges for distributed algorithm development and convergence analysis.
First, practical objective functions may comprise multiple nonsmooth terms coupled with local linear mappings, making distributed proximal algorithms \cite{Li2019NIDS,Huang2025Unified,Hamedani2022Strongly} designed for single nonsmooth terms computationally demanding. Although subgradient methods handle nonsmoothness, their oscillatory behavior necessitates additional conditions to ensure convergence, such as subgradient boundedness \cite{Nedic2009Subgradient,Xing2025Randomized}, and local constraint boundedness \cite{Zhu2020Projected,Liu2025Continuous}.
Second, the (weighted) Laplacian matrix is widely adopted to handle consensus and global coupling constraints \cite{Huang2023Multiproximal,Liu2021Region,Liang2020Smooth,Huang2025Unified}. However, for convergence proofs, estimating the global Laplacian's largest eigenvalue becomes challenging in large-scale networks, which also forces conservative step-size selections. Additionally, many distributed methods rely on specifically tailored Lyapunov functions for convergence analysis \cite{Liu2021Region,Zhu2020Projected,Liu2025Continuous,Liang2018Modified}, hindering the development of a unified framework capable of accommodating diverse objective and constraint structures.
Third, while intricate optimization problems are tackled by efficient distributed algorithms, certain centralized computations remain unavoidable, such as solving local minimization subproblems \cite{Yang2020DCDOPF,Wu2023Distributed}. Moreover, consensus-based and non-consensus distributed problems are typically studied in isolation \cite{Huang2025Unified,Alaviani2025Parallel}. Algorithms designed for one cannot be readily reused for the other due to the presence of edge-based equality constraints, local nonlinear constraints, and global coupling constraints.
Finally, from a centralized perspective, general nonsmooth convex problems can be solved with sublinear convergence using first-order primal-dual methods \cite{Chambolle2011Imaging}, whereas NAMA \cite{Stella2019Newton} and SuperMann \cite{Themelis2019SuperMann} respectively achieve linear and superlinear rates under strong convexity or metric subregularity. However, these methods operate in a centralized manner and are incapable of directly offering a distributed solution for the entire constraint structure inherent in problem \eqref{OrigProblem}.}

Inspired by the aforementioned observations, this paper aims to develop a distributed optimization algorithm to solve the generic problem \eqref{OrigProblem}, accompanied by rigorous convergence analysis under mild assumptions. The main contributions are {threefold:

\begin{enumerate}
\item By integrating the Karush-Kuhn-Tucker (KKT) conditions with conjugate decomposition, a new edge-based local first-order condition is derived for problem \eqref{OrigProblem}. This condition unifies the first-order optimality conditions for consensus-based optimization  \cite{Huang2026Constrained,Huang2023Multiproximal,Hamedani2022Strongly,Liu2025Continuous}, resource allocation problems \cite{Liu2021Region,Sherson2019Separable,Carli2020Distributed,Falsone2020Tracking}, and tri-convex optimization \cite{Latafat2019TriPD,Li2022TPUS}, as they are all special cases of \eqref{OrigProblem}. Unlike conditions relying on Laplacian matrices \cite{Huang2023Multiproximal,Liu2021Region,Liang2020Smooth,Huang2025Unified,Lin2021Compensation,Alaviani2025Parallel}, our derived condition eliminates the necessity for global (weighted) Laplacians or weight-balancing strategies. Consequently, it promotes fully distributed algorithm execution and permits agents to manage their data independently.

\item The proposed fully distributed optimization algorithm (FDOA) tackles several technical challenges. To begin with, variable substitution is adopted to cope with the complex proximal operators arising from nonsmooth composite terms coupled with local linear mappings, which seamlessly accommodates objectives comprising multiple proximable nonsmooth terms. Next, variable decomposition is formulated to overcome the non-separability of particular indicator functions introduced by coupling constraints. More remarkably, FDOA can be configured to generate novel inexact distributed variants of classical centralized approaches \cite{Popov1980Arrow,Esser2010General} and existing distributed schemes \cite{Latafat2019TriPD}. This bypasses the difficulty of lacking analytical solutions for proximal mappings.

\item Analyzed from a variational inequality perspective, we establish the $o(1/k)$ convergence rate for the first-order optimality residual. This theoretical result is strictly independent of feasible region boundedness \cite{Xing2025Randomized,Lin2021Compensation}, subgradient boundedness \cite{Jiang2025Stochastic,Lin2021Compensation,Fang2025Gossip}, strongly convex quadratic structures \cite{Liu2021Region}, and Lipschitz continuity \cite{Carli2020Distributed,Huang2026Constrained}. Our theoretical result is stronger than the asymptotic convergence guarantees \cite{Liu2021Region,Sherson2019Separable,Alaviani2025Parallel} and exact iterative algorithms \cite{Carli2020Distributed,Wu2023Distributed,Huang2025Unified}. Furthermore, in contrast to the common step-sizes in \cite{Sherson2019Separable,Carli2020Distributed,Huang2026Constrained,Liu2025Continuous,Alaviani2025Parallel,Huang2025Unified}, the explicit upper bounds for locally constant step-sizes derived in this article are decoupled from the largest eigenvalue of the Laplacian matrix and other global parameters. This avoids conservative step-size selections and offers broader applicability for large-scale systems.
\end{enumerate}
}

\emph{Notations}: Let $0$ and $I$ be a zero vector and identity matrix with appropriate dimensions, respectively. Given vectors ${a_1}, \cdots ,{a_m}$ (or matrices ${A_1}, \cdots ,{A_m}$), denote their stacked form (or block diagonal form) by $col\{ {a_1}, \cdots ,{a_m}\} $/$stack\{ {a_1}, \cdots ,{a_m}\}$ (or $blkdiag\{ {A_1}, \cdots ,{A_m}\} $). Utilize $\langle {x,y} \rangle $ to denote the inner product for vectors $x,y \in {\mathbb{R}^n}$, and ${\| x \|_A} = \sqrt {\langle {x,Ax} \rangle } $ to represent the $A$-induced norm with a positive definite matrix $A \in {\mathbb{R}^{n \times n}}$. Considering a nonempty set $\Theta $, the function ${\mathcal{I}_\Theta }(x)$ is its indicator function when ${\mathcal{I}_\Theta }(x) = 0$ for $x \in \Theta$, and ${\mathcal{I}_\Theta }(x) =  + \infty $ for $x \notin \Theta $. {The number of edges of $\mathcal{G}$ is denoted by $\left| \mathcal{E} \right|$. For agent $i \in \mathcal{V}$, its neighbor set is denoted by ${\mathcal{N}_i} = \{ j|(i,j) \in \mathcal{E}\} $, and its number of neighbors is denoted by $\left| {{\mathcal{N}_i}} \right|$.}
Define the distance function ${\text{dis}}{{\text{t}}_A}\left( {x,\Theta } \right) = {\min _{y \in \Theta }}{\left\| {x - y} \right\|_A}$. For a function $f:{\mathbb{R}^n} \to {\mathbb{R}^s}$, let $\partial f$, ${f^ \star }(\omega ) = {\sup _{x \in {\mathbb{R}^n}}}\{ \langle {\omega ,x} \rangle  - f(x)\} $ and ${\rm{prox}}_f^{{A^{ - 1}}}(z) = {\rm{argmi}}{{\rm{n}}_{x \in {\mathbb{R}^n}}}\{ f(x) + \frac{1}{2}\left\| {x - z} \right\|_{{A^{ - 1}}}^2\} $ be its subgradient mapping, conjugation setting and proximal operator with the positive definite matrix $A$, respectively. In particular, when $A$ is a positive scalar, its proximal operator reduces to ${\rm{pro}}{{\text{x}}_{\alpha f}}(z) = {\rm{argmi}}{{\rm{n}}_{x \in {\mathbb{R}^n}}}\{ f(x) + \frac{1}{{2\alpha }}{\| {x - z} \|^2}\} $. Denote domain and relative interior of $f$ by $dom(f)$ and $ridom(f)$, respectively. Other mathematical notations involved in this paper are listed in Table \ref{tab:notation}.

\emph{Organization}: {The rest of this paper is sketched as follows. Section \ref{sec:problem} gives examples of applications, local first-order condition of problem \eqref{OrigProblem}, and presents the details of developing the fully distributed algorithm. The convergence analysis is provided in Section \ref{SectConAna}, and the performance evaluation is conducted in Section \ref{PerforEvalu}. Finally, Section \ref{SecConcl} remarks the conclusion and future works of this article.}

\section{Problem Development and Algorithm Design} \label{sec:problem}

{
For analysis, we consider the local function ${\mathcal{C}_i}:{x_i} \to ( {{A_i}{x_i} - {a_i},{C_i}( {{x_i}} )} )$, the set $\Omega  = 0 \times \mathbb{R}_ - ^c$, and the operator
\begin{flalign}\label{Edgemaps}
\tilde \Pi :x \to {({{\tilde \Pi }_{(i,j)}}x)_{(i,j) \in \mathcal{E}}} \in {\mathbb{R}^{(2l) \times n}}
\end{flalign}
with ${{\tilde \Pi }_{\left( {i,j} \right)}}:{\text{ }}x \to ({B_{ij}}{x_i},{\text{ }}{B_{ji}}{x_j}) \in {\mathbb{R}^{(2{l_{(i,j)}}) \times n}}$ for each edge ${( {i,j} ) \in \mathcal{E}}$, where $l = \sum\nolimits_{(i,j) \in \mathcal{E}} {{l_{(i,j)}}} $ and $n = \sum\nolimits_{i = 1}^m {{n_i}} $. Define the local set
\begin{flalign}\label{Edgeconst}
{\tilde D_{( {i,j} )}} = \{ {( {{z_1},{z_2}} ) \in {\mathbb{R}^{{l_{( {i,j} )}} }} \times {\mathbb{R}^{{l_{( {i,j} )}} }}| {{z_1} + {z_2} = {b_{( {i,j} )}}} } \}
\end{flalign}
Then, the original problem \eqref{OrigProblem} can be reformulated into
\begin{subequations}\label{RefProblem}
  \begin{equation}\label{RefProblema}
   \mathop {{\text{minimize}}}\limits_{{x_1}, \cdots ,{x_m}} {\text{  }}\sum\limits_{i = 1}^m {\left( {{f_i}\left( {{x_i}} \right) + {g_i}\left( {{R_i}{x_i}} \right)} \right)}  + {\mathcal{I}_{\tilde D}}( {\tilde \Pi x} )
  \end{equation}
  \begin{equation}\label{RefProblemb}
   {\text{subject to  }}{h_i}\left( {{x_i}} \right) \leq 0, i = 1, \cdots ,m \qquad\qquad\quad
  \end{equation}
  \begin{equation}\label{RefProblemc}
   \sum\limits_{i = 1}^m {{\mathcal{C}_i}\left( {{x_i}} \right)}  \in \Omega \qquad\qquad\quad
  \end{equation}
\end{subequations}
where ${\mathcal{I}_{\tilde D}}(\tilde \Pi x) = \sum\nolimits_{\left( {i,j} \right) \in \mathcal{E}} {{\mathcal{I}_{{{\tilde D}_{\left( {i,j} \right)}}}}({{\tilde \Pi }_{\left( {i,j} \right)}}x)} $, $\tilde D = \prod\nolimits_{\left( {i,j} \right) \in \mathcal{E}} {{{\tilde D}_{\left( {i,j} \right)}}} $. To ensure meaningful subsequent analysis, we impose the following assumption on problem \eqref{RefProblem}.

{
\begin{assumption}\label{AssumpNonempty}
(i) The undirected network $\mathcal{G}$ is connected. (ii) The local functions $f_i$, $g_i$, $h_i$, and $C_i$ are proper, closed, convex functions, the mappings $B_{ij}$, $B_{ji}$, $R_i$, and $A_i$ are linear and bounded. (iii) The optimal solution set of problem \eqref{OrigProblem} is nonempty. (iv) The Slater's constraint qualification condition holds, i.e., there exists a point ${x_{{\text{in}}}} = col\left\{ {{x_{{\text{in}},1}}, \cdots ,{x_{{\text{in}},m}}} \right\} \in {\mathbb{R}^n}$ such that ${R_i}{x_{{\text{in}},i}} \in ridom\left( {{g_i}} \right)$, $h_i(x_{\mathrm{in},i})<0$ for all $i\in\mathcal{V}$, $B_{ij}x_{\mathrm{in},i}+B_{ji}x_{\mathrm{in},j}=b_{(i,j)}$ for all $(i,j)\in\mathcal{E}$, $\sum_{i=1}^m(A_i x_{\mathrm{in},i}-a_i)=0$ and $\sum_{i=1}^m C_i(x_{\mathrm{in},i})<0$.
\end{assumption}

\begin{remark}
Under the above basic premise, problem \eqref{OrigProblem} is established as a convex constrained optimization problem. Because only standard convexity is required for both the objectives and constraints, this formulation encompasses and extends the specific problems previously explored in \cite{Hamedani2022Strongly,Wu2023Distributed,Li2022TPUS}, circumventing the need for strong/strict convexity or Lipschitzian continuity. Meanwhile, Slater's condition ensures the feasible set is non-empty \cite{Boyd2004Convex}. In cases where obtaining a unique optimal solution is desired, it is sufficient to assume that $f_i$ (or $g_i$) exhibits strongly convex on ${{dom(}}{f_i})$ (or ${{dom(}}{g_i})$).
\end{remark}
}

\subsection{Examples of Applications}
Numerous real-world engineering applications can be formulated within the framework of \eqref{OrigProblem}. Two representative examples are presented below.

\emph{1) Direct Current Dynamic Optimal Power Flow Problem (DC-DOPF) \cite{Yang2020DCDOPF}}: In a power network comprising buses ${S^{{\text{Bus}}}} = \{ 1,2, \cdots ,m\} $, generators ${S^{{\text{Ge}}}} \subseteq {S^{{\text{Bus}}}}$ , and transmission lines ${S^{{\text{Line}}}} \subseteq {S^{{\text{Bus}}}} \times {S^{{\text{Bus}}}}$, the DC-DOPF aims to optimize power dispatch to fulfill load requirements, subject to generation capacities, emission caps, and line limits. The cardinalities $| {{S^{{\text{Ge}}}}} |$ and $| {{S^{{\text{Line}}}}} |$ denote the number of generators and branches, respectively. The total cost aggregates the generator fuel cost ${C^{{\text{fuel}}}}({P^{\text{G}}}) = \sum\nolimits_{i \in {S^{{\text{Ge}}}}} {\sum\nolimits_{t \in {S^{{\text{TP}}}}} {C_{i,t}^{{\text{fuel}}}(P_{i,t}^{\text{G}})} } $ and the carbon emission cost ${C^{{\text{car}}}}(\Delta E) = (1/|{S^{{\text{Bus}}}}|)\sum\nolimits_{i \in {S^{{\text{Bus}}}}} {( {{\pi _{{\text{bou}}}}\Delta {E_b} - {\pi _{{\text{sold}}}}\Delta {E_s}} )} $ with $\Delta E = col\{ \Delta {E_b},\Delta {E_s}\} $, which are possibly nonsmooth. Here, time periods are indexed by ${S^{{\text{TP}}}} = \left\{ {1,2, \cdots ,T} \right\}$, ${P^{\text{G}}} = col\{ P_1^{\text{G}}, \cdots ,P_{\left| {{S^{{\text{Ge}}}}} \right|}^{\text{G}}\}$ denotes the power output with $P_i^{\text{G}} = col\{ P_{i,1}^{\text{G}}, \cdots ,P_{i,T}^{\text{G}}\} $, ${\pi _{{\text{bou}}}}$ and ${\pi _{{\text{sold}}}}$ are purchasing and selling prices for emission allowances, and $\Delta E$ is the allowance capacity. The physical restrictions are formulated as:
\begin{subequations}\label{DOPFGeneCon}
  \begin{equation}\label{DOPFGeneCona}
   P_{i,t}^{\text{G}} - P_{i,t}^{\text{D}} = \sum\nolimits_{j \in {S^{{\text{Bus}}}}} {{\mathcal{B}_{ij}}{\theta _{j,t}}} ,\forall t \in {S^{{\text{TP}}}},\forall i \in {S^{{\text{Bus}}}}
  \end{equation}
  \begin{equation}\label{DOPFGeneConz}
   \sum\nolimits_{i \in {S^{{\text{Bus}}}}} {\sum\nolimits_{j \in {S^{{\text{Bus}}}}} {{\mathcal{B}_{ij}}{\theta _{j,t}}} }  = 0 \qquad\qquad\qquad\qquad\,\,\,
  \end{equation}
  \begin{equation}\label{DOPFGeneConb}
  P_{i,t}^{\text{G}} = 0,\forall i \in {S^{{\text{Bus}}}} - {S^{{\text{Ge}}}},\forall t \in {S^{{\text{TP}}}} \qquad \qquad \qquad\quad
  \end{equation}
  \begin{equation}\label{DOPFGeneConc}
  {\bar P}_{i }^{\text{G},\min} \leq P_{i,t}^{\text{G}} \leq {\bar P}_{i }^{\text{G},\max},\forall i \in {S^{{\text{Ge}}}},\forall t \in {S^{{\text{TP}}}}{\text{ }}  \qquad\,\,\,
  \end{equation}
  \begin{equation}\label{DOPFGeneCond}
  P_{i,t}^{\text{G}} - P_{i,t - 1}^{\text{G}} \leq P_i^{{\text{G,up}}},\forall i \in {S^{{\text{Ge}}}},\forall t \in {S^{{\text{TP}}}}  \qquad \qquad \,\,
  \end{equation}
  \begin{equation}\label{DOPFGeneCone}
  P_{i,t - 1}^{\text{G}} - P_{i,t}^{\text{G}} \leq P_i^{{\text{G,down}}},\forall i \in {S^{{\text{Ge}}}},\forall t \in {S^{{\text{TP}}}} \qquad \quad\,\,\,
  \end{equation}
  \begin{equation}\label{DOPFGeneConf}
  {0 \leq \Delta {E_b} \leq \Delta E_b^{\max },0 \leq \Delta {E_s} \leq \Delta E_s^{\max }} \qquad \qquad \,\,
  \end{equation}
  \begin{equation}\label{DOPFGeneCong}
  \sum\nolimits_{i \in {S^{{\text{Ge}}}}} {\sum\nolimits_{t \in {S^{{\text{TP}}}}} {{E_i}\left( {P_{i,t}^{\text{G}}} \right)} }  \leq {E_0} + \Delta {E_b} - \Delta {E_s} \,\,\,
  \end{equation}
  \begin{equation}\label{DOPFAngCona}
   {\text{ }}\left| {\frac{{{\theta _{i,t}} - {\theta _{j,t}}}}{{{X_{ij}}}}} \right| \leq F_{ij}^{\max },\forall t \in {S^{{\text{TP}}}},\forall ij \in {S^{{\text{Line}}}}  \qquad\qquad
  \end{equation}
  \begin{equation}\label{DOPFAngConb}
  {\theta _{i,t}} = 0,\forall i \in {S^{{\text{Ref}}}}{\text{ }},\forall t \in {S^{{\text{TP}}}} \qquad \qquad \qquad \qquad \quad \,\,\,\,\,\,
  \end{equation}
\end{subequations}
In detail, \eqref{DOPFGeneCona} \eqref{DOPFGeneCona} handles nodal power balance, \eqref{DOPFGeneConz} imposes the global transmission constraint, and \eqref{DOPFGeneConb} addresses non-generator buses. Generator capacities are limited by \eqref{DOPFGeneConc}, while ramp-up/down constraints are given in \eqref{DOPFGeneCond}/\eqref{DOPFGeneCone}. Constraints \eqref{DOPFGeneConf} and \eqref{DOPFGeneCong} respectively bound the tradable emission allowances and enforce the globally coupled carbon emission limit. Line capacity and reference bus requirements are captured by \eqref{DOPFAngCona} and \eqref{DOPFAngConb}. Space constraints preclude listing all symbol definitions, the reader is referred to \cite{Yang2020DCDOPF}. Decoupling \eqref{DOPFGeneCona} and \eqref{DOPFAngCona} requires the introduction of a local variable ${\theta _{ij}}$ to estimate ${\theta _j}$. Each bus $i$ stores the phase angle variable ${\theta _{{\mathcal{N}_i}}} = col\{ {{{\{ {{\theta _{ij}}} \}}_{j \in {\mathcal{N}_i}}},{\theta _{ii}}} \} \in {\mathbb{R}^{\left( {\left| {{\mathcal{N}_i}} \right| + 1} \right)T}}$, with ${\theta _{ii}}$ being its local phase angle ${\theta _{i}}$. The decision variable for bus $i$ is defined as ${x_i} = col\{ {P_i^{\rm{G}},{\theta _{{\mathcal{N}_i}}},\Delta {E_i}} \} \in {\mathbb{R}^{{n_i}}}$, where ${n_i} = ( {| {{\mathcal{N}_i}} | + 2} )T + 2$ and $\Delta {E_i}$ estimates $\Delta {E}$. This allows the DC-DOPF objective to be recast as
\begin{flalign}\label{DOPFFi}
{f_i}( {{x_i}} ) = \sum\nolimits_{t \in {S^{{\text{TP}}}}} {C_i^{{\text{fuel}}}(P_{i,t}^{\text{G}})}  + \frac{1}{m}{\pi _{{\text{bs}}}}\Delta {E_i}
\end{flalign}
with ${\pi _{{\text{bs}}}} = [{\pi _{{\text{bou}}}}, - {\pi _{{\text{sold}}}}]$. By defining linear mappings ${S_{{\text{P}}{{\text{G}}_i}}}:{x_i} \to P_i^{\text{G}}$, ${S_{{\theta _{{\mathcal{N}_i}}}}}:{x_i} \to {\theta _{{\mathcal{N}_i}}}$, ${S_{\Delta {E_i}}}:{x_i} \to \Delta {E_i}$ and ${\tilde B_i}:{\theta _{{\mathcal{N}_i}}} \to \sum\nolimits_{j = 1}^m {{B_{ij}}{\theta _{ij}}} $, \eqref{DOPFGeneCona} takes the local form
\begin{flalign}\label{DOPFhi}
h({x_i}) = \left[ {\begin{array}{*{20}{c}}
   {{S_{{\text{P}}{{\text{G}}_i}}} - {{\tilde B}_i}{S_{{\theta _{{\mathcal{N}_i}}}}}}  \\
   {{{\tilde B}_i}{S_{{\theta _{{\mathcal{N}_i}}}}} - {S_{{\text{P}}{{\text{G}}_i}}}}  \\
 \end{array} } \right] \cdot {x_i} + \left[ {\begin{array}{*{20}{c}}
   { - P_i^{\text{D}}}  \\
   {P_i^{\text{D}}}  \\
 \end{array} } \right] \leq 0
\end{flalign}
where $P_i^{\text{D}} = col\{ P_{i,1}^{\text{D}}, \cdots ,P_{i,T}^{\text{D}}\}  \in {\mathbb{R}^T}$. Meanwhile, \eqref{DOPFGeneConz} decouples to $\sum\nolimits_{i \in {S^{{\text{Bus}}}}} {{{ \mathcal{\tilde B}}_i}{S_{{\theta _{{\mathcal{N}_i}}}}}{x_i}}  = 0$. Setting $P_i^{{\text{G,min}}} = 0,P_i^{{\text{G,max}}} = 0$ if $i \in {S^{{\text{Bus}}}} - {S^{{\text{Ge}}}}$, and $P_i^{{\text{G,min}}} = \bar P_i^{{\text{G,min}}} \otimes {1_T},P_i^{{\text{G,max}}} = \bar P_i^{{\text{G,max}}} \otimes {1_T}$ for $i \in {S^{{\text{Ge}}}}$, we can merge \eqref{DOPFGeneConb} and \eqref{DOPFGeneConc} into $P_i^{{\text{G,min}}} \leq P_i^{\text{G}} \leq P_i^{{\text{G,max}}},\forall i \in {S^{{\text{Bus}}}}$. Similarly, \eqref{DOPFGeneConf} simplifies to $0 \leq \Delta {E_i} \leq \Delta {E^{\max }}$ with $\Delta {E^{\max }} = col\{ \Delta E_b^{\max },\Delta E_s^{\max }\} $. Introducing the mapping ${M_{{\text{ramp}}}}:P_i^{\text{G}} \to col\{ P_{i,1}^{\text{G}} - P_{i,2}^{\text{G}},P_{i,2}^{\text{G}} - P_{i,3}^{\text{G}} \cdots ,P_{i,T - 1}^{\text{G}} - P_{i,T}^{\text{G}}\} $, we can rewrite \eqref{DOPFGeneCond} and \eqref{DOPFGeneCone} as $ - P_i^{{\text{G,down}}} \otimes {1_{T - 1}} \leq {M_{{\text{ramp}}}}P_i^{\text{G}} \leq P_i^{{\text{G,up}}} \otimes {1_{T - 1}}$ for $i \in {S^{{\text{Bus}}}}$. Letting $F_{{X_i}}^{\max } = stack{\{ {F_{ij}^{\max }{X_{ij}}} \}_{j \in {\mathcal{N}_i}}},{\tilde L_i} = stack{\{ {{{\tilde L}_{ij}}} \}_{j \in {\mathcal{N}_i}}}$ and $F_{{X_i}}^{\max } = stack{\{ {F_{ij}^{\max }{X_{ij}}} \}_{j \in {\mathcal{N}_i}}},{\tilde L_i} = stack{\{ {{{\tilde L}_{ij}}} \}_{j \in {\mathcal{N}_i}}}$ with the linear mapping ${\tilde L_{ij}}:{\theta _{{\mathcal{N}_i}}} \to {\theta _{ii}} - {\theta _{ij}}$, constraint \eqref{DOPFAngCona} is reformulated as $ - F_{{X_i}}^{\max } \leq {\tilde L_i}{\theta _{{\mathcal{N}_i}}} \leq F_{{X_i}}^{\max }$ for $i \in {S^{{\text{Bus}}}}$. By further defining ${S_{ii}}:{\theta _{{\mathcal{N}_i}}} \to {\theta _{ii}}$ and letting $\theta _i^{\min } = 0,\theta _i^{\max } = 0$ for $i \in {S^{{\text{Ref}}}}$, \eqref{DOPFAngConb} is compactly expressed as $\theta _i^{\min } \leq {S_{ii}}{\theta _{{\mathcal{N}_i}}} \leq \theta _i^{\max }$. Utilizing the local profiles ${\theta _{ij}}$ and $\Delta {E_i}$, the linear mapping is constructed as
\begin{flalign}\label{DOPFBij}
{B_{ij}} = stack\{ {{\mathcal{\tilde A}}_{ij}}{S_{{\theta _{{\mathcal{N}_i}}}}},{{\mathcal{\bar A}}_{ij}}{S_{\Delta {E_i}}}\}
\end{flalign}
where ${{\mathcal{\tilde A}}_{ij}}{\theta _{{\mathcal{N}_i}}} = col\{ {\theta _{ii}}, - {\theta _{ij}}\} $ for $i<j$ and ${{ \mathcal{\tilde A}}_{ij}}{\theta _{{\mathcal{N}_i}}} = col\{  - {\theta _{ij}},{\text{ }}{\theta _{ii}}\} $ otherwise, while ${{\mathcal{\bar A}}_{ij}}\Delta {E_i} = \Delta {E_i}$ for $i<j$ and ${{\mathcal{\bar A}}_{ij}}\Delta {E_i} =  - \Delta {E_i}$ otherwise. With these, we define ${\mathcal{Z}_i} = \{ {y_i} \in {\mathbb{R}^{\left( {\left| {{\mathcal{N}_i}} \right| + 3} \right)T + 1}}|x_i^{\min } \leq {y_i} \leq x_i^{\max }\} $ where $x_i^{\min } = col\{  - F_{{X_i}}^{\max },\theta _i^{\min },P_i^{{\text{G,min}}}, - P_i^{{\text{G,down}}} \otimes {1_{T - 1}},0\} $, $x_i^{\max } = col\{ F_{{X_i}}^{\max },\theta _i^{\max },P_i^{{\text{G,max}}},P_i^{{\text{G,up}}} \otimes {1_{T - 1}}{\text{ }},\Delta {E^{\max }}\} $ and ${R_i} = stack\{ {\tilde L_i}{S_{{\theta _{{\mathcal{N}_i}}}}},{S_{ii}}{S_{{\theta _{{\mathcal{N}_i}}}}},{S_{{\text{P}}{{\text{G}}_i}}},{M_{{\text{ramp}}}}{S_{{\text{P}}{{\text{G}}_i}}},{S_{\Delta {E_i}}}\} $. This yields
\begin{flalign}\label{DOPFgi}
{g_i}\left( {{R_i}{x_i}} \right) = {\mathcal{I}_{{\mathcal{Z}_i}}}\left( {{R_i}{x_i}} \right)
\end{flalign}
We specify the function
\begin{flalign}\label{DOPFCi}
{C_i}\left( {{x_i}} \right) = \sum\nolimits_{t \in {S^{{\text{TP}}}}} {{E_i}(P_{i,t}^{\text{G}})}  + ([ - 1{\text{ }}1] \cdot \Delta {E_i} - {E_0})/m
\end{flalign}
for $i \in {S^{{\text{Ge}}}}$, and ${C_i}\left( {{x_i}} \right) = ([ - 1{\text{ }}1] \cdot \Delta {E_i} - {E_0})/m$ for $i \in {S^{{\text{Bus}}}} - {S^{{\text{Ge}}}}$. Ultimately, through \eqref{DOPFFi}, \eqref{DOPFhi}, \eqref{DOPFgi}, \eqref{DOPFCi}, and ${A_i} = {{ \mathcal{\tilde B}}_i}{S_{{\theta _{{\mathcal{N}_i}}}}},{a_i} = 0$, the DC-DOPF problem conforms to our general framework \eqref{OrigProblem}.

\emph{2) Multi-Vehicle Formation Stabilization Problem (MVFS) \cite{Wang2026Globally}}: We now formulate the MVFS problem, which aims to cooperatively maneuver  $m$ planar vehicles to a target point. Let the state of vehicle $i$ at time $t$ be ${{\rm{\varsigma }}_{i,t}} = col\{ p_{i,t}^o,{v_{i,t}}\}$ (position and velocity) and its control input be ${\mu _{i,t}}$. The finite-horizon optimization over $t = 0, \cdots ,T$ is defined for all $i = 1, \cdots ,m$ as:
\begin{subequations}\label{MVFSProblem}
  \begin{equation}\label{MVFSProblema}
   \mathop {{\mathop{\rm minimize}\nolimits} }\limits_{\scriptstyle {{\rm{\varsigma }}_1}, \cdots ,{{\rm{\varsigma }}_m} \hfill \atop
  \scriptstyle {\mu _1}, \cdots ,{\mu _m} \hfill} \sum\limits_{t = 1}^{T - 1} {\sum\limits_{i = 1}^m {{{\tilde f}_{i,t}}\left( {{\mu _i},{\rm{\varsigma }}} \right)} }  \qquad\,\,
  \end{equation}
  \begin{equation}\label{MVFSProblemb}
   {\text{subject to  }} {{\rm{\varsigma }}_{i,t + 1}} = {\Phi _i}{{\rm{\varsigma }}_{i,t}} + {\Delta _i}{\mu _{i,t}}
  \end{equation}
  \begin{equation}\label{MVFSProblemc}
  {{\rm{\varsigma }}_{i,t}} \in {\cal X}_i^{\rm{\varsigma }},{\mu _{i,t}} \in {\cal X}_i^\mu \quad\,\,\,\,\,
  \end{equation}
  \begin{equation}\label{MVFSProblemd}
   \sum\limits_{i = 1}^m {\left[ {\begin{array}{*{20}{c}}
   {A_i^{\rm{\varsigma }}{{\rm{\varsigma }}_{i,t}} - a_i^{\rm{\varsigma }}}  \\
   {A_i^\mu {\mu _{i,t}} - a_i^\mu }  \\
\end{array}} \right]}  \le {\rm{0}}
  \end{equation}
\end{subequations}
In the above formulation, the stacked variables are ${\rm{\varsigma }} = col\{ {{\rm{\varsigma }}_1}, \cdots ,{{\rm{\varsigma }}_m}\} $, ${{\rm{\varsigma }}_i} = col\{ {{\rm{\varsigma }}_{i,1}}, \cdots ,{{\rm{\varsigma }}_{i,T - 1}}\} $, ${\mu _i} = col\{ {\mu _{i,0}}, \cdots ,{\mu _{i,T - 1}}\} $. The stage cost is given by ${{\tilde f}_{i,t}}( {{\mu _i},{\rm{\varsigma }}} ) = {\vartheta _{i,v}}{\| {{v_{i,t}}} \|^2} + {\vartheta _{i,\mu }}{\| {{\mu _{i,t}}} \|^2} + {\vartheta _{i,p}}\sum\nolimits_{j \in {{\cal N}_i}} {{{\| {p_{i,t}^o - p_{j,t}^o + {d_{ij}}} \|}^2}} $, where ${{d_{ij}}}$ denotes the target equilibrium spacing between vehicles $i$ and $j$. While constraint \eqref{MVFSProblemb} describes the vehicle dynamics with matrices ${\Phi _i}$ and ${\Delta _i}$, constraint \eqref{MVFSProblemc} imposes local state/input bounds. Constraint \eqref{MVFSProblemd} captures both global resource limits and network-wide kinematic requirements. To fully decouple the problem \eqref{MVFSProblem}, we introduce $p_{ij,t}^o$ to estimate ${p_{j,t}^o}$, denoting the self-variables as ${{\rm{\varsigma }}_{ii,t}},{\mu _{ii,t}}$. We then construct the augmented private variable ${x_i} = col\{ {{\rm{\varsigma }}_{{{\cal N}_i}}},{\mu _{ii}}\} $ with ${{\rm{\varsigma }}_{{{\cal N}_i}}} = col{\{ {{\rm{\varsigma }}_{ij}}\} _{j \in {{\cal N}_i} \cup \{ i\} }}$, and the linear mapping ${R_i}:{x_i} \to col\{ {{\rm{\varsigma }}_{ii}},{\mu _{ii}}\} $. This allows \eqref{MVFSProblemb} to be rewritten as ${\mathfrak{A}_i}{R_i}{x_i} = {\upsilon _i}$, and equivalently as $h\left( {{x_i}} \right) = [{\mathfrak{A}_i}{R_i}; - {\mathfrak{A}_i}{R_i}] \cdot {x_i} + col\{  - {\upsilon _i},{\upsilon _i}\} $. Here ${\mathfrak{A}_i}$ and ${\upsilon _i}$ are obtained from ${\Phi _i},{\Delta _i}$ and initial conditions. The local bounds \eqref{MVFSProblemc} are transformed into an indicator function ${g_i}( {{R_i}{x_i}} ) = {\mathcal{I}_{{\mathcal{X}_i}}}( {{R_i}{x_i}} )$, where ${\mathcal{X}_i}$ is determined by ${\cal X}_i^{\rm{\varsigma }}$ and $\mathcal{X}_i^\mu $. Meanwhile, \eqref{MVFSProblemd} is represented via $C_i(x_i)$ as defined in \eqref{OrigProblem}. By specifying the local objective as ${f_i}( {{x_i}} ) = \sum\nolimits_{t = 0}^{T - 1} {( {\vartheta _{i,p}}\sum\nolimits_{j \in {{\cal N}_i}} {{{|| {p_{ii,t}^o - p_{ij,t}^o + {d_{ij}}} ||}^2}} } + {{\vartheta _{i,v}}{{\| {{v_{ii,t}}} \|}^2} + {\vartheta _{i,\mu }}{{\| {{\mu _{ii,t}}} \|}^2}  )} $, the MVFS problem aligns perfectly with the proposed general framework \eqref{OrigProblem}.

\begin{remark} \label{RemarProbel}
A wide variety of optimization problems can be formulated as special cases of \eqref{OrigProblem}. For instant, the consensus (nonsmooth) problems without local constraints \cite{Jakovetic2019Unification,Li2019NIDS,Bastianello2021Asynchronous,Nguyen2023Accelerated} emerge when ${R_i} = I$, ${B_{ij}} = I$, ${B_{ji}} = -I$, $b_{(i,j)} = 0$, while simultaneously eliminating \eqref{OrigProblemc} and \eqref{OrigProblemd}. Similarly, distributed economic dispatch and resource allocation models \cite{Sherson2019Separable,Liu2021Region,Carli2020Distributed,Alaviani2025Parallel,Falsone2020Tracking} can be derived by omitting \eqref{OrigProblemb} and \eqref{OrigProblemc}. When setting ${B_{ij}} = I$, ${B_{ji}} = -I$ and $b_{(i,j)} = 0$, the framework \eqref{OrigProblem} captures consensus-based optimization featuring globally coupling constraints \cite{Huang2025Unified,Huang2023Multiproximal,Lin2021Compensation}. Furthermore, recent edge-coupled optimization formulations \cite{Li2022TPUS,Latafat2019TriPD} fall within this framework \eqref{OrigProblem} without consideration of \eqref{OrigProblemc} and \eqref{OrigProblemd}. In addition, the inclusion of two practical examples (i.e., DC-DOPF and MVFS) verifies that \eqref{OrigProblem} is not a mere mathematical abstraction but is well grounded in practical applications. Although existing literature \cite{Bastianello2021Asynchronous,Huang2025Unified,Alaviani2025Parallel,Lin2021Compensation} has explored related domains, critical difficulties persist in managing multi-different local constraints, nonseparable costs (see \eqref{MVFSProblema}), and coupled local constraints (see \eqref{DOPFAngCona} and \eqref{DOPFGeneCona}). Unlike approaches that treat local structures as purely nonsmooth terms \cite{Li2019NIDS,Bastianello2021Asynchronous,Alaviani2025Parallel,Huang2023Multiproximal}, our inclusion of the edge-based constraint \eqref{OrigProblemb} and the mapping $R_i$ in \eqref{OrigProblema} substantially empowers the model to handle composite constraints and streamlines subsequent algorithmic iterations. Therefore, \eqref{OrigProblem} constitutes a more comprehensive framework than existing structures.
\end{remark}
}

\subsection{Local First-order Condition}

For the local inequality constraint \eqref{RefProblemb} and the global coupled one \eqref{RefProblemc}, we introduce dual profiles ${u_i} \in \mathbb{R}_ + ^{{p_i}}$ and ${s_g} \in {\Omega _{{\rm{dual}}}}$ with ${\Omega _{{\rm{dual}}}} = {\mathbb{R}^q} \times \mathbb{R}_ + ^c$, respectively. By setting $\Theta  = \prod\nolimits_{i = 1}^m {\mathbb{R}_ + ^{{p_i}}} $, $R = blkdiag\left\{ {{R_1}, \cdots ,{R_m}} \right\}$, $f\left( x \right) = \sum\nolimits_{i = 1}^m {{f_i}\left( {{x_i}} \right)} $, $g\left( {Rx} \right) = \sum\nolimits_{i = 1}^m {{g_i}\left( {{R_i}{x_i}} \right)} $, and $h\left( x \right) = col\left\{ {{h_1}\left( {{x_1}} \right), \cdots ,{h_m}\left( {{x_m}} \right)} \right\}$, ${x^ * } = col\left\{ {x_1^ * , \cdots ,x_m^ * } \right\}$ solves problem \eqref{RefProblem} under Assumption \ref{AssumpNonempty} if and only if ${x^ * },{u^ * } = col\left\{ {u_1^ * , \cdots ,u_m^ * } \right\},s_g^ * $ satisfy the following KKT conditions:
\begin{subequations}\label{OrigKKT}
  \begin{equation}\label{OrigKKTa}
   0 \in \mathcal{\tilde Q}({x^ * },{u^ * },s_g^ * ) \qquad \qquad \quad\,\,\,
  \end{equation}
  \begin{equation}\label{OrigKKTb}
   0 \in h\left( {{x^ * }} \right) - \partial {\mathcal{I}_\Theta }\left( {{u^ * }} \right)  \qquad \quad
  \end{equation}
  \begin{equation}\label{OrigKKTc}
   0 \in \sum\limits_{i = 1}^m {{\mathcal{C}_i}\left( {x_i^ * } \right)}  - \partial {\mathcal{I}_{{\Omega _{{\text{dual}}}}}}\left( {s_g^ * } \right)
  \end{equation}
\end{subequations}
where $\mathcal{\tilde Q}:({x^ * },{u^ * },s_g^ * ) \to \partial f({x^ * }) + {R^{\text{T}}}\partial g(R{x^ * }) + {{\tilde \Pi }^{\text{T}}}\partial {\mathcal{I}_{\tilde D}}(\tilde \Pi {x^ * }) + {(\partial h({x^ * }))^{\text{T}}}{u^ * } + {(\partial \mathcal{C}({x^ * }))^{\text{T}}}({1_m} \otimes s_g^ * )$ with $\partial \mathcal{C}\left( x \right) = col\left\{ {\partial {\mathcal{C}_1}\left( {{x_1}} \right), \cdots ,\partial {\mathcal{C}_m}\left( {{x_m}} \right)} \right\}$. {While proximal operators are effective tools for handling nonsmooth objectives \cite{Li2019NIDS,Huang2025Unified,Hamedani2022Strongly,Wu2023Distributed}, their direct application to designing distributed algorithms based on the KKT system \eqref{OrigKKT} is hindered by three technical obstacles. First, even though the proximal mapping for $g(x)$ is readily computable \cite{Li2019NIDS,Huang2025Unified,Hamedani2022Strongly,Wu2023Distributed}, a straightforward proximal expression is unattainable for the composite term $R^{\mathrm{T}}\partial g(Rx)$, due to the absence of a closed-form solution for the proximal mapping of $g_i\circ R_i$ under a generic linear operator $R_i$. Second, the term $\tilde{\Pi}^{\mathrm{T}}\partial\mathcal{I}_{\tilde D}(\tilde{\Pi}x)$ is nonseparable across agents, given that \eqref{OrigProblemb} couples the variables of neighboring agents. Third, the global coupling constraint \eqref{OrigProblemd}  prevents a direct decomposition of the dual condition into local updates. To navigate around these obstacles, a conjugate decomposition is formulated below, serving to convert the nonsmooth components in \eqref{OrigKKT} into forms that are both locally separable and amenable to proximal operations.

Under Assumption \ref{AssumpNonempty}, the convexity of $g_i$ implies the convexity of its conjugate setting $g_i^ \star ({w_i}) = {\sup _{{y_i} \in {\mathbb{R}^{{{\tilde r}_i}}}}}\{ \left\langle {{y_i},{w_i}} \right\rangle  - {g_i}({y_i})\} $ \cite[Proposition 13.13]{Bauschke2011Convex}. Hence, for each} $\partial {g_i}({R_i}x_i^ * )$ in ${R^{\text{T}}}\partial g(R{x^ * }) = col\{ R_1^{\rm T}\partial {g_1}({R_1}x_1^ * ), \cdots ,R_m^{\rm T}\partial {g_m}({R_m}x_m^ * )\} $, there exists a $w_i^ * $ such that $w_i^ *  \in \partial {g_i}\left( {y_i^ * } \right){|_{y_i^ *  = {R_i}x_i^ * }}$. Then, one follows from \cite[Theorem 16.29]{Bauschke2011Convex} that $w_i^ *  \in \partial {g_i}(y_i^ * ){|_{y_i^ *  = {R_i}x_i^ * }} \Leftrightarrow y_i^ *  \in {(\partial {g_i})^{ - 1}}w_i^ *  \Leftrightarrow 0 \in (\partial g_i^ \star (w_i^ * ) - y_i^ * ){|_{y_i^ *  = {R_i}x_i^ * }}$, meaning that ${w^ * } \in \partial g\left( {R{x^ * }} \right) \Leftrightarrow 0 \in \partial {g^ \star }\left( {{w^ * }} \right) - R{x^ * }$, where ${w^ * } = col\{ w_1^ * , \cdots ,w_m^ * \} $, ${g^ \star }({w^ * }) = \sum\nolimits_{i = 1}^m {g_i^ \star (w_i^ * )} $. Therefore, ${R^{\rm{T}}}\partial g(R{x^ * })$ in \eqref{OrigKKTa} is equivalently decomposed into
\begin{flalign}\label{Gxstar}
{R^{\rm{T}}}\partial g(R{x^ * }) \Leftrightarrow \left\{ \begin{gathered}
  {R^{\rm{T}}}{w^ * } \hfill \\
  {w^ * } \in \partial g(R{x^ * }) \hfill \\
\end{gathered}  \right. \Leftrightarrow \left\{ \begin{gathered}
  {R^{\rm{T}}}{w^ * } \hfill \\
  0 \in \partial {g^ \star }({w^ * }) - R{x^ * } \hfill \\
\end{gathered}  \right.
\end{flalign}
On { the other hand, the set ${\tilde D_{\left( {i,j} \right)}}$ for each $\left( {i,j} \right) \in \mathcal{E}$ is nonempty, closed and convex under Assumption \ref{AssumpNonempty}, which implies the proper, closed, convex properties of its indicator function ${\mathcal{I}_{{{\tilde D}_{\left( {i,j} \right)}}}}$ \cite[Theorem 8.3]{Bauschke2011Convex}. Then, the conjugate function $\mathcal{I}_{{{\tilde D}_{\left( {i,j} \right)}}}^ \star ({\lambda _{\left( {i,j} \right)}}) = {\sup _{{y_{\left( {i,j} \right)}} \in {\mathbb{R}^{2{l_{\left( {i,j} \right)}}}}}}\{ \langle {{y_{\left( {i,j} \right)}},{\lambda _{\left( {i,j} \right)}}} \rangle  - {\mathcal{I}_{{{\tilde D}_{\left( {i,j} \right)}}}}({y_{\left( {i,j} \right)}})\} $ is convex \cite[Proposition 13.13]{Bauschke2011Convex}. Hence, for each} $\tilde \Pi _{\left( {i,j} \right)}^{\rm T}\partial {\mathcal{I}_{{{\tilde D}_{\left( {i,j} \right)}}}}({\tilde \Pi _{\left( {i,j} \right)}}{x^ * })$ in ${\tilde \Pi ^{\text{T}}}\partial {\mathcal{I}_{\tilde D}}(\tilde \Pi {x^ * }) = col{\{ \tilde \Pi _{\left( {i,j} \right)}^{\rm T}\partial {\mathcal{I}_{{{\tilde D}_{\left( {i,j} \right)}}}}({\tilde \Pi _{\left( {i,j} \right)}}{x^ * })\} _{\left( {i,j} \right) \in \mathcal{E}}}$, there exists $\lambda _{\left( {i,j} \right)}^ *  = col\{ \lambda _{\left( {i,j} \right),i}^ * ,\lambda _{\left( {i,j} \right),j}^ * \} $ such that $\lambda _{\left( {i,j} \right)}^ *  \in \partial {\mathcal{I}_{{{\tilde D}_{\left( {i,j} \right)}}}}(y_{\left( {i,j} \right)}^ * ){|_{y_{\left( {i,j} \right)}^ *  = {{\tilde \Pi }_{\left( {i,j} \right)}}{x^ * }}}$, and it has the similar equivalence $\lambda _{\left( {i,j} \right)}^ *  \in \partial {\mathcal{I}_{{{\tilde D}_{\left( {i,j} \right)}}}}({\tilde \Pi _{\left( {i,j} \right)}}{x^ * }) \Leftrightarrow 0 \in \partial \mathcal{I}_{{{\tilde D}_{\left( {i,j} \right)}}}^ \star (\lambda _{\left( {i,j} \right)}^ * ) - {\tilde \Pi _{\left( {i,j} \right)}}{x^ * }$. {Considering $\mathcal{I}_{\tilde D}^ \star ({\lambda ^*}) = \sum\nolimits_{\left( {i,j} \right) \in \mathcal{E}} {\mathcal{I}_{{{\tilde D}_{\left( {i,j} \right)}}}^ \star (\lambda _{\left( {i,j} \right)}^*)}$} with ${\lambda ^ * } = col{\{ \lambda _{\left( {i,j} \right)}^ * \} _{\left( {i,j} \right) \in \mathcal{E}}}$, we thus have the following relationship associated with ${\tilde \Pi ^{\text{T}}}\partial {\mathcal{I}_{\tilde D}}(\tilde \Pi {x^ * })$:
\begin{flalign}\label{Ipaixstar}
{\tilde \Pi ^{\text{T}}}\partial {\mathcal{I}_{\tilde D}}(\tilde \Pi {x^ * }) \! \Leftrightarrow \! \left\{ \begin{gathered}
  {{\tilde \Pi }^{\text{T}}}{\lambda ^ * } \hfill \\
  {\lambda ^ * } \! \in \! \partial {\mathcal{I}_{\tilde D}}(\tilde \Pi {x^ * }) \hfill \\
\end{gathered}  \right. \! \Leftrightarrow \! \left\{ \begin{gathered}
  {{\tilde \Pi }^{\text{T}}}{\lambda ^ * } \hfill \\
  0 \! \in \! \partial \mathcal{I}_{\tilde D}^ \star ({\lambda ^ * }) \!-\! \tilde \Pi {x^ * } \hfill \\
\end{gathered}  \right.
\end{flalign}
Summarizing the formulations \eqref{Gxstar} and \eqref{Ipaixstar}, the condition system \eqref{OrigKKT} takes the following equivalent form of
\begin{subequations}\label{EquivaKKT}
  \begin{equation}\label{EquivaKKTa}
   0 \in \mathcal{\bar Q}({x^ * },{w^ * },{\lambda ^ * },{u^ * },s_g^ * )  \qquad
  \end{equation}
  \begin{equation}\label{EquivaKKTb}
   0 \in \partial {g^ \star }\left( {{w^ * }} \right) - R{x^ * }  \qquad\qquad
  \end{equation}
  \begin{equation}\label{EquivaKKTc}
   0 \in \partial \mathcal{I}_{\tilde D}^ \star \left( {{\lambda ^ * }} \right) - \tilde \Pi {x^ * } \qquad\qquad
  \end{equation}
  \begin{equation}\label{EquivaKKTd}
  0 \in \partial {\mathcal{I}_\Theta }\left( {{u^ * }} \right) - h\left( {{x^ * }} \right) \qquad\quad
  \end{equation}
  \begin{equation}\label{EquivaKKTe}
  0 \in \sum\limits_{i = 1}^m {{\mathcal{C}_i}\left( {x_i^ * } \right)}  - \partial {\mathcal{I}_{{\Omega _{{\text{dual}}}}}}\left( {s_g^ * } \right)
  \end{equation}
\end{subequations}
where $\mathcal{\bar Q}({x^ * },{w^ * },{\lambda ^ * },{u^ * },s_g^ * ) \to \partial f({x^ * }) + {R^{\text{T}}}{w^ * } + {{\tilde \Pi }^{\text{T}}}{\lambda ^ * } + {(\partial h({x^ * }))^{\text{T}}}{u^ * } + {(\partial \mathcal{C}({x^ * }))^{\text{T}}}({1_m} \otimes s_g^ * )$. To further decouple \eqref{EquivaKKTe}, we introduce the variable $s = col\left\{ {{s_1}, \cdots ,{s_m}} \right\}$ to estimate $s_g$, and construct the following edge-based consensus constraint:
\begin{flalign}\label{Edgeconss}
{E_{ij}}{s_i} + {E_{ji}}{s_j} = 0,{\text{ }}\left( {i,j} \right) \in \mathcal{E}
\end{flalign}
where $E_{ij} = I$ if $i<j$, and $E_{ij} = -I$ for otherwise. Similar to \eqref{Edgemaps} and \eqref{Edgeconst}, we define the mapping $\bar \Pi :s \to {( {{{\bar \Pi }_{\left( {i,j} \right)}}s} )_{\left( {i,j} \right) \in \mathcal{E}}}$ with ${\bar \Pi _{\left( {i,j} \right)}}:{\text{ }}s \to \left( {{E_{ij}}{s_i},{\text{ }}{E_{ji}}{s_j}} \right)$, and the set ${\bar D_{\left( {i,j} \right)}} = \{ {\left( {{z_1},{z_2}} \right) \in {\mathbb{R}^{\left( {c + q} \right)}} \times {\mathbb{R}^{\left( {c + q} \right)}}\left| {{z_1} + {z_2} = 0} \right.} \}$. Moreover, considering the instrumental variable $z = col{\{ {{z_{\left( {i,j} \right)}}} \}_{\left( {i,j} \right) \in \mathcal{E}}}$ with ${z_{\left( {i,j} \right)}} = col\{ {{z_{\left( {i,j} \right),i}},{z_{\left( {i,j} \right),j}}} \} \in {\mathbb{R}^{2( {c + q} )}}$, we can move the condition \eqref{EquivaKKT} of solving problem \eqref{OrigProblem} to a new extended first-order formulation via the following lemma.

\begin{lemma}\label{DistriMN}
Let Assumption \ref{AssumpNonempty} hold. Consider a point $\left( {{x^ * },{w^ * },{\lambda ^ * },{u^ * },{s^ * },{z^ * }} \right)$, where ${x^ * } = col\left\{ {x_1^ * , \cdots ,x_m^ * } \right\}$, ${w^ * } = col\left\{ {w_1^ * , \cdots ,w_m^ * } \right\}$, ${s^ * } = col\left\{ {s_1^ * , \cdots ,s_m^ * } \right\}$, ${u^ * } = col\left\{ {u_1^ * , \cdots ,u_m^ * } \right\}$, ${\lambda ^ * } = col{\{ {\lambda _{( {i,j} )}^ * } \}_{( {i,j} ) \in \mathcal{E}}}$ with $\lambda _{( {i,j} )}^ *  = col\{ {\lambda _{\left( {i,j} \right),i}^ * ,\lambda _{( {i,j} ),j}^ * } \}$, and ${z^ * } = col{\{ {z_{( {i,j} )}^ * } \}_{( {i,j} ) \in \mathcal{E}}}$ with $z_{( {i,j} )}^ *  = col\{ {z_{( {i,j} ),i}^ * ,z_{( {i,j} ),j}^ * } \}$. There exists $s_g^ * $ such that $s_1^ *  =  \cdots  = s_m^ *  = s_g^ * $, and $x^*$ solves problem \eqref{OrigProblem} provided that for each $i \in \mathcal{V}$, the local components $x_i^ * ,w_i^ * ,u_i^ * ,s_i^ * ,\lambda _{\left( {i,j} \right)}^ * ,z_{\left( {i,j} \right)}^ * ,j \in {\mathcal{N}_i}$ satisfy
\begin{subequations}\label{DisMNcondit}
  \begin{equation}\label{DisMNcondita}
   0 \in {\mathcal{A}_i}(x_i^ * ,w_i^ * ,{\{ \lambda _{\left( {i,j} \right),i}^ * \} _{j \in {\mathcal{N}_i}}},u_i^ * ,s_i^ * )  \qquad\qquad
  \end{equation}
  \begin{equation}\label{DisMNconditb}
  0 \in \partial g_i^ \star \left( {w_i^ * } \right) - {R_i}x_i^ *  \qquad\qquad \qquad\qquad \qquad\quad
  \end{equation}
  \begin{equation}\label{DisMNconditc}
   0 \in \partial \mathcal{I}_{{{\tilde D}_{\left( {i,j} \right)}}}^ \star \left( {\lambda _{\left( {i,j} \right)}^ * } \right) - {{\tilde \Pi }_{\left( {i,j} \right)}}{x^ * }  \qquad\qquad \qquad\,\,\,
  \end{equation}
  \begin{equation}\label{DisMNconditd}
  0 \in \partial {\mathcal{I}_{\mathbb{R}_ + ^{{p_i}}}}\left( {u_i^ * } \right) - {h_i}\left( {x_i^ * } \right)  \qquad\qquad \qquad\qquad \quad\,\,\,
  \end{equation}
  \begin{equation}\label{DisMNcondite}
   0 \in \partial {\mathcal{I}_{{\Omega _{{\text{dual}}}}}}\left( {s_i^ * } \right) - {\mathcal{C}_i}\left( {x_i^ * } \right) + \sum\nolimits_{j \in {\mathcal{N}_i}} {E_{ij}^{\rm T}z_{\left( {i,j} \right),i}^ * }
  \end{equation}
  \begin{equation}\label{DisMNconditf}
   0 \in \partial \mathcal{I}_{{{\bar D}_{\left( {i,j} \right)}}}^ \star \left( {z_{\left( {i,j} \right)}^ * } \right) - {{\bar \Pi }_{\left( {i,j} \right)}}{s^ * }  \qquad\qquad \qquad\quad
  \end{equation}
\end{subequations}
where ${\mathcal{A}_i}:(x_i^ * ,w_i^ * ,{\{ \lambda _{\left( {i,j} \right),i}^ * \} _{j \in {\mathcal{N}_i}}},u_i^ * ,s_i^ * ) \to \partial {f_i}(x_i^ * ) + R_i^{\rm T}w_i^ *  + \sum\nolimits_{j \in {\mathcal{N}_i}} {B_{ij}^{\rm T}\lambda _{\left( {i,j} \right),i}^ * }  + {(\partial {h_i}(x_i^ * ))^{\rm{T}}}u_i^ *  + {(\partial {\mathcal{C}_i}(x_i^ * ))^{\rm{T}}}s_i^ * $.
\end{lemma}

\begin{proof}
See Appendix \ref{ProofDistriMN}.
\end{proof}

\begin{remark}
{ It is noteworthy that condition \eqref{OrigKKT} characterizes the optimality of problem \eqref{OrigProblem} under Assumption \ref{AssumpNonempty}, with \eqref{EquivaKKT} serving as its conjugate equivalent and \eqref{DisMNcondit} representing its distributed local reformulation. Given that ${\tilde \Pi _{\left( {i,j} \right)}}{x^ * } = col\{ {B_{ij}}x_i^ * ,{B_{ji}}x_j^ * \} $ and ${\bar \Pi _{\left( {i,j} \right)}}{s^ * } = col\{ {E_{ij}}s_i^ * ,{E_{ji}}s_j^ * \} $,  the provided first-order system \eqref{DisMNcondit} is entirely local. This contrasts with Laplacian-based conditions \cite{Huang2023Multiproximal,Liu2021Region,Liang2020Smooth,Huang2025Unified,Lin2021Compensation}, which encounter inherent difficulties in deriving local formulations due to the global coupling nature of Laplacian matrices. Motivated by Remark \ref{RemarProbel}, novel first-order conditions for consensus optimization \cite{Jakovetic2019Unification,Li2019NIDS,Bastianello2021Asynchronous,Nguyen2023Accelerated} and edge-based constraint-coupled optimization \cite{Li2022TPUS,Latafat2019TriPD} correspond to \eqref{DisMNcondita}-\eqref{DisMNconditd} and \eqref{DisMNcondita}-\eqref{DisMNconditc}, respectively. Meanwhile, the combination of \eqref{DisMNcondita}, \eqref{DisMNconditb}, \eqref{DisMNcondite} and \eqref{DisMNconditf} constitutes the condition for distributed resource allocation problems \cite{Sherson2019Separable,Liu2021Region,Carli2020Distributed,Alaviani2025Parallel,Falsone2020Tracking}. Consequently, \eqref{DisMNcondit} can be interpreted as a unified first-order optimality framework for all the aforementioned optimization problems. Beyond that, framework \eqref{DisMNcondit} can be seamlessly generalized to} handle objective functions comprising multiple proximable nonsmooth terms. If $g_i(R_ix_i)$ is represented as the sum of nonsmooth convex functions ${{\tilde g}_{i,1}}({R_{i,1}}{x_i}), \cdots ,{{\tilde g}_{i,\tilde m}}({R_{i,\tilde m}}{x_i})$, meaning ${g_i}( {{R_i}{x_i}} ) = \sum\nolimits_{l = 1}^{\tilde m} {{{\tilde g}_{i,l}}({R_{i,l}}{x_i})} $ with ${R_i} = stack\{ {R_{i,1}}, \cdots ,{R_{i,\tilde m}}\} $, then implies that expression \eqref{DisMNconditb} can be expanded into $\tilde m$ parts. That is $\partial g_i^ \star (w_i^ * ) - {R_i}x_i^ *  = col\{ \partial g_{i,1}^ \star (w_{i,1}^ * ) - {R_{i,1}}x_i^ * , \cdots ,\partial g_{i,\tilde m}^ \star (w_{i,\tilde m}^ * ) - {R_{i,\tilde m}}x_i^ * \} $ with $w_i^ *  = col\{ w_{i,1}^ * , \cdots ,w_{i,\tilde m}^ * \} $. {This crucial enhancement considerably expands the range of solvable problems, overcoming the restrictions found in \cite{Li2019NIDS,Liang2018Modified,Huang2025Unified,Huang2023Multiproximal,Hamedani2022Strongly,Wu2023Distributed}.}
\end{remark}

\subsection{Distributed Algorithm Design} \label{sec:algorithm}
{
In this section, the FDOA is formulated to tackle \eqref{RefProblem}. A key observation driving our design is that existing proximal algorithms \cite{Huang2025Unified,Huang2023Multiproximal} typically demand the proximal mapping ${\text{pro}}{{\text{x}}_{{g_i}}}$ to be explicitly computable. Consequently, they fail to handle the composite structure $g_i(R_ix_i)$ arising in our problem, as the proximal operator for $g_i\circ R_i$ usually lacks a closed-form representation. Additionally, the inherent coupling within the indicator functions $\mathcal{I}_{{{\tilde D}_{\left( {i,j} \right)}}}$ and $\mathcal{I}_{\bar D_{(i,j)}}$ prevents these conventional techniques from yielding fully distributed updates. To overcome these hurdles, we leverage the newly derived local system \eqref{DisMNcondit} to resolve the intractability of these composite nonsmooth terms through conjugate decomposition and proximal splitting. More precisely, an equivalent proximal splitting formulation can be directly derived from the local sufficient system. To be specific, the local sufficient system \eqref{DisMNcondit} can be transformed into the equivalent  proximal splitting formulations, i.e., $w_i^* = {\text{pro}}{{\text{x}}_{{\beta _i}g_i^ \star }}(w_i^* + {\beta _i}{R_i}x_i^*)$, $u_i^* = {\mathcal{P}_{\mathbb{R}_ + ^{{p_i}}}}(u_i^* + {\delta _i}{h_i}(x_i^*))$, $\lambda _{\left( {i,j} \right)}^* = {\text{pro}}{{\text{x}}_{{\chi _{\left( {i,j} \right)}}\mathcal{I}_{{{\tilde D}_{\left( {i,j} \right)}}}^ \star }}(\lambda _{\left( {i,j} \right)}^* + {\chi _{\left( {i,j} \right)}}{{\tilde \Pi }_{\left( {i,j} \right)}}{x^*})$, $z_{\left( {i,j} \right)}^* = {\text{pro}}{{\text{x}}_{{\chi _{\left( {i,j} \right)}}\mathcal{I}_{{{\bar D}_{\left( {i,j} \right)}}}^ \star }}(z_{\left( {i,j} \right)}^* + {\chi _{\left( {i,j} \right)}}{{\bar \Pi }_{\left( {i,j} \right)}}{s^*})$, $s_i^* = {\mathcal{P}_{{\Omega _{{\text{dual}}}}}}(s_i^* - {\eta _i}\sum\nolimits_{j \in {\mathcal{N}_i}} {E_{ij}^{\text{T}}z_{\left( {i,j} \right),i}^*}  + {\eta _i}{\mathcal{C}_i}(x_i^*))$, and $x_i^* = {\text{pro}}{{\text{x}}_{{\alpha _i}{f_i}}}(x_i^* - {\alpha _i}(R_i^{\text{T}}w_i^* + \sum\nolimits_{j \in {\mathcal{N}_i}} {A_{ij}^{\text{T}}\lambda _{\left( {i,j} \right),i}^*} + {(\partial {h_i}(x_i^*))^{\text{T}}}u_i^* + {(\partial {\mathcal{C}_i}(x_i^*))^{\text{T}}}s_i^*))$, where proximal operators ${\text{pro}}{{\text{x}}_{{\delta _i}{\mathcal{I}_{\mathbb{R}_ + ^{{p_i}}}}}}$, ${\text{pro}}{{\text{x}}_{{\eta _i}{\mathcal{I}_{{\Omega _{{\text{dual}}}}}}}}$ are equivalent to the projections ${\mathcal{P}_{{\Omega _{{\text{dual}}}}}}$, ${\mathcal{P}_{\mathbb{R}_ + ^{{p_i}}}}$,} and ${\beta _i},{\chi _{\left( {i,j} \right)}},{\delta _i},{\eta _i},{\alpha _i}$ are positive step-sizes. To reach the above goal, we firstly design the following prediction rules:
\begin{subequations}\label{LocalPredict}
  \begin{equation}\label{LocalPredicta}
   {{\tilde w}_{i,k}} = {\text{pro}}{{\text{x}}_{{\beta _i}g_i^ \star }}({w_{i,k}} + {\beta _i}{R_i}{x_{i,k}}) \qquad\qquad\qquad\qquad
  \end{equation}
  \begin{equation}\label{LocalPredictb}
  {{\tilde \lambda }_{\left( {i,j} \right),k}} = {\text{pro}}{{\text{x}}_{{\chi _{\left( {i,j} \right)}}\mathcal{I}_{{{\tilde D}_{\left( {i,j} \right)}}}^ \star }}({\lambda _{\left( {i,j} \right),k}} + {\chi _{\left( {i,j} \right)}}{{\tilde \Pi }_{\left( {i,j} \right)}}{x_k})
  \end{equation}
  \begin{equation}\label{LocalPredictc}
   {{\tilde z}_{\left( {i,j} \right),k}} = {\text{pro}}{{\text{x}}_{{\chi _{\left( {i,j} \right)}}\mathcal{I}_{{{\bar D}_{\left( {i,j} \right)}}}^ \star }}({z_{\left( {i,j} \right),k}} + {\chi _{\left( {i,j} \right)}}{{\bar \Pi }_{\left( {i,j} \right)}}{s_k}) \,\,\,
  \end{equation}
  \begin{equation}\label{LocalPredictd}
  {{\tilde u}_{i,k}} = {\mathcal{P}_{\mathbb{R}_ + ^{{p_i}}}}({u_{i,k}} + {\delta _i}{h_i}({{\tilde x}_{i,k}})) \qquad\qquad\qquad\qquad\quad
  \end{equation}
  \begin{equation}\label{LocalPredicte}
   \begin{gathered}
  {{\tilde s}_{i,k}} = {\mathcal{P}_{{\Omega _{{\text{dual}}}}}}({s_{i,k}} + {\eta _i}{\mathcal{C}_i}({{\tilde x}_{i,k}}) \hfill \\
  \qquad\qquad\qquad - {\eta _i}\sum\nolimits_{j \in {\mathcal{N}_i}} {E_{ij}^{\rm{T}}(2{{\tilde z}_{\left( {i,j} \right),i,k}} - {z_{\left( {i,j} \right),i,k}})} ) \hfill \\
\end{gathered}
  \end{equation}
\end{subequations}
where ${{\tilde w}_{i,k}},{{\tilde \lambda }_{\left( {i,j} \right),k}} = col\{ {{\tilde \lambda }_{\left( {i,j} \right),i,k}},{{\tilde \lambda }_{\left( {i,j} \right),j,k}}\} ,{{\tilde z}_{\left( {i,j} \right),k}} = col\{ {{\tilde z}_{\left( {i,j} \right),i,k}},{{\tilde z}_{\left( {i,j} \right),j,k}}\} ,{{\tilde u}_{i,k}},{{\tilde s}_{i,k}}$ and ${{\tilde x}_{i,k}}$ are the immediate variables with
\begin{flalign}\label{Localtildx}
\frac{1}
{{{\alpha _i}}}{x_{i,k}} \!-\!\! \frac{1}{{{\alpha _i}}}{{\tilde x}_{i,k}} \! \in \! {\mathcal{A}_i}({{\tilde x}_{i,k}},{{\tilde w}_{i,k}},{\{ {{\tilde \lambda }_{\left( {i,j} \right),i,k}}\} _{j \in {\mathcal{N}_i}}},{{\tilde u}_{i,k}},{{\tilde s}_{i,k}})
\end{flalign}

Employing \cite[Theorem 14.3]{Bauschke2011Convex}, one can derive the equivalent forms of \eqref{LocalPredicta}-\eqref{LocalPredictc}, i.e.,
\begin{flalign}\label{Localtildwi}
{{\tilde w}_{i,k}} = {w_{i,k}} + {\beta _i}{R_i}{x_{i,k}} - {\beta _i}{\text{pro}}{{\text{x}}_{\beta _i^{ - 1}{g_i}}}(\beta _i^{ - 1}{w_{i,k}} + {R_i}{x_{i,k}})
\end{flalign}
and ${{\tilde \lambda }_{\left( {i,j} \right),k}} = {\lambda _{\left( {i,j} \right),k}} + {\chi _{\left( {i,j} \right)}}{{\tilde \Pi }_{\left( {i,j} \right)}}{x_k} - {\chi _{\left( {i,j} \right)}}{\mathcal{P}_{{{\tilde D}_{\left( {i,j} \right)}}}}{\text{(}}\chi _{\left( {i,j} \right)}^{ - 1}{\lambda _{\left( {i,j} \right),k}} + {{\tilde \Pi }_{\left( {i,j} \right)}}{x_k}{\text{)}}$, ${{\tilde z}_{\left( {i,j} \right),k}} = {z_{\left( {i,j} \right),k}} + {\chi _{\left( {i,j} \right)}}{{\bar \Pi }_{\left( {i,j} \right)}}{s_k} - {\chi _{\left( {i,j} \right)}}{\mathcal{P}_{{{\bar D}_{\left( {i,j} \right)}}}}{\text{(}}\chi _{\left( {i,j} \right)}^{ - 1}{z_{\left( {i,j} \right),k}} + {{\bar \Pi }_{\left( {i,j} \right)}}{s_k}{\text{)}}$. Similar to the projection computations of ${{{\bar D}_{\left( {i,j} \right)}}}$ in Appendix \ref{ProofDistriMN}, ${{\tilde \lambda }_{\left( {i,j} \right),k}}$ and ${{\tilde z}_{\left( {i,j} \right),k}}$ can be further decomposed into ${{\tilde \lambda }_{\left( {i,j} \right),i,k}} = {{\tilde \lambda }_{\left( {i,j} \right),j,k}},{{\tilde z}_{\left( {i,j} \right),i,k}} = {{\tilde z}_{\left( {i,j} \right),j,k}}$ with
\begin{flalign}\label{Localtillamzij}
\begin{gathered}
  {{\tilde \lambda }_{\left( {i,j} \right),i,k}} = \frac{1}
{2}({\lambda _{\left( {i,j} \right),i,k}} + {\lambda _{\left( {i,j} \right),j,k}}) \hfill \\
  \qquad\qquad\qquad + \frac{{{\chi _{\left( {i,j} \right)}}}}{2}({B_{ij}}{x_{i,k}} + {B_{ji}}{x_{j,k}} - {b_{\left( {i,j} \right)}}) \hfill \\
  {{\tilde z}_{\left( {i,j} \right),i,k}} = \frac{1}{2}({z_{\left( {i,j} \right),i,k}} + {z_{\left( {i,j} \right),j,k}}) \hfill \\
  \qquad\qquad\qquad + \frac{{{\chi _{\left( {i,j} \right)}}}}
{2}({E_{ij}}{s_{i,k}} + {E_{ji}}{s_{j,k}}) \hfill \\
\end{gathered}
\end{flalign}

Recalling the operator ${\mathcal{A}_i}$ in \eqref{DisMNcondita}, transforming \eqref{Localtildx} into a proximal iterative form of $f_i$  remains challenging, primarily because $h_i$ and ${{\mathcal{C}_i}}$ not only depend on $\tilde x_{i,k}$, but may also be inherently nonlinear. {To overcome this technical issue, we integrate the prediction rules \eqref{LocalPredictc} and \eqref{LocalPredictd} with the aforementioned proximal splitting formulations. We then demonstrate that the implicit update in \eqref{Localtildx} is equivalent to solving a locally strongly convex subproblem, as formalized in the following proposition. The rigorous proof is detailed in Appendix \ref{ProofEquivTilexi}.
}

\begin{proposition}\label{EquivTilexi}
Let Assumption \ref{AssumpNonempty} hold. Consider the rules \eqref{LocalPredict} and \eqref{Localtillamzij}. At $k$-th iteration, we define the local function ${F_{i,k}}({x_i},{u_i},{s_i}) = {f_i}\left( {{x_i}} \right) + \frac{1}{{2{\alpha _i}}}||{x_i} - {{\overset{\lower0.5em\hbox{$\smash{\scriptscriptstyle\frown}$}}{x} }_{i,k}}|{|^2} + \frac{1}
{{2{\delta _i}}}||{\mathcal{P}_{\mathbb{R}_ + ^{{p_i}}}}({u_i} + {\delta _i}{h_i}({x_i}))|{|^2} + \frac{1}{{2{\eta _i}}}||{\mathcal{P}_{{\Omega _{{\rm{dual}}}}}}({s_i} + {\eta _i}{\mathcal{C}_i}({x_i}))|{|^2}$ with ${{\overset{\lower0.5em\hbox{$\smash{\scriptscriptstyle\frown}$}}{x} }_{i,k}} = {x_{i,k}} - {\alpha _i}(R_i^{\rm{T}}{{\tilde w}_{i,k}} + \sum\nolimits_{j \in {\mathcal{N}_i}} {B_{ij}^{\rm{T}}{{\tilde \lambda }_{\left( {i,j} \right),i,k}}} )$. Then, the rule \eqref{Localtildx} is equivalent to the following minimization problem:
\begin{flalign} \label{LOvaslTildxi}
{\tilde x_{i,k}} = {\text{argmi}}{{\text{n}}_{{x_i} \in {\mathbb{R}^{{n_i}}}}}{F_{i,k}}({x_i},{u_{i,k}},{\overset{\lower0.5em\hbox{$\smash{\scriptscriptstyle\frown}$}}{s} _{i,k}})
\end{flalign}
where ${{\overset{\lower0.5em\hbox{$\smash{\scriptscriptstyle\frown}$}}{s} }_{i,k}} = {s_{i,k}} - {\eta _i}\sum\nolimits_{j \in {\mathcal{N}_i}} {E_{ij}^{\text{T}}(2{{\tilde z}_{\left( {i,j} \right),i,k}} - {z_{\left( {i,j} \right),i,k}})} $.
\end{proposition}

{Observe that directly updating ${\tilde x_{i,k}}$ via $\mathcal{A}_i$  is infeasible, given the potential nonlinearity of $h_i$ and $\mathcal{C}_i$, as well as the possible nonsmoothness of $f_i$. Proposition \ref{EquivTilexi} establishes an equivalent transformation of \eqref{Localtildx} into a local optimization problem solvable independently by each agent. This equivalence forms the cornerstone of the algorithm design. However, obtaining an explicit solution or achieving high accuracy for \eqref{LOvaslTildxi} poses a significant challenge. To mitigate computational burdens, we adopt an absolutely summable error criterion for \eqref{LOvaslTildxi} and directly assign its acceptable inexact solution to $x_{i,k+1}$. Subsequently, building upon \eqref{LocalPredictd}, \eqref{LocalPredicte} and \eqref{Localtillamzij}, we set ${z_{(i,j),i,k + 1}} = {{\tilde z}_{\left( {i,j} \right),i,k}}$. This variable, along with ${{\overset{\lower0.5em\hbox{$\smash{\scriptscriptstyle\frown}$}}{s} }_{i,k}}$ and ${{\tilde \lambda }_{(i,j),i,k}}$ then leveraged to drive the updates of ${s_{i,k + 1}}$, ${w_{i,k + 1}}$, ${u_{i,k + 1}}$ and ${\lambda _{(i,j),i,k + 1}}$, as detailed in Algorithm \ref{alg:1}.}

\begin{algorithm}
	\renewcommand{\algorithmicrequire}{\textbf{Initialization:}}
	\renewcommand{\algorithmicensure}{\textbf{For} $k = 0,{\text{ }}1,{\text{ }} \cdots $
                                      \textbf{do:}\qquad\qquad\qquad\qquad\qquad\qquad\qquad}
	\caption{\textbf{:} FDOA}
	\label{alg:1}
	\begin{algorithmic}[1]
		\REQUIRE {Each agent $i \in \mathcal{V}$ initializes $x_{i,0} \in {\mathbb{R}^{{n_i}}}$, ${w_{i,0}} \in {\mathbb{R}^{{{\tilde r}_i}}}$, ${u_{i,0}} \in {\mathbb{R}^{{p_i}}}$, ${s_{i,0}} \in {\Omega _{{\rm{dual}}}}$, ${\lambda _{( {i,j} ),i,0}} \in {\mathbb{R}^{{e_{( {i,j} )}}}}$, and ${z_{( {i,j} ),i,0}} \in {\mathbb{R}^{( {c + q} )}}$.
		\ENSURE
		\STATE Agent $i$ computes auxiliary variables ${{\tilde w}_{i,k}}$ and ${{\tilde \lambda }_{\left( {i,j} \right),i,k}}$ following from \eqref{Localtildwi} and \eqref{Localtillamzij}, respectively.

		\STATE Update edge-based variables: ${z_{(i,j),i,k + 1}} = \frac{1}{2}({z_{(i,j),i,k}} + {z_{(i,j),j,k}}) + \frac{{{\chi _{(i,j)}}}}{2}({E_{ij}}{s_{i,k}} + {E_{ji}}{s_{j,k}})$ for all $j \in {\mathcal{N}_i}$.

         \STATE Collect ${{\overset{\lower0.5em\hbox{$\smash{\scriptscriptstyle\frown}$}}{x} }_{i,k}} = {x_{i,k}} - {\alpha _i}(R_i^{\rm{T}}{{\tilde w}_{i,k}} + \sum\nolimits_{j \in {\mathcal{N}_i}} {B_{ij}^{\text{T}}{{\tilde \lambda }_{( {i,j} ),i,k}}} )$ and ${{\overset{\lower0.5em\hbox{$\smash{\scriptscriptstyle\frown}$}}{s} }_{i,k}} = {s_{i,k}} - {\eta _i}\sum\nolimits_{j \in {\mathcal{N}_i}} {E_{ij}^{\rm{T}}(2{z_{(i,j),i,k + 1}} - {z_{(i,j),i,k}})} $.

         \STATE Calculate the primal variable
         \begin{flalign} \label{InexactIter}
           {x_{i,k + 1}} \approx {\rm{argmi}}{{\rm{n}}_{{x_i} \in {\mathbb{R}^{{n_i}}}}}{F_{i,k}}({x_i},{u_{i,k}},{{\overset{\lower0.5em\hbox{$\smash{\scriptscriptstyle\frown}$}}{s} }_{i,k}})
         \end{flalign}
         such that there exists a vector ${\zeta _{i,k}} \in {\partial _{{x_i}}}{F_{i,k}}({x_i},{u_{i,k}},{{\overset{\lower0.5em\hbox{$\smash{\scriptscriptstyle\frown}$}}{s} }_{i,k}})$ satisfying $\| {{\zeta _{i,k}}} \| \leq {\varepsilon _{i,k}}$, where the local function ${F_{i,k}}$ is formulated in Proposition \ref{EquivTilexi}.

         \STATE Update dual variables:
         \begin{flalign*}
           \begin{gathered}
             {s_{i,k + 1}} = {\mathcal{P}_{{\Omega _{{\text{dual}}}}}}({{\overset{\lower0.5em\hbox{$\smash{\scriptscriptstyle\frown}$}}{s} }_{i,k}} + {\eta _i}{\mathcal{C}_i}({x_{i,k + 1}})) \hfill \\
             {w_{i,k + 1}} \!=\! {w_{i,k}} \!\!+\! {\beta _i}{R_i}{x_{i,k + 1}} \!\!-\! {\beta _i}{\text{pro}}{{\text{x}}_{\beta _i^{ \!- 1}{g_i}}}(\beta _i^{ \!- 1}{w_{i,k}} \!\!+\! {R_i}{x_{i,k}}) \hfill \\
             {u_{i,k + 1}} = {u_{i,k}} \!-\! {\gamma _i}{u_{i,k}} \!+\! {\gamma _i}{\mathcal{P}_{\mathbb{R}_ + ^{{p_i}}}}({u_{i,k}} \!+\! {\delta _i}{h_i}({x_{i,k + 1}})) \hfill \\
             {{\lambda _{(i,j),i,k + 1}} = {{\tilde \lambda }_{(i,j),i,k}} + {\chi _{(i,j)}}{B_{ij}}({x_{i,k + 1}} - {x_{i,k}})} \hfill \\
           \end{gathered}
         \end{flalign*}
}
	\end{algorithmic}
\textbf{End.}
\end{algorithm}

{

\begin{remark}
Three core design principles underpin the proposed algorithm. First, proximal splitting techniques are leveraged to handle the three nonsmooth components within problem \eqref{RefProblem}, specifically the composite term $g_i\circ R_i$, alongside the indicator functions $\mathcal{I}_{{{\tilde D}_{\left( {i,j} \right)}}}$ and $\mathcal{I}_{\bar D_{(i,j)}}$. Second, the Moreau decomposition \cite[Theorem 14.3]{Bauschke2011Convex}  is employed to recast the proximal operators of the associated conjugate functions (i.e., $g_i^ \star ,\mathcal{I}_{{{\tilde D}_{\left( {i,j} \right)}}}^ \star, \mathcal{I}_{{{\bar D}_{\left( {i,j} \right)}}}^ \star $) into tractable projection or proximal evaluations. This circumvents the necessity of directly calculating the proximal mapping for $g_i\circ R_i$, $\mathcal{I}_{{{\tilde D}_{\left( {i,j} \right)}}}$ and $\mathcal{I}_{\bar D_{(i,j)}}$. Third, obtaining exact solutions for local subproblems is impractical; Algorithm \ref{alg:1} permits inexact solutions characterized by summable errors while rigorously preserving the convergence guarantee.
\end{remark}

\begin{remark} \label{RemarDisFulDsitri}
Algorithm \ref{alg:1} exhibits several advantages compared to existing literature. First, by appropriately specializing \eqref{OrigProblemb} (see Remark \ref{RemarProbel}), our unified framework directly accommodates both consensus-based and non-consensus problems. This avoids the necessity of designing two separate algorithms as done in \cite{Huang2025Unified,Alaviani2025Parallel}, and bypasses the reliance on multiple auxiliary iterates found in \cite{Alaviani2025Parallel}, resulting in a much more streamlined implementation. Second, complete decomposition is guaranteed since each agent retains full control over its own state variables $x_i$, $s_i$, $w_i$, $u_i$, and $\lambda_{(i,j),i}$, $z_{(i,j),i}$. Third, the proposed updates operate purely on local functions (e.g. ${f_i}$, $g_i$ and $h_i$), local matrices (e.g., ${R_i}$ and $A_i$), and neighbor information (e.g., ${z_{\left( {i,j} \right),j,k}}$ and ${E_{ji}}{s_{j,k}}$. Consequently, no centralized coordinator is required to aggregate data or dictate decisions, which contrasts with \cite{Sherson2019Separable,Yi2025Semi}. Forth, our approach demands less private information than \cite{Li2019NIDS,Wang2026Globally}, as it does not necessitate access to neighbors' objective or constraint functions. Finally, unlike existing fully distributed algorithms \cite{Falsone2020Tracking,Liu2025Continuous,Huang2023Multiproximal,Alaviani2025Parallel} that enforce a common step-size, our framework supports local step-sizes for independent agent managing. These features collectively substantiate the designation of FDOA as a fully distributed algorithm.
\end{remark}
}

\section{Convergence Analysis} \label{SectConAna}

We define the integrated vector $\psi  = col\{ x,w,\lambda ,u,s,z\} $, and the operator {$\mathcal{T}:\psi  \to (\partial f(x) + {R^{\rm{T}}}w + {{\tilde \Pi }^{\text{T}}}\lambda  + {(\partial h(x))^{\rm{T}}}u + {(\partial \mathcal{C}(x))^{\rm{T}}}s,\partial {g^ \star }(w) - Rx,\partial \mathcal{I}_{\tilde D}^ \star (\lambda ) - \tilde \Pi x,\partial {\mathcal{I}_\Theta }(u) - h(x),\partial {\mathcal{I}_{{\Omega _{\text{d}}}}}(s) - \mathcal{C}(x) + {{\bar \Pi }^{\rm{T}}}z,\partial \mathcal{I}_{\bar D}^ \star (z) - \bar \Pi s)$ with ${\Omega _{\text{d}}} = \prod\nolimits_{i = 1}^m {{\Omega _{{\text{dual}}}}} $ and $\bar D = {\prod _{(i,j) \in \mathcal{E}}}{{\bar D}_{\left( {i,j} \right)}}$}. Then, one can write the local condition system \eqref{DisMNcondit} as the compact form $0 \in \mathcal{T}{\psi ^ * }$ with $\psi^*  = col\{ x^*, w^*,\lambda^* ,u^*,s^*,z^*\} $. It is easy to check that $0 \in \mathcal{T}{\psi ^ * }$ also is the KKT condition of the following max-min problem:
\begin{flalign}\label{MinMaxPro}
\mathop {{\text{maximize}}}\limits_{\varpi  \in {\Omega _\varpi }} \mathop {{\text{minimize}}}\limits_{x \in {\mathbb{R}^n},z \in {\Omega _z}} {\text{ }}\mathcal{L}(x,z,\varpi )
\end{flalign}
where $\varpi  = col\{ w,\lambda ,u,s\}$, ${\Omega _\varpi } = {\mathbb{R}^r} \times {\mathbb{R}^{2l}} \times {\mathbb{R}^p} \times {\mathbb{R}^{m(c + q)}}$, $p = \sum\nolimits_{i = 1}^m {{p_i}} $, {${\Omega _z} = {\mathbb{R}^{2\left| \mathcal{E} \right|(p + c)}}$, and $\mathcal{L}(x,z,\varpi ) = f(x) + \langle {u,h(x)} \rangle  + \langle {s,\mathcal{C}(x)} \rangle  + \langle {Rx,w} \rangle  + \langle {\lambda ,\tilde \Pi x} \rangle  - \langle {z,\bar \Pi s} \rangle  - {g^ \star }(w) - \mathcal{I}_\Theta (u) - \mathcal{I}_{\tilde D}^ \star (\lambda ) - {\mathcal{I}_{{\Omega _{\text{d}}}}}(s) + \mathcal{I}_{\bar D}^ \star (z)$}. Since Assumption \ref{AssumpNonempty} holds and the conjugate function $\mathcal{I}_{\bar D}^ \star ( z )$ is convex, $\mathcal{L}(x,z,\varpi )$ then is convex at $(x,z)$, and also is concave at $\varpi$. Therefore, one confirms that $\mathcal{T}:\psi  \to ({\partial _x}\mathcal{L}(x,z,\varpi ),-{\partial _\varpi }\mathcal{L}(x,z,\varpi ),{\partial _z}\mathcal{L}(x,z,\varpi ))$, and $\mathcal{T}$ is maximally monotone due to the maximal monotonicity of ${\partial _x}\mathcal{L}(x,z,\varpi ),-{\partial _\varpi }\mathcal{L}(x,z,\varpi ),{\partial _z}\mathcal{L}(x,z,\varpi )$ \cite[Theorem 20.25]{Bauschke2011Convex}. Let $\mathcal{S} \subseteq {\mathbb{R}^n} \times {\Omega _\varpi } \times {\Omega _z}$ be the solution set of problem \eqref{MinMaxPro}. Then, a point $({x^ * },{\varpi ^ * },{z^ * }) \in \mathcal{S}$ can be viewed as a saddle point if and only if for any $ x \in {\mathbb{R}^n},z \in {\Omega _z},\varpi  \in {\Omega _\varpi }$, it holds that
\begin{flalign}\label{InequalSadd}
\mathcal{L}({x^ * },{z^ * },\varpi ) \leq \mathcal{L}({x^ * },{z^ * },{\varpi ^ * }) \leq \mathcal{L}(x,z,{\varpi ^ * })
\end{flalign}
which can be alternatively reformulated as two critical variational inequalities, {i.e., for any $\varpi  \in {\Omega _\varpi }$,
\begin{flalign}\label{Firsvariaineq}
\begin{gathered}
  0 \leq \mathcal{L}({x^ * },{z^ * },{\varpi ^ * }) - \mathcal{L}({x^ * },{z^ * },\varpi ) \hfill \\
    \,\,\,\,= ({g^ \star }(w) + \mathcal{I}_\Theta (u) + \mathcal{I}_{\tilde D}^ \star (\lambda ) + {\mathcal{I}_{{\Omega _{\text{d}}}}}(s)) \hfill \\
  \qquad - ({g^ \star }({w^ * }) + \mathcal{I}_\Theta ({u^ * }) + \mathcal{I}_{\tilde D}^ \star ({\lambda ^ * }) + {\mathcal{I}_{{\Omega _{\text{d}}}}}({s^ * })) \hfill \\
  \qquad + \langle {{s^ * } - s,\mathcal{C}({x^ * })} \rangle  + \langle {{u^ * } - u,h({x^ * })} \rangle  \hfill \\
  \qquad + \langle {{w^ * } - w,R{x^ * }} \rangle  + \langle {{\lambda ^ * } - \lambda ,\tilde \Pi {x^ * }} \rangle  - \langle {{z^ * },\bar \Pi ({s^ * } - s)} \rangle  \hfill \\
\end{gathered}
\end{flalign}
and for} any $x \in {\mathbb{R}^n},z \in {\Omega _z}$,
\begin{flalign}\label{Secdvariaineq}
\begin{gathered}
  0 \leq \mathcal{L}(x,z,{\varpi ^ * }) - \mathcal{L}({x^ * },{z^ * },{\varpi ^ * }) \hfill \\
   \,\,\,\,= (f( x ) + \mathcal{I}_{\bar D}^ \star (z)) - (\mathcal{I}_{\bar D}^ \star ({z^ * }) + f({x^ * })) \hfill \\
  \qquad + \langle {{u^ * },h(x) - h({x^ * })} \rangle  + \langle {{s^ * },\mathcal{C}(x) - \mathcal{C}({x^ * })} \rangle  \hfill \\
  \qquad + \langle {R(x - {x^ * }),{w^ * }} \rangle  + \langle {{\lambda ^ * },\tilde \Pi (x - {x^ * })} \rangle  + \langle {{z^ * } - z,\bar \Pi {s^ * }} \rangle  \hfill \\
\end{gathered}
\end{flalign}
To proceed, define the {function $\mathcal{Y}\left( \psi  \right) = f\left( x \right) + \mathcal{I}_{\bar D}^ \star \left( z \right) + {g^ \star }\left( w \right) + \mathcal{I}_\Theta \left( u \right) + \mathcal{I}_{\tilde D}^ \star \left( \lambda  \right) + {\mathcal{I}_{{\Omega _{\text{d}}}}}\left( s \right)$,} and new operators ${\mathcal{K}_1}:\psi  \to ({R^{\text{T}}}w + {{\tilde \Pi }^{\text{T}}}\lambda , - Rx, - \tilde \Pi x,0,{{\bar \Pi }^{\text{T}}}z, - \bar \Pi s)$, ${\mathcal{K}_2}:\psi  \to (h(x),\mathcal{C}(x),u,s)$ and ${\mathcal{K}_3}:\psi  \to (u,s, - h(x), - \mathcal{C}(x))$. Then, for any $\psi  \in {\mathbb{R}^n} \times {\Omega _\varpi } \times {\Omega _z}$, summing \eqref{Firsvariaineq} and \eqref{Secdvariaineq} yields the following variational inequality:
\begin{flalign}\label{Totlalvariaineq}
\mathcal{Y}( \psi  ) \!-\!\! \mathcal{Y}( {{\psi ^ * }} ) \!+\! \langle {\psi  \!-\! {\psi ^ * },{\mathcal{K}_1}{\psi ^ * }} \rangle  \!+\! \langle {{\mathcal{K}_2}\psi  \!-\! {\mathcal{K}_2}{\psi ^ * },{\mathcal{K}_3}{\psi ^ * }} \rangle  \ge 0
\end{flalign}

\begin{remark} \label{RemEquSolt}
It holds that \eqref{Firsvariaineq} $+$ \eqref{Secdvariaineq} $\Rightarrow $ \eqref{Totlalvariaineq}, and in return, \eqref{Totlalvariaineq} can be split into \eqref{Firsvariaineq} and \eqref{Secdvariaineq} by setting $\psi  = col({x^ * },w,\lambda ,u,s,{z^ * })$ and $\psi  = col(x,{w^ * },{\lambda ^ * },{u^ * },{s^ * },z)$, respectively. Therefore, one can conclude that ${\psi ^ * } \in \mathcal{S}$ if and only if \eqref{Totlalvariaineq} holds for any $\psi  \in {\mathbb{R}^n} \times {\Omega _\varpi } \times {\Omega _z}$.
\end{remark}

Define step-size matrices $\Psi  = blkdiag\{ {\alpha _1}{I_{{n_1}}}, \cdots ,{\alpha _m}{I_{{n_m}}}\}$, $\Gamma  = blkdiag\{ {\beta _1}{I_{{{\tilde r}_1}}}, \cdots ,{\beta _m}{I_{{{\tilde r}_m}}}\} $, $\Upsilon  = blkdiag\{ {\eta _1}{I_{( {c + q} )}}, \cdots ,{\eta _m}{I_{( {c + q} )}}\}$ , $\Sigma  = blkdiag\{ {\delta _1}{I_{{p_1}}}, \cdots ,{\delta _m}{I_{{p_m}}}\} $, {$\Lambda  = blkdiag\{ {\gamma _1}{I_{{p_1}}} \cdots ,{\gamma _m}{I_{{p_m}}}\} $}, $\Xi  = blkdiag{\{ {\chi _{( {i,j} )}}{I_{2{l_{( {i,j} )}}}}\} _{( {i,j} ) \in \mathcal{E}}}$, and $\Phi  = blkdiag{\{ {\chi _{( {i,j} )}}{I_{2( {c + q} )}}\} _{( {i,j} ) \in \mathcal{E}}}$. We introduce an auxiliary variable ${\tilde \psi _k} = col\{ {\tilde x_k},{\tilde w_k},{\tilde \lambda _k},{\tilde u_k},{\tilde s_k},{\tilde z_k}\} $ with ${\tilde x_k} = col\{ {\tilde x_{1,k}}, \cdots ,{\tilde x_{m,k}}\}$, ${\tilde w_k} = col\{ {\tilde w_{1,k}}, \cdots ,{\tilde w_{m,k}}\}$, ${\tilde u_k} = col\{ {\tilde u_{1,k}}, \cdots ,{\tilde u_{m,k}}\}$, ${\tilde s_k} = col\{ {\tilde s_{1,k}}, \cdots ,{\tilde s_{m,k}}\} $, ${{\tilde \lambda }_k} = col{\{ {{\tilde \lambda }_{( {i,j} ),k}}\} _{( {i,j} ) \in \mathcal{E}}}$, and ${{\tilde z}_k} = col{\{ {{\tilde z}_{( {i,j} ),k}}\} _{( {i,j} ) \in \mathcal{E}}}$. Recalling Algorithm \ref{alg:1}, we then derive its equivalent compact forms via the following lemma, whose detailed proof is presented in Appendix \ref{ProofLemInetqua}.

\begin{lemma} \label{LemInetqua}
Let Assumption \ref{AssumpNonempty} hold. Consider mappings $\mathcal{W}:\psi  \to ({\Psi ^{ - 1}}x,{\Gamma ^{ - 1}}w + Rx,{\Xi ^{ - 1}}\lambda  + \tilde \Pi x,{\Sigma ^{ - 1}}u,{\Upsilon ^{ - 1}}s + {{\bar \Pi }^{\text{T}}}z,{\Phi ^{ - 1}}z + \bar \Pi s)$ and $\mathcal{M}:\psi  \to (x,w + \Gamma Rx,\lambda  + \Xi \tilde \Pi x,\Lambda u,s,z)$. Algorithm \ref{alg:1} can be reformulated as
\begin{flalign}\label{AlgoriOne}
{\psi _{k + 1}} = {\psi _k} + \mathcal{M}( {{{\tilde \psi }_k} - {\psi _k}} )
\end{flalign}
with ${{\tilde \psi }_k}$ such that for any $\psi  \in {\mathbb{R}^n} \times {\Omega _\varpi } \times {\Omega _z}$,
\begin{flalign}\label{AlgoriVariaIneq}
\begin{gathered}
  \quad \langle {\psi  - {{\tilde \psi }_k},\mathcal{W}( {{\psi _k} - {{\tilde \psi }_k}} )} \rangle  + \langle {x - {{\tilde x}_k},{\zeta _k}} \rangle  \hfill \\
   \leq \mathcal{Y}( \psi  ) \!-\! \mathcal{Y}( {{{\tilde \psi }_k}} ) \!+\! \langle {\psi  \!-\! {{\tilde \psi }_k},{\mathcal{K}_1}{{\tilde \psi }_k}} \rangle  \!+\! \langle {{\mathcal{K}_2}\psi  \!-\! {\mathcal{K}_2}\tilde \psi ,{\mathcal{K}_3}{{\tilde \psi }_k}} \rangle  \hfill \\
\end{gathered}
\end{flalign}
\end{lemma}

Using the above result, we have the following critical inequalities.

\begin{lemma} \label{LemFeryarInqua}
Let Assumption \ref{AssumpNonempty} hold. Consider the sequence ${\{ {\psi _k}\} _{k \geq 0}}$ generated by Algorithm \ref{alg:1}, and the operator $\mathcal{R}:\psi  \to ({\Psi ^{ - 1}}x,{\Gamma ^{ - 1}}w,{\Xi ^{ - 1}}\lambda ,{\Sigma ^{ - 1}}{\Lambda ^{ - 1}}u,{\Upsilon ^{ - 1}}s + {\bar \Pi ^{\text{T}}}z,{\Phi ^{ - 1}}z + \bar \Pi s)$. For any $\psi  \in {\mathbb{R}^n} \times {\Omega _\varpi } \times {\Omega _z}$, it holds that
\begin{flalign}\label{GDgeghy}
\begin{gathered}
  \quad ||\psi  - {\psi _{k + 1}}||_\mathcal{R}^2 + ||{{\tilde \psi }_k} - {\psi _k}||_\mathcal{R}^2 - ||{{\tilde \psi }_k} - {\psi _{k + 1}}||_\mathcal{R}^2 \hfill \\
   \leq \| {\psi  - {\psi _k}} \|_\mathcal{R}^2 + 2\langle {\psi  - {{\tilde \psi }_k},{\mathcal{K}_1}{{\tilde \psi }_k}} \rangle  + 2(\mathcal{Y}(\psi ) - \mathcal{Y}({{\tilde \psi }_k})) \hfill \\
  \quad - 2\langle {x - {{\tilde x}_k},{\zeta _k}} \rangle  + 2\langle {{\mathcal{K}_2}\psi  - {\mathcal{K}_2}\tilde \psi ,{\mathcal{K}_3}{{\tilde \psi }_k}} \rangle  \hfill \\
\end{gathered}
\end{flalign}
and {for ${\psi ^ * } \in \mathcal{S}$, it holds that}
\begin{flalign}\label{GDgRGghWy}
\begin{gathered}
  \| {{\psi _{k + 1}} - {\psi ^ * }} \|_\mathcal{R}^2 \leq \| {{\psi _k} - {\psi ^ * }} \|_\mathcal{R}^2 + 2\langle {{{\tilde x}_k} - {x^ * },{\zeta _k}} \rangle  \hfill \\
  \qquad\qquad\qquad\qquad - \| {{{\tilde \psi }_k} - {\psi _k}} \|_{{\mathcal{W}^{\rm T}} + \mathcal{W} - {\mathcal{M}^{\rm T}}\mathcal{R}\mathcal{M}}^2 \hfill \\
\end{gathered}
\end{flalign}
\end{lemma}

\begin{proof}
See Appendix \ref{ProofLemFeryarInqua}
\end{proof}

\begin{theorem} \label{TheoreConverge}
Let Assumption \ref{AssumpNonempty} hold. Consider the sequence ${\{ {\psi _k}\} _{k \geq 0}}$ generated by Algorithm \ref{alg:1}. If $0 < {\gamma _i} < 2$, $\alpha _i^{ - 1} > {\beta _i}\| {R_i^{\rm{T}}{R_i}} \| + \sum\nolimits_{j \in {\mathcal{N}_i}} {{\chi _{( {i,j} )}}\| {B_{ij}^{\rm{T}}{B_{ij}}} \|} $, $\eta _i^{ - 1} > \sum\nolimits_{j \in {\mathcal{N}_i}} {{\chi _{( {i,j} )}}\| {E_{ij}^{\rm{T}}{E_{ij}}} \|} $ with ${\beta _i}$, ${\delta _i}$, ${\chi _{\left( {i,j} \right)}} >0$, and $\sum\nolimits_{k = 0}^\infty  {{\varepsilon _{i,k}}}  <  + \infty $, we have the following two statements:

(i) {for any ${\psi ^ * } = col({x^ * },{\varpi ^ * },{z^ * }) \in \mathcal{S}$, it holds that $||{\psi _{k + 1}} - {\psi ^ * }|{|_\mathcal{R}} \leq ||{\psi _k} - {\psi ^ * }|{|_\mathcal{R}} + \kappa \| {{\zeta _k}} \|$ with $\kappa >0$. Moreover, ${\{ {{\psi _k}} \}_{k \geq 0}}$ converges to a point ${\psi _\infty} = col({x_\infty },{\varpi _{\infty} },{z_{\infty} }) \in \mathcal{S}$ with ${x_\infty }$ being a solution to the problem \eqref{OrigProblem};}

(ii) when ${\varepsilon _{i,k}} = O({({\rho _i})^k})$ with $0 < {\rho _i} < 1$, there exists a positive constant $\xi $ such that $\| {{\psi _k} - {\psi _{k + 1}}} \|_\mathcal{R}^2 + \xi {({\rho _{\max }})^k}/(1 - {\rho _{\max }}) = o( {1/k} )$, where ${\rho _{\max }} = \max \{ {\rho _1}, \cdots ,{\rho _m}\} $. {Moreover, the first-order optimality residual ${\text{dis}}{{\text{t}}^2}( {0,\mathcal{T}{{\tilde \psi }_k}} ) = o(1/k) + O({({\rho _{\max }})^k})$ holds.}
\end{theorem}

\begin{proof}
See Appendix \ref{ProofTheoreConverge}
\end{proof}

{
\begin{remark}
The above theorem verifies the quasi-Fej{\'{e}}r monotonicity of the sequence ${\{ ||{\psi _k} - {\psi ^ * }|{|_\mathcal{R}}\} _{k \geq 0}}$, which ensures the convergence of the proposed inexact algorithm. Notably, this guarantee holds without necessitating strictly bounded subgradients, bounded feasible sets \cite{Lin2021Compensation}, or Lipschitz-continuous inequality constraints functions \cite{Huang2023Multiproximal,Huang2025Unified}. Furthermore, our convergence analysis framework accommodates the problems investigated in \cite{Lin2021Compensation,Huang2023Multiproximal,Huang2025Unified,Alaviani2025Parallel}, and delivers an $o(1/k)$ convergence rate that outperforms the asymptotic results in \cite{Alaviani2025Parallel}. While dedicated algorithms exist for directed or unbalanced networks \cite{Nguyen2023Accelerated,Hamedani2022Strongly}, they are restricted from solving problem \eqref{OrigProblem} due to specific objective function assumptions and the necessity for intricate dual agreement analysis under global coupling constraints. Given the space limitation, generalizing our method to broader network topologies is left as a direction for future work.
\end{remark}

\section{Discussions}
This section elaborates on the locally constant step-sizes, the inexact distributed variants of existing methods, and the convergence analysis framework, while comparing them with related works.

\emph{ 1) Discussions on step-sizes}: The selection of locally fixed step-sizes for each agent does not necessitate global coordinators or common parameters shared across agents. Instead, it relies solely on local matrix norms (e.g., $\| {R_i^{\text{T}}{R_i}} \|$, $\| {B_{ij}^{\text{T}}{B_{ij}}} \|$) and network step-sizes ${\chi _{( {i,j} )}}$. As discussed in Remark \ref{RemarDisFulDsitri}, each agent has access to its own matrices $R_i$, $B_{ij}$, while network step-sizes are regarded as inherent parameters of the communication graph. Consequently, these quantities are locally available, allowing admissible step-sizes to be computed from local information without requiring the global Laplacian or other global spectral radius as in \cite{Alaviani2025Parallel,Wang2026Globally,Huang2023Multiproximal,Huang2025Unified}. For instance, the upper bounds of $\alpha_i$ and $\eta_i$ in Theorem \ref{TheoreConverge} depend exclusively on local information (e.g., $R_i$, $B_{ij}$, $E_{ij}$, and $\chi_{(i,j)}$), enabling agents to select their own admissible step-sizes independently of global spectral information. Furthermore, while the established step-size condition is sufficient for the convergence of FDOA, alternative selections are theoretically viable. For example, by imposing additional requirements ${\beta _i}\| {R_i^{\text{T}}{R_i}} \| \leq {\mu _i}{\alpha _i}$ and ${\chi _{( {i,j} )}}\| {B_{ij}^{\text{T}}{B_{ij}}} \| \leq {\mu _i}{\alpha _i}$ with $0<{\mu _i}<1$, the upper bound on $\alpha_i$ simplifies to $0 < {\alpha _i} < 1/(\sqrt {( {1 + | {{\mathcal{N}_i}} |} ){\mu _i}} )$, where $| {{\mathcal{N}_i}} |$ denotes the number of neighbors of $i$. This alternative selection eliminates the dependence on the quantities ${\beta _i}\| {R_i^{\text{T}}{R_i}} \|$ and ${\chi _{( {i,j} )}}\| {B_{ij}^{\text{T}}{B_{ij}}} \|$. Conversely, although decaying or globally coordinated step-sizes might accelerate convergence, they require global coordination, contradicting our objective of fully distributed algorithms. In contrast, locally fixed step-sizes offer less restrictive and more flexible choices than their global counterparts \cite{Alaviani2025Parallel,Wang2026Globally,Huang2023Multiproximal,Huang2025Unified}. Therefore, appropriate step-size selection ensures satisfactory practical convergence performance.

\emph{ 2) Discussions on solvability and existence of inexact solutions}: Under Assumption \ref{AssumpNonempty}, the quadratic proximal term ${\| {{x_i} - {{\overset{\lower0.5em\hbox{$\smash{\scriptscriptstyle\frown}$}}{x} }_{i,k}}} \|^2}$ renders ${F_{i,k}}({x_i},{u_{i,k}},{\overset{\lower0.5em\hbox{$\smash{\scriptscriptstyle\frown}$}}{s} _{i,k}})$ proper, closed, and strongly convex with respect to $x_i$, thereby admitting a unique minimizer $x_{i,k+1}^*$. For any ${x_i} \ne x_{i,k + 1}^ * $, we have $0 \notin \partial {F_{i,k}}({x_i},{u_{i,k}},{{\overset{\lower0.5em\hbox{$\smash{\scriptscriptstyle\frown}$}}{s} }_{i,k}})$, as $0 \in \partial {F_{i,k}}(x_{i,k + 1}^ * ,{u_{i,k}},{{\overset{\lower0.5em\hbox{$\smash{\scriptscriptstyle\frown}$}}{s} }_{i,k}})$ uniquely characterizes the global minimizer. Consequently, non-zero elements exist in $\partial {F_{i,k}}({x_i},{u_{i,k}},{{\overset{\lower0.5em\hbox{$\smash{\scriptscriptstyle\frown}$}}{s} }_{i,k}})$ whenever ${x_i} \ne x_{i,k + 1}^ * $. Thus, if the local solver yields ${x_{i,k+1}} \ne x_{i,k + 1}^ * $  along with a subgradient ${\zeta _{i,k}} \in \partial {F_{i,k}}({x_{i,k + 1}},{u_{i,k}},{{\overset{\lower0.5em\hbox{$\smash{\scriptscriptstyle\frown}$}}{s} }_{i,k}})$ satisfying $\| {{\zeta _{i,k}}} \| \leq {\varepsilon _{i,k}}$, then ${\zeta _{i,k}} \ne 0$ automatically, yielding the desired non-trivial inexact solution. Algorithmically, the condition ${\text{dist}}(0,\partial {F_{i,k}}({x_i},{u_{i,k}},{{\overset{\lower0.5em\hbox{$\smash{\scriptscriptstyle\frown}$}}{s} }_{i,k}})) \leq {\varepsilon _{i,k}}$ serves as a standard optimality-residual stopping criterion \cite{Zhang2023Primal}. The per-iteration arithmetic complexity of \eqref{InexactIter} is $\mathcal{O}( {{n_i}({{\tilde r}_i} + \sum\nolimits_{j \in {\mathcal{N}_i}} {{l_{\left( {i,j} \right)}}} )} $, depending solely on local dimensions and the number of neighbors. In contrast, for exact methods \cite{Huang2025Unified,Alaviani2025Parallel}, the cost of computing matrix-vector products is $\mathcal{O}( {2{n_i}(c + q)} )$, and the method in \cite{Wu2023Distributed} requires $\mathcal{O}( {n_i^2 + {n_i}(c + q)}) $, which potentially incurs higher computational costs due to the dimension (i.e., $c+q$) of global coupling constraints. Moreover, the absolutely summable error criterion $\sum\nolimits_{k = 0}^{ + \infty } {{\varepsilon _{i,k}}}  <  + \infty $ permits rough inner solves without error amplification, thereby preventing inner-loop explosion compared to exact schemes \cite{Huang2025Unified,Alaviani2025Parallel}. Hence, the proposed inexact scheme is well-posed and exhibits superior scalability over existing distributed methods regarding the per-iteration computational complexity of local subproblems, particularly for high-dimensional global coupling constraints.

\emph{ 3) Discussions on inexact distributed extensions}: Appropriate modifications to FDOA yield significant inexact extensions of existing methods. By omitting \eqref{OrigProblemc} and \eqref{OrigProblemd}, the iterations for ${u_{i,k}}$, ${{s_{i,k}}}$, ${z_{\left( {i,j} \right),i,k}}$ vanish, causing FDOA to naturally encompass the inexact distributed variant of TriPD-Dist \cite{Latafat2019TriPD} in the absence of smooth terms. In this case, ${x_{i,k + 1}} \approx {\text{argmi}}{{\text{n}}_{{x_i} \in {\mathbb{R}^{{n_i}}}}}\{ {f_i}\left( {{x_i}} \right) + \frac{1}
{{2{\alpha _i}}}||{x_i} - ({x_{i,k}} - {\alpha _i}(R_i^{\text{T}}{{\tilde w}_{i,k}} + \sum\nolimits_{j \in {\mathcal{N}_i}} {B_{ij}^{\text{T}}{{\tilde \lambda }_{\left( {i,j} \right),i,k}}} ))|{|^2}\} $ can be equivalently reformulated into ${x_{i,k + 1}} \approx {\text{pro}}{{\text{x}}_{{\alpha _i}{f_i}}}({x_{i,k}} - {\alpha _i}(R_i^{\text{T}}{{\tilde w}_{i,k}} + \sum\nolimits_{j \in {\mathcal{N}_i}} {B_{ij}^{\text{T}}{{\tilde \lambda }_{\left( {i,j} \right),i,k}}} ))$. Under such consideration without \eqref{OrigProblemc} and \eqref{OrigProblemd}, by directly assigning ${\tilde w_{i,k}}, {\tilde \lambda _{\left( {i,j} \right),i,k}}$ to ${w_{i,k + 1}} ,{\lambda _{\left( {i,j} \right),i,k + 1}}$ in Algorithm \ref{alg:1}, FDOA reduces to the recursions ${w_{i,k + 1}} = {w_{i,k}} + {\beta _i}{R_i}{x_{i,k}} - {\beta _i}{\rm{pro}}{{\rm{x}}_{\beta _i^{ - 1}{g_i}}}(\beta _i^{ - 1}{w_{i,k}} + {R_i}{x_{i,k}})$, ${\lambda _{\left( {i,j} \right),i,k + 1}} = \frac{1}{2}({\lambda _{\left( {i,j} \right),i,k}} + {\lambda _{\left( {i,j} \right),j,k}}) + \frac{{{\chi _{\left( {i,j} \right)}}}}{2}({B_{ij}}{x_{i,k}} + {B_{ji}}{x_{j,k}} - {b_{\left( {i,j} \right)}})$ and ${x_{i,k + 1}} \approx {\text{pro}}{{\text{x}}_{{\alpha _i}{f_i}}}({x_{i,k}} - {\alpha _i}(R_i^{\text{T}}{w_{i,k + 1}} + \sum\nolimits_{j \in {\mathcal{N}_i}} {B_{ij}^{\text{T}}{\lambda _{\left( {i,j} \right),i,k + 1}}} ))$ for $i \in {\cal V}$, $j \in {\mathcal{N}_i}$ at $k$-th iteration. This corresponds to the inexact distributed extension of the centralized Arrow-Hurwicz method \cite{Popov1980Arrow}. Notably, if all step-sizes are set to a common scalar, the aforementioned iteration further yields the inexact distributed extension of PDHG \cite{Esser2010General}. While methods \cite{Esser2010General,Popov1980Arrow,Latafat2019TriPD} require the first nonsmooth terms to be proximable (i.e., admitting analytical solutions), the inexact distributed versions generated by FDOA removes this assumption, proving more practical than the exact algorithms \cite{Esser2010General,Popov1980Arrow,Latafat2019TriPD}.

\emph{ 4) Discussions on convergence analysis framework}: Our convergence analysis framework (i.e., Lemma \ref{LemInetqua} and Theorem \ref{TheoreConverge}) applies to the aforementioned inexact distributed variant of TriPD-Dist \cite{Latafat2019TriPD} without smooth terms, establishing an $o(1/k)$ convergence rate under general convexity assumptions. To the best of our knowledge, only asymptotic convergence has been provided under general assumptions, and it is only applicable to the exact iterations in \cite{Latafat2019TriPD}. Therefore, the provided framework of this paper generalizes theoretical result of \cite{Latafat2019TriPD} without smooth terms, and enhance the convergence analysis including exact and inexact iterations.

\section{Performance Evaluation} \label{PerforEvalu}

This section provides a numerical example to validate the performance of Algorithm \ref{alg:1}. Consider the constrained optimization problem
\begin{subequations}\label{SimuProblem}
  \begin{equation}\label{SimuProblema}
   \mathop {{\text{minimize}}}\limits_{{x_1}, \cdots ,{x_m}} {\text{ }}\sum\limits_{i = 1}^m {( {\frac{1}{2}{{\left\| {{\Phi _i}{x_i} - {\phi _i}} \right\|}^2} + {{\| {{R_i}{x_i}} \|}_1}} )} \qquad\qquad\quad\,\,
  \end{equation}
  \begin{equation}\label{SimuProblemb}
  {\text{subject to  }}{x_i} - {x_j} = 0,\left( {i,j} \right) \in \mathcal{E} \qquad\qquad\qquad\quad\qquad\quad
  \end{equation}
  \begin{equation}\label{SimuProblemc}
  {x_i} \in {{\cal X}_i} \cap {\Omega _i}, i \in \mathcal{V} \qquad\qquad\qquad\qquad\quad\,\,\,
  \end{equation}
  \begin{equation}\label{SimuProblemd}
   \sum\limits_{i = 1}^m {\left( {{A_i}{x_i} - {a_i}} \right)}  = 0,\sum\limits_{i = 1}^m {( {{{\left\| {{x_i}} \right\|}^2} - {\tilde c_i}} )}  \leq 0
  \end{equation}
\end{subequations}
with ${\mathcal{X}_i} = \{ {{x_i} \in {\mathbb{R}^{{\tilde n}}}|{x_{i,\min }} \leq {x_i} \leq {x_{i,\max }}} \}$ and ${\Omega _i} = \{ {{x_i} \in {\mathbb{R}^{{\tilde n}}}|{{\| {{x_i} - {{\tilde d}_i}} \|}_1} \leq {{\tilde q}_i}} \}$, where ${{\Phi _i}}$ is a known matrix, ${{{\tilde d}_i}}$, $\phi_i$, ${{{\tilde q}_i}}$, ${{a_i}}$, and $c_i$ are known vectors, and ${x_{i,\min }}/{x_{i,\max }}$ is the lower/upper bound of $x_i$. It can be checked that problem \eqref{SimuProblem} can be reformulated into the considered framework \eqref{OrigProblem}, by setting ${f_i}\left( {{x_i}} \right) = \frac{1}{2}{\left\| {{\Phi _i}{x_i} - {\phi _i}} \right\|^2} + {{\cal I}_{{{\cal X}_i}}}\left( {{x_i}} \right)$, ${g_i}\left( {{R_i}{x_i}} \right) = {\| {{R_i}{x_i}} \|_1}$, ${h_i}( {{x_i}} ) = {\| {{x_i} - {{\tilde d}_i}} \|_1} - {{\tilde q}_i}$, ${B_{ij}} = I$, ${B_{ji}} =  - I$, and ${C_i}\left( {{x_i}} \right) = {\left\| {{x_i}} \right\|^2} - {c_i}$. Hence, $B_{ij}^{\text{T}}{B_{ij}} = I$ holds for all $( {i,j} ) \in \mathcal{E}$. In this simulation, we consider $m=20$, ${\tilde n} = 5$, the entries of ${\Phi _i} \in {\mathbb{R}^{3 \times 5}}$ and ${R _i} \in {\mathbb{R}^{1 \times 5}}$ are generated randomly from the interval $( { - 15,15} )$. To ensure the existence of solution of problem \eqref{SimuProblem}, we randomly generate a shared decision ${x_{\text{s}}}$ in $( { - 3,3} )$ to give ${\phi _i} = {\Phi _i}{x_{\text{s}}} + {\phi _{i,{\text{s}}}}$, ${\tilde q_i} = {\| {{x_i} - {\tilde d_i}} \|_1} + {q_{i,{\text{s}}}}$, ${a_i} = {A_i}{x_i} + {a_{i,{\text{s}}}}$, and ${{\tilde c}_i} = {\| {{x_i}} \|^2} + {{\tilde c}_{i,{\text{s}}}}$, with instant vectors ${\phi _{i,{\text{s}}}}$, ${\tilde d_i}$, ${q_{i,{\text{s}}}}$, ${a_{i,{\text{s}}}}$ and ${{\tilde c}_{i,{\text{s}}}}$ generated randomly from $(0,1)$. The communication network $\mathcal{G}$ is generated at random to satisfy Assumption \ref{AssumpNonempty}(i). Under the above settings, problem \eqref{SimuProblem} satisfies Assumption \ref{AssumpNonempty}(ii)-(iv). All step-sizes for the proposed algorithm strictly follow the theoretical conditions in Theorem \ref{TheoreConverge}. Specifically, we take ${\gamma _i} = 1$, ${\delta _i} = 0.5$, ${\beta _i} = 0.8/\| {R_i^{\text{T}}{R_i}} \|$, ${\alpha _i} = 0.95/({\beta _i}\| {R_i^{\text{T}}{R_i}} \| + \sum\nolimits_{j \in {\mathcal{N}_i}} {{\chi _{( {i,j} )}}} )$, ${\eta _i} = 0.98/\sum\nolimits_{j \in {\mathcal{N}_i}} {{\chi _{( {i,j} )}}} $ with ${\chi _{( {i,j} )}} = \max \{ | {{\mathcal{N}_i}} |,| {{\mathcal{N}_j}} |\} $.

For the inexact computation \eqref{InexactIter} in FDOA, we apply the proximal-point algorithm \cite[Theorem 28.1]{Bauschke2011Convex} to get the inner iteration rule ${x_{i,k,\varsigma  + 1}} = {\text{pro}}{{\text{x}}_{{F_i}}}({x_{i,k,\varsigma }})$, where $\varsigma $ is the inner iteration index. By the above inner iteration, we have ${x_{i,k,\varsigma }} - {x_{i,k,\varsigma  + 1}} \in {\partial _{{x_i}}}{F_i}({x_{i,k,\varsigma  + 1}},{u_{i,k}},{{\overset{\lower0.5em\hbox{$\smash{\scriptscriptstyle\frown}$}}{s} }_{i,k}})$, which allows us to define the computable residual as ${\zeta _{i,k,\varsigma  + 1}} = {x_{i,k,\varsigma }} - {x_{i,k,\varsigma  + 1}}$. Therefore, we stop the inner loop and set ${x_{i,k + 1}} = {x_{i,k,\varsigma  + 1}}$ when the stopping criterion holds, i.e., $||{x_{i,k,\varsigma }} - {x_{i,k,\varsigma  + 1}}|| \le {\varepsilon _{i,k}}$ or $\varsigma  > 500$. In order to compare the effect of inexact iteration, we execute FDOA with different values of ${\varepsilon _{i,k}} \in \{ 0,{10^{ - 8}},{k^{ - 2}},{k^{ - 1.5}}\} $, and introduce related comparative methods, i.e., distributed ADMM \cite{Yang2020DCDOPF}, distributed compensation algorithm \cite{Lin2021Compensation}, fully parallel algorithm \cite{Alaviani2025Parallel}. We use $\left\| {{x_k} - {1_m} \otimes {{\hat x}^ * }} \right\|$, $|f( {{x_k}} ) + g(R{x_k}) - ({f^ * } + {g^ * })|/|{f^ * } + {g^ * }|$, $||\sum\nolimits_{i = 1}^m {( {{A_i}{x_{i,k}} - {a_i}} )} || + \max \{ \sum\nolimits_{i = 1}^m {(||{x_{i,k}}|{|^2} - {{\tilde c}_i})} ,0\} $ to denote the solution error, objective error and constraint violation, respectively, where the solution ${{{\hat x}^*}}$ and optimal value ${f^ * } + {g^ * }$ are provided by the centralized ADMM \cite{Chen2021ALM}. The simulation results are reported in Figs. \ref{PerforCompa} and \ref{figOutPutPo}. Note that Fig. \ref{PerforCompa} shows the convergence performance of the solution error and objective error in terms of time cost. We observe that: i) the exact FDOA (i.e., ${\varepsilon _{i,k}} = 0$) spend more time than other inexact versions; ii) the proposed FDOA outperforms the centralized ADMM, and exhibits a prominent advantage in the evolution of both solution error and objective error than these introduced exact algorithms. Fig. \ref{figOutPutPo} shows that FDOA achieves higher accuracy in terms of constraint violation. Therefore, it can be concluded that the inexact computation in FDOA reduces computational cost without losing the convergence accuracy, and also enjoys superiority compared with related exact algorithms.

\begin{figure}[!t]
   \centering
   \subfigure[Evolution of solution error.]{\includegraphics[width=2.5in]{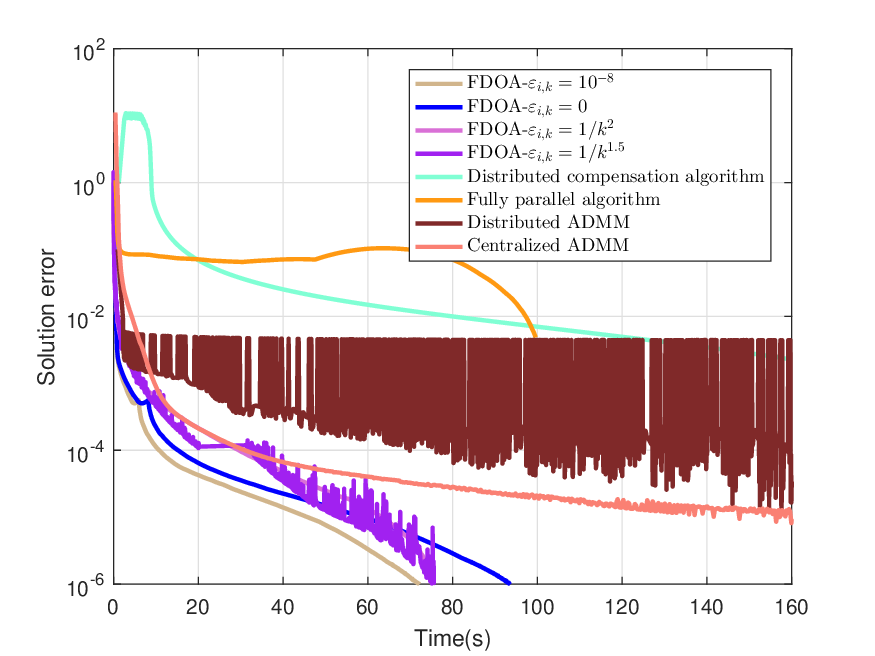} \label{Evolusolu}}
   \hfil
   \subfigure[Evolution of objective error.]{\includegraphics[width=2.5in]{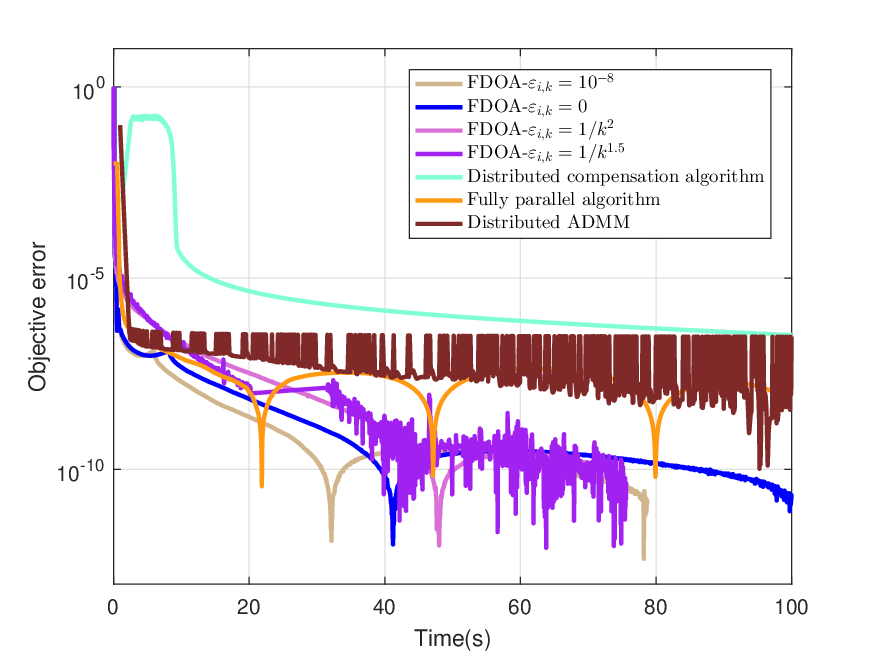} \label{Evoluobje}}
   \caption{Performance comparison with state-of-the-arts.}\label{PerforCompa}
\end{figure}

\begin{figure}[!t]
\centerline{\includegraphics[width=2.5in]{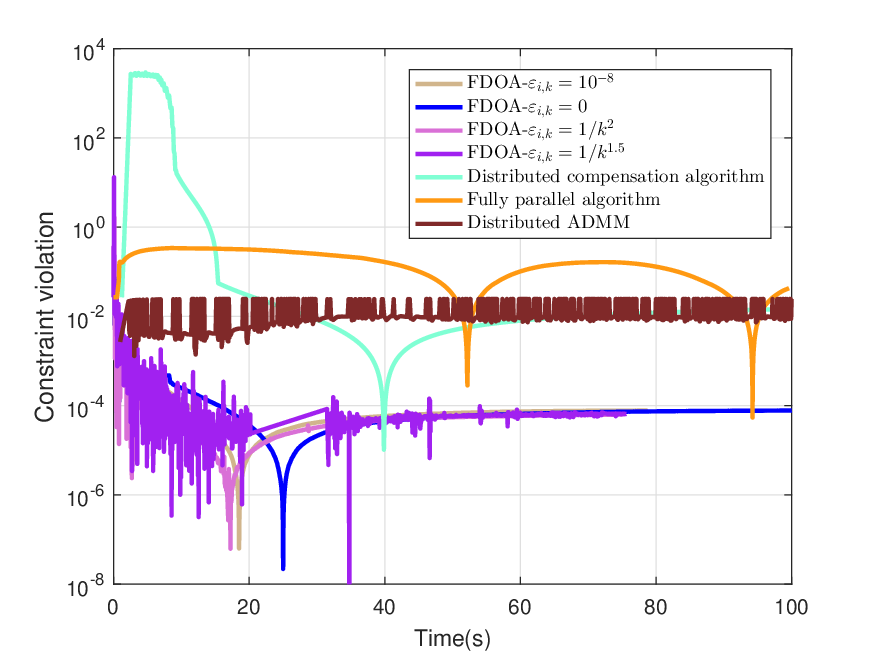}}
\caption{Evolution of constraint violation.} \label{figOutPutPo}
\end{figure}

}

\section{Conclusion} \label{SecConcl}

This paper developed a fully distributed inexact algorithm for solving a nonsmooth composite convex optimization problem with locally nonlinear constraints and globally coupling nonlinear ones. It is more practical than existing methods as the proposed algorithm can address many studied optimization problems without restrictive assumption on objective functions or constraints. Theoretical analysis showed that our {algorithm has an $o(1/k)$ convergence rate even by means} of inexact iterations. {Future research will extend the proposed algorithm to asynchronous communication networks (e.g., asynchronous lock-free algorithms \cite{Cannelli2016Parallel}) and directed/weight-unbalanced graphs, derive its (super)linear convergence rate under mild assumptions.}

\appendices

The other mathematical notations are listed in Table \ref{tab:notation}.

\begin{table}[!t]
\centering
\caption{Definitions and dimensions of key symbols}
\label{tab:notation}
\small
\renewcommand{\arraystretch}{1.35}
\setlength{\tabcolsep}{4pt}
\newcommand{\RowStrut}{\rule[-0.8ex]{0pt}{4.2ex}}
\begin{tabular}{@{}p{0.15\linewidth}p{0.18\linewidth}p{0.55\linewidth}@{}}
\hline
\textbf{Symbol} & \textbf{Index} & \textbf{Dimension / Domain} \\ \hline
\RowStrut $f_i$ & cf. \eqref{OrigProblema} & ${\mathbb{R}^{{n_i}}} \to \mathbb{R}$ \\
\RowStrut $g_i$ & cf. \eqref{OrigProblema} & ${\mathbb{R}^{{{\tilde r}_i}}} \to \mathbb{R} \cup \{  + \infty \} $ \\
\RowStrut $R_i$ & cf. \eqref{OrigProblema} & $\in {\mathbb{R}^{{{\tilde r}_i} \times {n_i}}}$ \\
\RowStrut $B_{ij}$, $B_{ji}$ & cf. \eqref{OrigProblemb} & $\in {\mathbb{R}^{{l_{\left( {i,j} \right)}} \times {n_i}}}, \in {\mathbb{R}^{{l_{\left( {i,j} \right)}} \times {n_j}}}$ \\
\RowStrut $h_i$ & cf. \eqref{OrigProblemc} & ${\mathbb{R}^{{{\tilde r}_i}}} \to {\mathbb{R}^{{p_i}}} $ \\
\RowStrut $A_i$ & cf. \eqref{OrigProblemd} & $\in {\mathbb{R}^{q \times {n_i}}}$ \\
\RowStrut $C_i$ & cf. \eqref{OrigProblemd} & ${\mathbb{R}^{{n_i}}} \to {\mathbb{R}^c}$ \\

\RowStrut $l$, $n$ & cf. \eqref{Edgemaps} & $l = \sum\nolimits_{(i,j) \in \mathcal{E}} {{l_{(i,j)}}} ,n = \sum\nolimits_{i = 1}^m {{n_i}} $ \\
\RowStrut $\tilde\Pi_{(i,j)}$ &  cf. \eqref{Edgemaps}& $\in {\mathbb{R}^{(2{l_{(i,j)}}) \times n}}$ \\
\RowStrut $\tilde\Pi$ &  cf. \eqref{Edgemaps}& $\in {\mathbb{R}^{(2{l}) \times n}}$ \\
\RowStrut ${\tilde D_{( {i,j} )}}$ &  cf. \eqref{Edgeconst} & Edge constraint set of \eqref{OrigProblemb} \\
\RowStrut $\tilde D$&  cf. \eqref{RefProblema} & $\prod\nolimits_{(i,j) \in \mathcal{E}} {{{\tilde D}_{\left( {i,j} \right)}}}$ \\
\RowStrut $\mathcal{\tilde Q}$ & cf. \eqref{OrigKKTa} & $(x,u,{s_g}) \to \partial f(x) + {R^{\text{T}}}\partial g(Rx) + {{\tilde \Pi }^{\text{T}}}\partial {\mathcal{I}_{\tilde D}}(\tilde \Pi x) + {(\partial h(x))^{\text{T}}}u + {(\partial \mathcal{C}(x))^{\text{T}}}({1_m} \otimes {s_g})$ \\
\RowStrut ${\Omega _{{\rm{dual}}}}$& cf. \eqref{OrigKKTc} & ${\mathbb{R}^q} \times \mathbb{R}_ + ^c$ \\
\RowStrut $\Theta$& cf. \eqref{OrigKKTb}  & $\prod\nolimits_{i = 1}^m {\mathbb{R}_ + ^{{p_i}}} $ \\
\RowStrut $\mathcal{\bar Q}$& cf. \eqref{EquivaKKTa} & $(x,w,\lambda ,u,{s_g}) \to \partial f(x) + {R^{\text{T}}}w + {{\tilde \Pi }^{\text{T}}}\lambda  + {(\partial h(x))^{\text{T}}}u + {(\partial \mathcal{C}(x))^{\text{T}}}({1_m} \otimes {s_g})$ \\
\RowStrut ${{\bar \Pi }_{\left( {i,j} \right)}}$ & cf. \eqref{DisMNconditf} & $s \to ({E_{ij}}{s_i},{\text{ }}{E_{ji}}{s_j}) \in {\mathbb{R}^{(2(c + q)) \times (m(c + q))}}$ \\
\RowStrut ${\bar \Pi }$ & After \eqref{Edgeconss} & $s \to {({{\bar \Pi }_{\left( {i,j} \right)}}s)_{(i,j) \in \mathcal{E}}} \in {\mathbb{R}^{(2\left| \mathcal{E} \right|(c + q)) \times (m(c + q))}}$ \\
\RowStrut ${{\bar D}_{\left( {i,j} \right)}}$ & After \eqref{Edgeconss} & Edge constraint set of \eqref{Edgeconss} \\
\RowStrut ${\bar D}$ & Before \eqref{MinMaxPro} & $\prod\nolimits_{(i,j) \in \mathcal{E}} {{{\bar D}_{\left( {i,j} \right)}}}$  \\
\RowStrut ${\mathcal{A}_i}$ & cf. \eqref{DisMNcondita} & $({x_i},{w_i},{\{ {\lambda _{\left( {i,j} \right),i}}\} _{j \in {\mathcal{N}_i}}},{u_i},{s_i}) \to \partial {f_i}({x_i}) + R_i^{\text{T}}{w_i} + \sum\nolimits_{j \in {\mathcal{N}_i}} {B_{ij}^{\text{T}}{\lambda _{\left( {i,j} \right),i}}}  + {(\partial {h_i}({x_i}))^{\text{T}}}{u_i} + {(\partial {\mathcal{C}_i}({x_i}))^{\text{T}}}{s_i}$ \\
\RowStrut ${\Omega _{\text{d}}} $ & Before \eqref{MinMaxPro} & $ \prod\nolimits_{i = 1}^m {{\Omega _{{\text{dual}}}}} $ \\
\RowStrut ${\mathcal{K}_1} $ & cf. \eqref{Totlalvariaineq} & $\psi  \to ({R^{\text{T}}}w + {{\tilde \Pi }^{\text{T}}}\lambda , - Rx, - \tilde \Pi x,0,{{\bar \Pi }^{\text{T}}}z, - \bar \Pi s) $ \\
\RowStrut ${\mathcal{K}_2} $ & cf. \eqref{Totlalvariaineq} & $\psi  \to (h(x),\mathcal{C}(x),u,s) $ \\
\RowStrut ${\mathcal{K}_3} $ & cf. \eqref{Totlalvariaineq} & $\psi  \to (u,s, - h(x), - \mathcal{C}(x))$ \\ \hline
\end{tabular}
\end{table}

\subsection{Proof of Lemma \ref{DistriMN}} \label{ProofDistriMN}
\begin{proof}
It is straightforward that \eqref{DisMNconditb}-\eqref{DisMNconditd} are the local formulations of \eqref{EquivaKKTb}-\eqref{EquivaKKTd}, respectively. Then, we show that \eqref{DisMNcondite} together with \eqref{DisMNconditf} contributes to \eqref{EquivaKKTe}. It follows from \cite[Theorem 14.3]{Bauschke2011Convex} that \eqref{DisMNconditf} $ \Leftrightarrow $ $z_{\left( {i,j} \right)}^ *  = {\text{pro}}{{\text{x}}_{\mathcal{I}_{{{\bar D}_{\left( {i,j} \right)}}}^ \star }}(z_{\left( {i,j} \right)}^ *  + {\bar \Pi _{\left( {i,j} \right)}}{s^ * })$ $ \Leftrightarrow $ $z_{\left( {i,j} \right)}^ *  = z_{\left( {i,j} \right)}^ *  + {{\bar \Pi }_{\left( {i,j} \right)}}{s^ * } - {\text{pro}}{{\text{x}}_{{\mathcal{I}_{{{\bar D}_{\left( {i,j} \right)}}}}}}(z_{\left( {i,j} \right)}^ *  + {{\bar \Pi }_{\left( {i,j} \right)}}{s^ * })$ $ \Leftrightarrow {{\bar \Pi }_{\left( {i,j} \right)}}{s^ * } = {\mathcal{P}_{{{\bar D}_{\left( {i,j} \right)}}}}( {z_{\left( {i,j} \right)}^ *  + {{\bar \Pi }_{\left( {i,j} \right)}}{s^ * }} )$. Recalling the set ${\bar D_{\left( {i,j} \right)}}$ and the mapping ${\bar \Pi _{\left( {i,j} \right)}}$, one can calculate the projection ${\mathcal{P}_{{{\bar D}_{\left( {i,j} \right)}}}}:z_{\left( {i,j} \right)}^ *  + {{\bar \Pi }_{\left( {i,j} \right)}}{s^ * } \to \frac{1}{2}(z_{\left( {i,j} \right),i}^ *  + {E_{ij}}s_i^ *  - z_{\left( {i,j} \right),j}^ *  - {E_{ji}}s_j^ * ,z_{\left( {i,j} \right),j}^ *  + {E_{ji}}s_j^ *  - z_{\left( {i,j} \right),i}^ *  - {E_{ij}}s_i^ * )$, which is combined with aforementioned results to have
\begin{flalign}\label{ProofDistab}
\left[ {\begin{array}{*{20}{c}}
   {{E_{ij}}s_i^ * }  \\
   {{E_{ji}}s_j^ * }  \\
 \end{array} } \right] = \frac{1}
{2}\left[ {\begin{array}{*{20}{c}}
   {z_{(i,j),i}^ *  + {E_{ij}}s_i^ *  - (z_{(i,j),j}^ *  + {E_{ji}}s_j^ * )}  \\
   {z_{(i,j),j}^ *  + {E_{ji}}s_j^ *  - (z_{(i,j),i}^ *  + {E_{ij}}s_i^ * )}  \\
 \end{array} } \right]
\end{flalign}
Summing both sides of \eqref{ProofDistab} gives ${E_{ij}}s_i^ *  + {E_{ji}}s_j^ *  = 0$. Hence, recalling $E_{ij}$ and $E_{ji}$ in \eqref{Edgeconss} results in $s_i^ *  - s_j^ *  = 0$, i.e., there exists $s_g^ * $ such that $s_1^ *  =  \cdots  = s_m^ *  = s_g^ * $. In return, plugging ${E_{ij}}s_i^ *  + {E_{ji}}s_j^ *  = 0$ into \eqref{ProofDistab} yields $z_{\left( {i,j} \right),i}^ *  = z_{\left( {i,j} \right),j}^ * $ for any $\left( {i,j} \right) \in \mathcal{E}$. Thus, substituting $s_i^*$ of \eqref{DisMNcondita} and \eqref{DisMNcondite} with $s_g^*$, we can find that ${\mathcal{A}_i}(x_i^ * ,w_i^ * ,{\{ \lambda _{\left( {i,j} \right),i}^ * \} _{j \in {\mathcal{N}_i}}},u_i^ * ,s_g^ * )$ is the local component of \eqref{EquivaKKTa}, and there exists $\upsilon _i^ *  \in \partial {\mathcal{I}_{{\Omega _{{\text{dual}}}}}}(s_g^ * )$ such that $0 = \upsilon _i^ *  - {\mathcal{C}_i}\left( {x_i^ * } \right) + \sum\nolimits_{j \in {\mathcal{N}_i}} {E_{ij}^{\rm T}z_{\left( {i,j} \right),i}^ * } $. In addition, sum \eqref{DisMNcondite} from $i=1$ to $i=m$ to have
\begin{flalign}\label{ProofDistbc}
0 \in \sum\limits_{i = 1}^m {\upsilon _i^ * }  - \sum\limits_{i = 1}^m {{\mathcal{C}_i}\left( {x_i^ * } \right)}  + \sum\limits_{i = 1}^m {\sum\limits_{j \in {\mathcal{N}_i}} {E_{ij}^{\rm T}z_{\left( {i,j} \right),i}^ * } }
\end{flalign}
Note that the set ${\Omega _{{\text{dual}}}}$ is convex, $\sum\nolimits_{i = 1}^m {\upsilon _i^ * }  \in \partial {\mathcal{I}_{{\Omega _{{\text{dual}}}}}}(s_g^ * )$ holds as well. Moreover, the obtained results $z_{\left( {i,j} \right),i}^ *  = z_{\left( {i,j} \right),j}^ * $ and ${E_{ij}} + {E_{ji}} = 0$ lead to $\sum\nolimits_{i = 1}^m {\sum\nolimits_{j \in {\mathcal{N}_i}} {E_{ij}^{\rm T}z_{\left( {i,j} \right),i}^ * } }  = 0$, \eqref{ProofDistbc} then reduces to \eqref{EquivaKKTe}. Consequently, it can be concluded that ${x^ * },{w^ * },{\lambda ^ * },{u^ * },{s^ * } = {{\mathbf{1}}_m} \otimes s_g^ * $ in \eqref{DisMNcondit} satisfy the condition \eqref{EquivaKKT} as well. Recalling the equivalence between \eqref{OrigKKT} and \eqref{EquivaKKT}, one can further confirm that $x^*$ solves \eqref{OrigProblem}.
\end{proof}

\subsection{Proof of Proposition \ref{EquivTilexi}} \label{ProofEquivTilexi}
\begin{proof}
According to \eqref{LocalPredictd} and \eqref{LocalPredicte}, they can be rewritten as
\begin{flalign*}
\begin{gathered}
  0 \in \partial {\mathcal{I}_{\mathbb{R}_ + ^{{p_i}}}}({{\tilde u}_{i,k}}) + \frac{1}{{{\delta _i}}}{{\tilde u}_{i,k}} - \frac{1}
{{{\delta _i}}}{u_{i,k}} - {h_i}({{\tilde x}_{i,k}}) \hfill \\
  0 \in \partial {\mathcal{I}_{{\Omega _{{\text{dual}}}}}}({{\tilde s}_{i,k}}) + \frac{1}{{{\eta _i}}}{{\tilde s}_{i,k}} - \frac{1}{{{\eta _i}}}{s_{i,k}} - {\mathcal{C}_i}({{\tilde x}_{i,k}}) \hfill \\
  \qquad + \sum\nolimits_{j \in {\mathcal{N}_i}} {E_{ij}^{\text{T}}(2{{\tilde z}_{\left( {i,j} \right),i,k}} - {z_{\left( {i,j} \right),i,k}})}  \hfill \\
\end{gathered}
\end{flalign*}
Under Assumption \ref{AssumpNonempty}, the above results and \eqref{Localtildx} are the KKT condition of the following min-max problem:
\begin{flalign} \label{ProofEquiab}
\mathop {\max }\limits_{{u_i} \in \mathbb{R}_ + ^{{p_i}},{s_i} \in {\Omega _{{\text{dual}}}}} \mathop {\min }\limits_{{x_i} \in {\mathbb{R}^{{n_i}}}} {{\bar F}_{i,k}}\left( {{x_i},{u_i},{s_i}} \right)
\end{flalign}
where ${{\bar F}_{i,k}}( {{x_i},{u_i},{s_i}} ) = {f_i}( {{x_i}} ) + \frac{1}{{2{\alpha _i}}}||{x_i} - {{\overset{\lower0.5em\hbox{$\smash{\scriptscriptstyle\frown}$}}{x} }_{i,k}}|{|^2} + \langle {{s_i},{\mathcal{C}_i}({x_i})} \rangle  - \frac{1}
{{2{\delta _i}}}||{u_i} - {u_{i,k}}|{|^2} + \langle {{u_i},{h_i}({x_i})} \rangle  - \frac{1}{{2{\eta _i}}}||{s_i} - {{\overset{\lower0.5em\hbox{$\smash{\scriptscriptstyle\frown}$}}{s} }_{i,k}}|{|^2}$. Since $f_i$, ${\mathcal{C}_i}$ and ${h_i}$ are convex functions, ${{\bar F}_{i,k}}( {{x_i},{u_i},{s_i}} )$ is strongly convex at $x_i$, while being strongly concave at $col({{u_i},{s_i}})$. Hence, problem \eqref{ProofEquiab} has the unique solution, which means that when ${u_i} = {\tilde u_{i,k}},{s_i} = {\tilde s_{i,k}}$, ${\tilde x_{i,k}}$ then can solve the equivalent minimization problem of \eqref{ProofEquiab}:
\begin{flalign} \label{ProofEquibc}
{\min _{{x_i} \in {\mathbb{R}^{{n_i}}}}}{\tilde F_{i,k}}({x_i},{u_{i,k}},{\overset{\lower0.5em\hbox{$\smash{\scriptscriptstyle\frown}$}}{s} _{i,k}})
\end{flalign}
where ${\tilde F_{i,k}}({x_i},{u_{i,k}},{\overset{\lower0.5em\hbox{$\smash{\scriptscriptstyle\frown}$}}{s} _{i,k}}) = {f_i}\left( {{x_i}} \right) + \frac{1}{{2{\alpha _i}}}||{x_i} - {\overset{\lower0.5em\hbox{$\smash{\scriptscriptstyle\frown}$}}{x} _{i,k}}|{|^2} + \tilde F_{i,k}^u({x_i},{u_{i,k}}) + \tilde F_{i,k}^s({x_i},{\overset{\lower0.5em\hbox{$\smash{\scriptscriptstyle\frown}$}}{s} _{i,k}})$ with $\tilde F_{i,k}^u({x_i},{u_{i,k}}) = \langle {{\mathcal{P}_{\mathbb{R}_ + ^{{p_i}}}}({u_{i,k}} + {\delta _i}{h_i}({x_i})),{h_i}({x_i})} \rangle  - \frac{1}
{{2{\delta _i}}}||{\mathcal{P}_{\mathbb{R}_ + ^{{p_i}}}}({u_{i,k}} + {\delta _i}{h_i}({x_i})) - {u_{i,k}}|{|^2}$ and $\tilde F_{i,k}^s({x_i},{{\overset{\lower0.5em\hbox{$\smash{\scriptscriptstyle\frown}$}}{s} }_{i,k}}) = \langle {{\mathcal{P}_{{\Omega _{{\text{dual}}}}}}({{\overset{\lower0.5em\hbox{$\smash{\scriptscriptstyle\frown}$}}{s} }_{i,k}} + {\eta _i}{\mathcal{C}_i}({x_i})),{\mathcal{C}_i}({x_i})} \rangle  - \frac{1}{{2{\eta _i}}}||{\mathcal{P}_{{\Omega _{{\text{dual}}}}}}({{\overset{\lower0.5em\hbox{$\smash{\scriptscriptstyle\frown}$}}{s} }_{i,k}} + {\eta _i}{\mathcal{C}_i}({x_i})) - {{\overset{\lower0.5em\hbox{$\smash{\scriptscriptstyle\frown}$}}{s} }_{i,k}}|{|^2}$. Note that the fact $\langle {\max \{ {0,r} \},r} \rangle  = {\| {\max \{ {0,r} \}} \|^2}$ gives $\langle {{\mathcal{P}_{\mathbb{R}_ + ^{{p_i}}}}( {{u_i}} ),{u_i}} \rangle  = {\| {{{[{u_i}]}_1}} \|^2} + {\| {\max \{ {0,{{[{u_i}]}_2}} \}} \|^2} = {\| {{\mathcal{P}_{\mathbb{R}_ + ^{{p_i}}}}( {{u_i}} )} \|^2}$. Based on these derivation, we expand $\tilde F_{i,k}^u({x_i},{u_{i,k}})$ to have
\begin{flalign} \label{ProofEquicd}
\begin{gathered}
  \quad \tilde F_{i,k}^u( {{x_i},{u_{i,k}}} ) \hfill \\
   =  - \frac{1}{{2{\delta _i}}}{\| {{\mathcal{P}_{\mathbb{R}_ + ^{{p_i}}}}( {{u_{i,k}} + {\delta _i}{h_i}( {{x_i}} )} )} \|^2} - \frac{1}{{2{\delta _i}}}{\| {{u_{i,k}}} \|^2} \hfill \\
  \quad + \frac{1}{{{\delta _i}}}\langle {{\mathcal{P}_{\mathbb{R}_ + ^{{p_i}}}}( {{u_{i,k}} + {\delta _i}{h_i}( {{x_i}} )} ),{u_{i,k}}} \rangle  \hfill \\
  \quad + \langle {{\mathcal{P}_{\mathbb{R}_ + ^{{p_i}}}}( {{u_{i,k}} + {\delta _i}{h_i}( {{x_i}} )} ),{h_i}( {{x_i}} )} \rangle  \hfill \\
   =  - \frac{1}{{2{\delta _i}}}{\| {{\mathcal{P}_{\mathbb{R}_ + ^{{p_i}}}}( {{u_{i,k}} + {\delta _i}{h_i}( {{x_i}} )} )} \|^2} - \frac{1}
{{2{\delta _i}}}{\| {{u_{i,k}}} \|^2} \hfill \\
  \quad + \frac{1}{{{\delta _i}}}\langle {{\mathcal{P}_{\mathbb{R}_ + ^{{p_i}}}}( {{u_{i,k}} + {\delta _i}{h_i}( {{x_i}} )} ),{u_{i,k}} + {\delta _i}{h_i}( {{x_i}} )} \rangle  \hfill \\
   = \frac{1}{{2{\delta _i}}}{\| {{\mathcal{P}_{\mathbb{R}_ + ^{{p_i}}}}( {{u_{i,k}} + {\delta _i}{h_i}( {{x_i}} )} )} \|^2} - \frac{1}{{2{\delta _i}}}{\| {{u_{i,k}}} \|^2} \hfill \\
\end{gathered}
\end{flalign}
where the last equality uses $\langle {{\mathcal{P}_{\mathbb{R}_ + ^{{p_i}}}}({u_{i,k}} + {\delta _i}{h_i}({x_i})),{u_{i,k}} + {\delta _i}{h_i}({x_i})} \rangle  = ||{\mathcal{P}_{\mathbb{R}_ + ^{{p_i}}}}({u_{i,k}} + {\delta _i}{h_i}({x_i}))|{|^2}$. As for $\tilde F_{i,k}^s({x_i},{\overset{\lower0.5em\hbox{$\smash{\scriptscriptstyle\frown}$}}{s} _{i,k}})$, we further get
\begin{flalign} \label{ProofEquide}
\begin{gathered}
  \quad \tilde F_{i,k}^s({x_i},{{\overset{\lower0.5em\hbox{$\smash{\scriptscriptstyle\frown}$}}{s} }_{i,k}}) \hfill \\
   = \langle {{\mathcal{P}_{{\Omega _{{\text{dual}}}}}}({{\overset{\lower0.5em\hbox{$\smash{\scriptscriptstyle\frown}$}}{s} }_{i,k}} + {\eta _i}{\mathcal{C}_i}({x_i})),{\mathcal{C}_i}({x_i})} \rangle  \hfill \\
  \quad  - \frac{1}{{2{\eta _i}}}||{\mathcal{P}_{{\Omega _{{\text{dual}}}}}}({{\overset{\lower0.5em\hbox{$\smash{\scriptscriptstyle\frown}$}}{s} }_{i,k}} + {\eta _i}{\mathcal{C}_i}({x_i}))|{|^2} \hfill \\
  \quad  - \frac{1}{{2{\eta _i}}}||{{\overset{\lower0.5em\hbox{$\smash{\scriptscriptstyle\frown}$}}{s} }_{i,k}}|{|^2} - \frac{1}{{{\eta _i}}}\langle {{\mathcal{P}_{{\Omega _{{\text{dual}}}}}}({{\overset{\lower0.5em\hbox{$\smash{\scriptscriptstyle\frown}$}}{s} }_{i,k}} + {\eta _i}{\mathcal{C}_i}({x_i})),{{\overset{\lower0.5em\hbox{$\smash{\scriptscriptstyle\frown}$}}{s} }_{i,k}}} \rangle  \hfill \\
   =  - \frac{1}{{2{\eta _i}}}||{\mathcal{P}_{{\Omega _{{\text{dual}}}}}}({{\overset{\lower0.5em\hbox{$\smash{\scriptscriptstyle\frown}$}}{s} }_{i,k}} + {\eta _i}{\mathcal{C}_i}({x_i}))|{|^2} - \frac{1}
{{2{\eta _i}}}||{{\overset{\lower0.5em\hbox{$\smash{\scriptscriptstyle\frown}$}}{s} }_{i,k}}|{|^2} \hfill \\
  \quad  + \frac{1}{{{\eta _i}}}\langle {{\mathcal{P}_{{\Omega _{{\text{dual}}}}}}({{\overset{\lower0.5em\hbox{$\smash{\scriptscriptstyle\frown}$}}{s} }_{i,k}} + {\eta _i}{\mathcal{C}_i}({x_i})),{{\overset{\lower0.5em\hbox{$\smash{\scriptscriptstyle\frown}$}}{s} }_{i,k}} + {\eta _i}{\mathcal{C}_i}({x_i})} \rangle  \hfill \\
   = \frac{1}{{2{\eta _i}}}||{\mathcal{P}_{{\Omega _{{\text{dual}}}}}}({{\overset{\lower0.5em\hbox{$\smash{\scriptscriptstyle\frown}$}}{s} }_{i,k}} + {\eta _i}{\mathcal{C}_i}({x_i}))|{|^2} - \frac{1}
{{2{\eta _i}}}||{{\overset{\lower0.5em\hbox{$\smash{\scriptscriptstyle\frown}$}}{s} }_{i,k}}|{|^2} \hfill \\
\end{gathered}
\end{flalign}
where the last equality follows from $\langle {{\mathcal{P}_{{\Omega _{{\text{dual}}}}}}({{\overset{\lower0.5em\hbox{$\smash{\scriptscriptstyle\frown}$}}{s} }_{i,k}} + {\eta _i}{\mathcal{C}_i}({x_i})),{{\overset{\lower0.5em\hbox{$\smash{\scriptscriptstyle\frown}$}}{s} }_{i,k}} + {\eta _i}{\mathcal{C}_i}({x_i})} \rangle  = ||{\mathcal{P}_{{\Omega _{{\text{dual}}}}}}({\overset{\lower0.5em\hbox{$\smash{\scriptscriptstyle\frown}$}}{s} _{i,k}} + {\eta _i}{\mathcal{C}_i}({x_i}))|{|^2}$. Combining \eqref{ProofEquicd} and \eqref{ProofEquide}, and eliminating the constant terms independent of $x_i$, problem \eqref{ProofEquibc} can be equivalently reformulated into the new minimization problem, i.e., ${\min _{{x_i} \in {\mathbb{R}^{{n_i}}}}}{F_{i,k}}({x_i},{u_{i,k}},{\overset{\lower0.5em\hbox{$\smash{\scriptscriptstyle\frown}$}}{s} _{i,k}})$. Therefore, the desired conclusion is obtained.
\end{proof}

\subsection{Proof of Lemma \ref{LemInetqua}} \label{ProofLemInetqua}
\begin{proof}
According to \eqref{LocalPredict}-\eqref{Localtillamzij} and Proposition \ref{EquivTilexi}, one can confirm that the compact updates of Algorithm \ref{alg:1} take the form of \eqref{AlgoriOne} with ${\tilde \psi _k} = col\{ {\tilde x_k},{\tilde w_k},{\tilde \lambda _k},{\tilde u_k},{\tilde s_k},{\tilde z_k}\} $ being
\begin{subequations}\label{ProofAlgoriV}
  \begin{equation}\label{ProofAlgoriVa}
   {{\tilde w}_k} = {\text{prox}}_{{g^ \star }}^{{\Gamma ^{ - 1}}}({w_k} + \Gamma R{x_k}) \qquad \qquad \qquad \qquad \,\,\,
  \end{equation}
  \begin{equation}\label{ProofAlgoriVb}
  {{\tilde \lambda }_k} = {\text{prox}}_{\mathcal{I}_{\tilde D}^ \star }^{{\Xi ^{ - 1}}}({\lambda _k} + \Xi \tilde \Pi {x_k}) \qquad \qquad \qquad \qquad \,\,\,
  \end{equation}
  \begin{equation}\label{ProofAlgoriVc}
   {{\tilde z}_k} = {\text{prox}}_{\mathcal{I}_{\bar D}^ \star }^{{\Phi ^{ - 1}}}({z_k} + \Phi \bar \Pi {s_k}) \qquad \qquad \qquad \qquad \quad
  \end{equation}
  \begin{equation}\label{ProofAlgoriVd}
  {{\tilde x}_k} \approx \arg {\min _{x \in {\mathbb{R}^n}}}{F_k}(x,{u_k},{s_k} - \Upsilon {{\bar \Pi }^{\text{T}}}(2{{\tilde z}_k} - {z_k})) \,\,
  \end{equation}
  \begin{equation}\label{ProofAlgoriVe}
   {{\tilde u}_k} = {\text{prox}}_{{\mathcal{I}_\Theta }}^{{\Sigma ^{ - 1}}}({u_k} + \Sigma h({{\tilde x}_k})) \qquad \qquad \qquad \qquad
  \end{equation}
  \begin{equation}\label{ProofAlgoriVf}
   {{\tilde s}_k} = {\text{prox}}_{{\mathcal{I}_{{\Omega _{\text{d}}}}}}^{{\Upsilon ^{ - 1}}}({s_k} - \Upsilon {{\bar \Pi }^{\text{T}}}(2{{\tilde z}_k} - {z_k}) + \Upsilon \mathcal{C}({{\tilde x}_k}))
  \end{equation}
\end{subequations}
where ${F_k}(x,{u_k},{s_k} - \Upsilon {\bar \Pi ^{\text{T}}}(2{\tilde z_k} - {z_k})) = \sum\nolimits_{i = 1}^m {{F_{i,k}}({x_i},{u_{i,k}},{s_{i,k}} - {\eta _i}\sum\nolimits_{j \in {\mathcal{N}_i}} {E_{ij}^{\text{T}}(2{{\tilde z}_{( {i,j} ),i,k}} - {z_{( {i,j} ),i,k}})} )} $. Inspired by the variational inequality of a minimization convex problem in \cite{He2020Optimal}, for any $w \in {\mathbb{R}^{\tilde r}}$, the rule \eqref{ProofAlgoriVa} can be interpreted into
\begin{flalign} \label{ProofAlgoriab}
  0 \leq {g^ \star }(w) \!-\! {g^ \star }({{\tilde w}_k}) \!+\! \langle {w \!-\! {{\tilde w}_k},{\Gamma ^{ - 1}}({{\tilde w}_k} \!-\! ({w_k} + \Gamma R{x_k}))} \rangle
\end{flalign}
Analogously, for any $\lambda  \in {\mathbb{R}^{2l}}$, $z \in {\mathbb{R}^{2( {q + c} )}}$, $u \in {\mathbb{R}^p}$ and $s \in {\mathbb{R}^{( {q + c} )}}$, these iteration formulas \eqref{ProofAlgoriVb}, \eqref{ProofAlgoriVc}, \eqref{ProofAlgoriVe} and \eqref{ProofAlgoriVf} are respectively expressed by
\begin{flalign} \label{ProofAlgoribc}
\begin{gathered}
  0 \leq \mathcal{I}_{\tilde D}^ \star (\lambda ) \!-\! \mathcal{I}_{\tilde D}^ \star ({{\tilde \lambda }_k}) \!+\! \langle {\lambda  - {{\tilde \lambda }_k},{\Xi ^{ - 1}}({{\tilde \lambda }_k} \!-\! ({\lambda _k} + \Xi \tilde \Pi {x_k}))} \rangle  \hfill \\
  0 \leq \mathcal{I}_{\bar D}^ \star (z) \!-\! \mathcal{I}_{\bar D}^ \star ({{\tilde z}_k}) \!+\! \langle {z - {{\tilde z}_k},{\Phi ^{ - 1}}({{\tilde z}_k} \!-\! ({z_k} + \Phi \bar \Pi {s_k}))} \rangle  \hfill \\
  { 0 \leq {\mathcal{I}_\Theta }(u) \!-\! {\mathcal{I}_\Theta }({{\tilde u}_k}) \!+\! \langle {u \!-\! {{\tilde u}_k},{\Sigma ^{ - 1}}({{\tilde u}_k} \!-\! ({u_k} \!+\! \Sigma h({{\tilde x}_k})))} \rangle}  \hfill \\
  0 \leq {\mathcal{I}_{{\Omega _{\text{d}}}}}(s) \!-\! {\mathcal{I}_{{\Omega _{\rm{d}}}}}({{\tilde s}_k}) \hfill \\
         \qquad + \langle {s \!-\! {{\tilde s}_k},{\Upsilon ^{ - 1}}({{\tilde s}_k} \!-\! ({s_k} \!-\! \Upsilon {{\bar \Pi }^{\rm{T}}}(2{{\tilde z}_k} - {z_k}) \!+\! \Upsilon \mathcal{C}({{\tilde x}_k})))} \rangle  \hfill \\
\end{gathered}
\end{flalign}
Through the derivation of Appendix \ref{ProofEquivTilexi}, one has that \eqref{ProofAlgoriVd} can be addressed if and only if the condition holds, i.e.,
\begin{flalign} \label{ProofAlgoriza}
\begin{gathered}
{\zeta _k} \in {\Psi ^{ - 1}}{\tilde x_k} - {\Psi ^{ - 1}}{x_k} + \partial f({\tilde x_k}) + {R^{\text{T}}}{\tilde w_k} + {\tilde \Pi ^{\rm{T}}}{\tilde \lambda _k} \hfill \\
 \qquad + {(\partial h({\tilde x_k}))^{\rm{T}}}{\tilde u_k} + {(\partial \mathcal{C}({\tilde x_k}))^{\rm{T}}}{\tilde s_k} \hfill \\
\end{gathered}
\end{flalign}
with ${\zeta _k} = col\{ {\zeta _{1,k}}, \cdots ,{\zeta _{m,k}}\} $. Based on this condition, \eqref{ProofAlgoriVd} is equivalent to the strongly convex problem ${\min _{x \in {\mathbb{R}^n}}}\{ f(x) + \frac{1}{2}||x - {x_k}||_{{\Psi ^{ - 1}}}^2 + \langle {h(x),{{\tilde u}_k}} \rangle  + \langle {\mathcal{C}(x),{{\tilde s}_k}} \rangle  + \langle {x,{R^{\rm{T}}}{{\tilde w}_k} + {{\tilde \Pi }^{\text{T}}}{{\tilde \lambda }_k} - {\zeta _k}} \rangle \} $, which thus results in the following variational inequality:
\begin{flalign} \label{ProofAlgoricd}
\begin{gathered}
  0 \leq f(x) \!-\! f({{\tilde x}_k}) \!+\! \langle {h(x) \!-\! h({{\tilde x}_k}),{{\tilde u}_k}} \rangle  \!+\! \langle {\mathcal{C}(x) \!-\! \mathcal{C}({{\tilde x}_k}),{{\tilde s}_k}} \rangle  \hfill \\
  \qquad + \langle {x - {{\tilde x}_k},{\Psi ^{ - 1}}({{\tilde x}_k} - {x_k}) + {R^{\rm{T}}}{{\tilde w}_k} + {{\tilde \Pi }^{\rm{T}}}{{\tilde \lambda }_k} - {\zeta _k}} \rangle  \hfill \\
\end{gathered}
\end{flalign}

Next, we only show the equivalence between \eqref{ProofAlgoriab}-\eqref{ProofAlgoricd} and \eqref{AlgoriVariaIneq}. Firstly, we prove that \eqref{ProofAlgoriab}-\eqref{ProofAlgoricd} lead to \eqref{AlgoriVariaIneq}. Summing these inequalities from \eqref{ProofAlgoriab} to \eqref{ProofAlgoricd}, one obtains that
\begin{flalign} \label{ProofAlgoride}
\begin{gathered}
  0 \leq \mathcal{Y}(\psi ) \!-\! \mathcal{Y}({{\tilde \psi }_k}) \!+\! \langle {h(x) \!-\! h({{\tilde x}_k}),{{\tilde u}_k}} \rangle  \!+\!\! \langle {\mathcal{C}(x) \!-\!\! \mathcal{C}({{\tilde x}_k}),{{\tilde s}_k}} \rangle  \hfill \\
  \qquad + \langle {x - {{\tilde x}_k},{\Psi ^{ - 1}}({{\tilde x}_k} - {x_k}) + {R^{\text{T}}}{{\tilde w}_k} + {{\tilde \Pi }^{\text{T}}}{{\tilde \lambda }_k} - {\zeta _k}} \rangle  \hfill \\
  \qquad + \langle {w - {{\tilde w}_k},{\Gamma ^{ - 1}}{{\tilde w}_k} - {\Gamma ^{ - 1}}{w_k} - R{x_k}} \rangle  \hfill \\
  \qquad + \langle {\lambda  - {{\tilde \lambda }_k},{\Xi ^{ - 1}}{{\tilde \lambda }_k} - {\Xi ^{ - 1}}{\lambda _k} - \tilde \Pi {x_k}} \rangle  \hfill \\
  \qquad + \langle {s \!-\! {{\tilde s}_k},{\Upsilon ^{ - 1}}{{\tilde s}_k} \!-\! {\Upsilon ^{ - 1}}{s_k} \!+\! {{\bar \Pi }^{\rm{T}}}(2{{\tilde z}_k} - {z_k}) \!-\! \mathcal{C}({{\tilde x}_k})} \rangle  \hfill \\
  \qquad + \langle {u - {{\tilde u}_k},{\Sigma ^{ - 1}}{{\tilde u}_k} - {\Sigma ^{ - 1}}{u_k} - h({{\tilde x}_k})} \rangle  \hfill \\
  \qquad + \langle {z - {{\tilde z}_k},{\Phi ^{ - 1}}{{\tilde z}_k} - {\Phi ^{ - 1}}{z_k} - \bar \Pi {s_k}} \rangle  \hfill \\
\end{gathered}
\end{flalign}
Using the definitions of operators ${\mathcal{K}_1}$, ${\mathcal{K}_2}$ and ${\mathcal{K}_3}$ in \eqref{Totlalvariaineq}, plugging $ \pm \langle {w - {{\tilde w}_k},R{{\tilde x}_k}} \rangle$, $\pm \langle {\lambda  - {{\tilde \lambda }_k},\tilde \Pi {{\tilde x}_k}} \rangle$, and $\pm \langle {z - {{\tilde z}_k},\bar \Pi {{\tilde s}_k}} \rangle $ into the above inequality, and rearranging \eqref{ProofAlgoride}, one then gets \eqref{AlgoriVariaIneq}. Finally, we prove that \eqref{AlgoriVariaIneq} can be split into \eqref{ProofAlgoriab}-\eqref{ProofAlgoricd}. Actually, replacing $\psi $ of \eqref{AlgoriVariaIneq} with $col({{\tilde x}_k},w,{{\tilde \lambda }_k},{{\tilde u}_k},{{\tilde s}_k},{{\tilde z}_k})$, $col({{\tilde x}_k},{{\tilde w}_k},\lambda ,{{\tilde u}_k},{{\tilde s}_k},{{\tilde z}_k})$, $col({{\tilde x}_k},{{\tilde w}_k},{{\tilde \lambda }_k},u,{{\tilde s}_k},{{\tilde z}_k})$, $col({{\tilde x}_k},{{\tilde w}_k},{{\tilde \lambda }_k},{{\tilde u}_k},s,{{\tilde z}_k})$, $col({{\tilde x}_k},{{\tilde w}_k},{{\tilde \lambda }_k},{{\tilde u}_k},{{\tilde s}_k},z)$, and $col(x,{{\tilde w}_k},{{\tilde \lambda }_k},{{\tilde u}_k},{{\tilde s}_k},{{\tilde z}_k})$ in sequence, these results \eqref{ProofAlgoriab}-\eqref{ProofAlgoricd} are generated.
\end{proof}

\subsection{Proof of Lemma \ref{LemFeryarInqua}} \label{ProofLemFeryarInqua}
\begin{proof}
It is easy to obtain the inverse operator of $\mathcal{M}$, i.e., ${\mathcal{M}^{ - 1}}:\psi  \to (x,w - \Gamma Rx,\lambda  - \Xi \tilde \Pi x,{\Lambda ^{ - 1}}u,s,z)$, and confirm that $\mathcal{R}\mathcal{M} = \mathcal{W}$. Then, for any $\psi  \in {\mathbb{R}^n} \times {\Omega _\varpi } \times {\Omega _z}$, one gets $\langle {\psi  - {{\tilde \psi }_k},\mathcal{W}( {{\psi _k} - {{\tilde \psi }_k}} )} \rangle  = \langle {\psi  - {{\tilde \psi }_k},\mathcal{R}\mathcal{M}({\psi _k} - {{\tilde \psi }_k})} \rangle  = \langle {\psi  - {{\tilde \psi }_k},\mathcal{R}({\psi _k} - {\psi _{k + 1}})} \rangle $, where the last equality follows from \eqref{AlgoriOne}. Note that $2\langle {\psi  - {{\tilde \psi }_k},\mathcal{R}({\psi _k} - {\psi _{k + 1}})} \rangle  = ||\psi  - {\psi _{k + 1}}||_\mathcal{R}^2 - ||\psi  - {\psi _k}||_\mathcal{R}^2 + ||{{\tilde \psi }_k} - {\psi _k}||_\mathcal{R}^2 - ||{{\tilde \psi }_k} - {\psi _{k + 1}}||_\mathcal{R}^2$, which combines with the previous equality to $2\langle {\psi  - {{\tilde \psi }_k},\mathcal{W}({\psi _k} - {{\tilde \psi }_k})} \rangle  = ||\psi  - {\psi _{k + 1}}||_\mathcal{R}^2 - ||\psi  - {\psi _k}||_\mathcal{R}^2 + ||{{\tilde \psi }_k} - {\psi _k}||_\mathcal{R}^2 - ||{{\tilde \psi }_k} - {\psi _{k + 1}}||_\mathcal{R}^2$. Based on this result, one directly obtains \eqref{GDgeghy} by using \eqref{AlgoriVariaIneq} in Lemma \ref{LemInetqua}. Next, we prove the second inequality. In \eqref{GDgeghy}, we set $\psi  = {\psi ^ * }$ to have
\begin{flalign}\label{ProofLemFera}
\begin{gathered}
  \quad ||{\psi ^ * } - {\psi _{k + 1}}||_\mathcal{R}^2 + ||{{\tilde \psi }_k} - {\psi _k}||_\mathcal{R}^2 - ||{{\tilde \psi }_k} - {\psi _{k + 1}}||_\mathcal{R}^2 \hfill \\
   \leq \| {{\psi ^ * } \!-\! {\psi _k}} \|_\mathcal{R}^2 \!+\! 2\langle {{\psi ^ * } \!-\! {{\tilde \psi }_k},{\mathcal{K}_1}{{\tilde \psi }_k}} \rangle  \!+\! 2(\mathcal{Y}({\psi ^ * }) \!-\! \mathcal{Y}({{\tilde \psi }_k})) \hfill \\
  \quad - 2\langle {{x^ * } - {{\tilde x}_k},{\zeta _k}} \rangle  + 2\langle {{\mathcal{K}_2}{\psi ^ * } - {\mathcal{K}_2}\tilde \psi ,{\mathcal{K}_3}{{\tilde \psi }_k}} \rangle  \hfill \\
\end{gathered}
\end{flalign}
By recalling these operators ${\mathcal{K}_1}$, ${\mathcal{K}_2}$, and ${\mathcal{K}_3}$, it always holds that $\langle {\psi  - {\psi ^ * },{\mathcal{K}_1}\psi  - {\mathcal{K}_1}{\psi ^ * }} \rangle  + \langle {{\mathcal{K}_2}\psi  - {\mathcal{K}_2}{\psi ^ * },{\mathcal{K}_3}\psi  - {\mathcal{K}_3}{\psi ^ * }} \rangle  = 0$ for any $\psi $. Thus, one concludes that $0 = \langle {{{\tilde \psi }_k} - {\psi ^ * },{\mathcal{K}_1}{{\tilde \psi }_k}} \rangle  - \langle {{{\tilde \psi }_k} - {\psi ^ * },{\mathcal{K}_1}{\psi ^ * }} \rangle  + \langle {{\mathcal{K}_2}{{\tilde \psi }_k} - {\mathcal{K}_2}{\psi ^ * },{\mathcal{K}_3}{{\tilde \psi }_k}} \rangle  - \langle {{\mathcal{K}_2}{{\tilde \psi }_k} - {\mathcal{K}_2}{\psi ^ * },{\mathcal{K}_3}{\psi ^ * }} \rangle $, which further yields that $\langle {{{\tilde \psi }_k} - {\psi ^ * },{\mathcal{K}_1}{{\tilde \psi }_k}} \rangle  + \langle {{\mathcal{K}_2}{{\tilde \psi }_k} - {\mathcal{K}_2}{\psi ^ * },{\mathcal{K}_3}{{\tilde \psi }_k}} \rangle  = \langle {{{\tilde \psi }_k} - {\psi ^ * },{\mathcal{K}_1}{\psi ^ * }} \rangle  + \langle {{\mathcal{K}_2}{{\tilde \psi }_k} - {\mathcal{K}_2}{\psi ^ * },{\mathcal{K}_3}{\psi ^ * }} \rangle $. Plugging the previous equality into \eqref{ProofLemFera}, and using \eqref{Totlalvariaineq} result in
\begin{flalign}\label{ProofLemFerdbc}
\begin{gathered}
  \| {{\psi _{k + 1}} - {\psi ^ * }} \|_\mathcal{R}^2 \leq \| {{\psi _k} - {\psi ^ * }} \|_\mathcal{R}^2 - \| {{{\tilde \psi }_k} - {\psi _k}} \|_\mathcal{R}^2 \hfill \\
  \qquad\qquad\qquad\qquad + \| {{{\tilde \psi }_k} - {\psi _{k + 1}}} \|_\mathcal{R}^2 - 2\langle {{x^ * } - {{\tilde x}_k},{\zeta _k}} \rangle  \hfill \\
\end{gathered}
\end{flalign}
It follows from \eqref{AlgoriOne} that
\begin{flalign}\label{ProofLemFerdcd}
\begin{gathered}
  \quad \| {{\psi _{k + 1}} - {{\tilde \psi }_k}} \|_\mathcal{R}^2 - \| {{{\tilde \psi }_k} - {\psi _k}} \|_\mathcal{R}^2 \hfill \\
   = \| {{\psi _k} - {{\tilde \psi }_k} + \mathcal{M}( {{{\tilde \psi }_k} - {\psi _k}} )} \|_\mathcal{R}^2 - \| {{{\tilde \psi }_k} - {\psi _k}} \|_\mathcal{R}^2 \hfill \\
   = \| {{\psi _k} - {{\tilde \psi }_k}} \|_\mathcal{R}^2 + \| {\mathcal{M}( {{{\tilde \psi }_k} - {\psi _k}} )} \|_\mathcal{R}^2 \hfill \\
  \quad + 2\langle {{\psi _k} - {{\tilde \psi }_k},\mathcal{R}\mathcal{M}( {{{\tilde \psi }_k} - {\psi _k}} )} \rangle  - \| {{{\tilde \psi }_k} - {\psi _k}} \|_\mathcal{R}^2 \hfill \\
   = \| {{{\tilde \psi }_k} - {\psi _k}} \|_{{\mathcal{M}^{\rm T}}\mathcal{R}\mathcal{M}}^2 - \langle {{{\tilde \psi }_k} - {\psi _k},( {{\mathcal{W}^{\rm T}} + \mathcal{W}} )( {{{\tilde \psi }_k} - {\psi _k}} )} \rangle  \hfill \\
   =  - \| {{{\tilde \psi }_k} - {\psi _k}} \|_{{\mathcal{W}^{\rm T}} + \mathcal{W} - {\mathcal{M}^{\rm T}}\mathcal{R}\mathcal{M}}^2 \hfill \\
\end{gathered}
\end{flalign}
which is combined with \eqref{ProofLemFerdbc} to get the desired result.
\end{proof}

\subsection{Proof of Theorem \ref{TheoreConverge}} \label{ProofTheoreConverge}
\begin{proof}
From Lemmas \ref{LemInetqua}-\ref{LemFeryarInqua}, one can derive $\mathcal{\tilde W} = \mathcal{W} + {\mathcal{W}^{\text{T}}} - {\mathcal{M}^{\rm T}}\mathcal{R}\mathcal{M} = diag\{ {\Psi ^{ - 1}} - ({R^{\text{T}}}\Gamma R + {\tilde \Pi ^{\text{T}}}\Xi \tilde \Pi ),{\Gamma ^{ - 1}},{\Xi ^{ - 1}},2{\Sigma ^{ - 1}} - \Lambda {\Sigma ^{ - 1}},Q\}$ with the linear mapping $Q:(s,z) \to ({\Upsilon ^{ - 1}}s + {{\bar \Pi }^{\text{T}}}z,{\Phi ^{ - 1}}z + \bar \Pi s)$. Since the step-size conditions hold, ${\Gamma ^{ - 1}},{\Xi ^{ - 1}},2{\Sigma ^{ - 1}} - \Lambda {\Sigma ^{ - 1}}$ are positive definite, and so do ${\Psi ^{ - 1}} - {R^{\text{T}}}\Gamma R - {\tilde \Pi ^{\text{T}}}\Xi \tilde \Pi $ and $Q$ by using the Schur complement lemma \cite{Horn2005Schur}. Then, $\mathcal{R}$, $\mathcal{\tilde W}$ and ${\mathcal{M}^{\rm T}}\mathcal{R}\mathcal{M}$ have positive definite properties naturally. Considering ${{\tilde \zeta }_k} = col\{ {\zeta _k},0,0,0,0,0\} $, one then can use $\langle {{{\tilde x}_k} - {x^ * },{\zeta _k}} \rangle  = \langle {{{\tilde \psi }_k} - {\psi ^ * },{{\tilde \zeta }_k}} \rangle = \langle {\sqrt \mathcal{R} ( {{\psi _k} - {\psi ^ * }} ),\sqrt {{\mathcal{R}^{ - 1}}} {{\tilde \zeta }_k}} \rangle  + \langle {\sqrt {\mathcal{\tilde W}} ( {{{\tilde \psi }_k} - {\psi _k}} ),\sqrt {{{\mathcal{\tilde W}}^{ - 1}}} {{\tilde \zeta }_k}} \rangle$ to derive
\begin{flalign*}
\begin{gathered}
  \langle {{{\tilde \psi }_k} - {\psi ^ * },{{\tilde \zeta }_k}} \rangle \leq \| {\sqrt \mathcal{R} ( {{\psi _k} - {\psi ^ * }} )} \|\| {\sqrt {{\mathcal{R}^{ - 1}}} {{\tilde \zeta }_k}} \| \hfill \\
  \qquad\qquad\qquad\qquad + \frac{1}{2}\| {{{\tilde \psi }_k} - {\psi _k}} \|_{\mathcal{\tilde W}}^2 + \frac{1}{2}{\| {\sqrt {{{\mathcal{\tilde W}}^{ - 1}}} {{\tilde \zeta }_k}} \|^2} \hfill \\
\end{gathered}
\end{flalign*}
which is plugged into \eqref{GDgRGghWy} to obtain
\begin{flalign*}
\begin{gathered}
  \quad \| {{\psi _{k + 1}} - {\psi ^ * }} \|_\mathcal{R}^2 \hfill \\
   \leq \| {{\psi _k} - {\psi ^ * }} \|_\mathcal{R}^2 + 2\| {\sqrt \mathcal{R} ( {{\psi _k} - {\psi ^ * }} )} \|\| {\sqrt {{\mathcal{R}^{ - 1}}} {{\tilde \zeta } _k}} \| + {\| {\sqrt {{{\mathcal{\tilde W}}^{ - 1}}} {{\tilde \zeta } _k}} \|^2} \hfill \\
   \leq {( {{{\| {{\psi _k} - {\psi ^ * }} \|}_\mathcal{R}} + \| {\sqrt {{\mathcal{R}^{ - 1}}} {{\tilde \zeta } _k}} \|} )^2} + {\| {\sqrt {{{\mathcal{\tilde W}}^{ - 1}}} {{\tilde \zeta } _k}} \|^2} \hfill \\
   \leq {( {{{\| {{\psi _k} - {\psi ^ * }} \|}_\mathcal{R}} + ( {\| {\sqrt {{\mathcal{R}^{ - 1}}} } \| + \| {\sqrt {{{\mathcal{\tilde W}}^{ - 1}}} } \|} )\| {{{ \zeta } _k}} \|} )^2} \hfill \\
\end{gathered}
\end{flalign*}
where $\| {{{\tilde \zeta }_k}} \| = \| {{\zeta _k}} \|$ is used. {Based on the above result, we have
\begin{flalign} \label{ProofTheorezs}
||{\psi _{k + 1}} - {\psi ^ * }|{|_\mathcal{R}} \leq {\| {{\psi _k} - {\psi ^ * }} \|_\mathcal{R}} + \kappa \| {{{ \zeta }_k}} \|
\end{flalign}
with $\kappa  = \| {\sqrt {{\mathcal{R}^{ - 1}}} } \| + \| {\sqrt {{{\mathcal{\tilde W}}^{ - 1}}} } \|$}. The condition $0 < \| {{\zeta _{i,k}}} \| \leq {\varepsilon _{i,k}}$ yields that $\| {{\zeta _k}} \| = \sqrt {\sum\nolimits_{i = 1}^m {{{\| {{\zeta _{i,k}}} \|}^2}} }  \leq \sqrt {\sum\nolimits_{i = 1}^m {\varepsilon _{i,k}^2} } \leq \sum\nolimits_{i = 1}^m {{\varepsilon _{i,k}}} $. Since $\sum\nolimits_{k = 0}^{ + \infty } {{\varepsilon _{i,k}}}  <  + \infty $ holds, one then has $\sum\nolimits_{k = 0}^{ + \infty } { \left\| {{\zeta _k}} \right\|}  \leq \sum\nolimits_{k = 0}^{ + \infty } {{\varepsilon _k}}  <  + \infty $, where ${\varepsilon _k} = \sum\nolimits_{i = 1}^m {{\varepsilon _{i,k}}} $. Therefore, we sum \eqref{ProofTheorezs} over $k$ from $0$ to $K$ to have
\begin{flalign}\label{ProofTheoreConab}
{\left\| {{\psi _{K + 1}} - {\psi ^ * }} \right\|_\mathcal{R}} \leq {\left\| {{\psi _0} - {\psi ^ * }} \right\|_\mathcal{R}} + \sum\limits_{k = 0}^K {\kappa \left\| {{\zeta _k}} \right\|}  <  + \infty
\end{flalign}
Since {$\mathcal{R}$ is positive definite, ${\{ ||{\psi _{k }} - {\psi ^ * }|{|_\mathcal{R}}\} _{k \geq 0}}$ is convergent \cite[Lemma 5.31]{Bauschke2011Convex}, ${\{ ||{\psi _{k}} - {\psi ^*}||\} _{k \ge 0}}$ and ${\{ {{\psi _k}} \}_{k \geq 0}}$ are bounded \cite[Theorem 5.33]{Bauschke2011Convex}. Recall the relation \eqref{AlgoriOne} to have ${\| {{{\tilde \psi }_k} - {\psi _k}} \|_{{\mathcal{M}^{\text{T}}}\mathcal{R}\mathcal{M}}} \le {\| {{\psi _{k + 1}} - {\psi ^ * }} \|_\mathcal{R}} + {\| {{\psi _k} - {\psi ^ * }} \|_\mathcal{R}}$. Hence, the boundedness of ${\{ {{\psi _k}} \}_{k \geq 0}}$ implies that ${\{ {{{\tilde \psi }_k}} \}_{k \geq 0}}$ is bounded as well. Summing \eqref{GDgRGghWy} over $k = 0,1, \cdots , + \infty $ yields that $\sum\nolimits_{k = 0}^{ + \infty } {||{{\tilde \psi }_k} - {\psi _k}||_{\mathcal{\tilde W}}^2}  \leq ||{\psi _0} - {\psi ^ * }||_\mathcal{R}^2 + \sum\nolimits_{k = 0}^{ + \infty } { ||{{\tilde x}_k} - {x^ * }|| {{\varepsilon _k}}}$. Note that ${\left\{ {{\varepsilon _k}} \right\}_{k \ge 0}}$ is summable, ${\{ {{\psi _k}} \}_{k \geq 0}}$ and ${\{ {{{\tilde \psi }_k}} \}_{k \ge 0}}$ with ${{\tilde \psi }_k} = col\{ {{\tilde x}_k},{{\tilde z}_k},{{\tilde \varpi }_k}\}$ are bounded, we then have
\begin{flalign}\label{ProofTheoregG}
\sum\limits_{k = 0}^{ + \infty } {\| {{{\tilde \psi }_k} - {\psi _k}} \|_{ \mathcal{\tilde W}}^2}  <  + \infty
\end{flalign}
Since ${ \mathcal{\tilde W}}$ is positive definite, it holds that ${\lim _{k \to  + \infty }}\| {{{\tilde \psi }_k} - {\psi _k}} \| = 0$. Let ${\psi _\infty }$ be a cluster point of ${\{ {{{\tilde \psi }_k}} \}_{k \geq 0}}$, and ${\{ {\tilde \psi _{{k_j}}}\} _{{k_j} \ge 0}}$ be a subsequence converging to ${\psi _\infty }$. Inspired by the proved result \eqref{AlgoriVariaIneq}, it holds that
\begin{flalign*}
\begin{gathered}
\quad \langle {\psi  - {{\tilde \psi }_{{k_j}}},\mathcal{W}({\psi _{{k_j}}} - {{\tilde \psi }_{{k_j}}}) + {\tilde \zeta _{{k_j}}}} \rangle \hfill \\
\le \mathcal{Y}(\psi ) \!-\! \mathcal{Y}({\tilde \psi _{{k_j}}}) \!+\! \langle {\psi  - {{\tilde \psi }_{{k_j}}},{\mathcal{K}_1}{{\tilde \psi }_{{k_j}}}} \rangle  + \langle {{\mathcal{K}_2}\psi  \!-\! {\mathcal{K}_2}{{\tilde \psi }_{{k_j}}},{\mathcal{K}_3}{{\tilde \psi }_{{k_j}}}} \rangle  \hfill \\
\end{gathered}
\end{flalign*}
Since $\mathcal{W}$ is nonsingular, ${\mathcal{K}_1}$, ${\mathcal{K}_2}$ and ${\mathcal{K}_3}$ are continuous, we take ${k_j} \to  + \infty $ in the above inequality to have
\begin{flalign*}
\mathcal{Y}(\psi ) \!-\! \mathcal{Y}({\psi _\infty }) \!+\! \langle {\psi  \!-\! {\psi _\infty },{\mathcal{K}_1}{\psi _\infty }} \rangle  \!+\! \langle {{\mathcal{K}_2}\psi  \!-\! {\mathcal{K}_2}{\psi _\infty },{\mathcal{K}_3}{\psi _\infty }} \rangle  \geq 0
\end{flalign*}
Consequently, it can be concluded from \eqref{Totlalvariaineq} that ${\psi _\infty } \in \mathcal{S}$. Therefore, by \eqref{ProofTheorezs}, we have
\begin{flalign} \label{ProofTheoreokq}
||{\psi _{k + 1}} - {\psi _\infty }|{|_\mathcal{R}} \leq {\| {{\psi _k} - {\psi _\infty }} \|_\mathcal{R}} + \kappa \| {{{ \zeta }_k}} \|
\end{flalign}
It follows from \cite[Lemma3.2]{Chen2021ALM} that ${\{ ||{\psi _k} - {\psi _\infty }|{|_\mathcal{R}}\} _{k \geq 0}}$ converges to a unique limit point. Next, let ${\{ {\psi _{{k_j}}}\} _{{k_j} \in \mathbb{N}}}$ be the induced sequence generated by ${\{ {{\psi _k}} \}_{k \geq 0}}$. By ${\lim _{j \to  + \infty }}{{\tilde \psi }_{{k_j}}} = {\psi _\infty }$ and ${\lim _{k \to  + \infty }}\| {{{\tilde \psi }_k} - {\psi _k}} \| = 0$, we have that ${\lim _{j \to  + \infty }}(||{\psi _{{k_j}}} - {{\tilde \psi }_{{k_j}}}|| + ||{{\tilde \psi }_{{k_j}}} - {\psi _\infty }||) = 0$, which results in $0 \leq {\lim _{j \to  + \infty }}||{\psi _{{k_j}}} - {\psi _\infty }|| \leq {\lim _{{k_j} \to  + \infty }}(||{\psi _{{k_j}}} - {{\tilde \psi }_{{k_j}}}|| + ||{{\tilde \psi }_{{k_j}}} - {\psi _\infty }||) = 0$, where the triangle inequality $||{\psi _{{k_j}}} - {\psi _\infty }|| \leq ||{\psi _{{k_j}}} - {{\tilde \psi }_{{k_j}}}|| + ||{{\tilde \psi }_{{k_j}}} - {\psi _\infty }||$ is used. Therefore, we have that
\begin{flalign*}
{\lim _{j \to  + \infty }}||{\psi _{{k_j}}} - {\psi _\infty }|| = 0
\end{flalign*}
which implies that the subsequence ${\{ {\psi _{{k_j}}}\} _{k \geq 0}}$ converges to ${\psi _\infty }$, and ${\psi _\infty }$ is also a cluster point of ${\{ {{{ \psi }_k}} \}_{k \geq 0}}$. It can be concluded that ${\lim _{k \to  + \infty }}{\psi _k} = {\psi _\infty }$ with ${\psi _\infty } = col\{ {x_\infty },{z_\infty },{\varpi _\infty }\} $. Consequently, one has that $x_\infty$ solves problem \eqref{OrigProblem} via Remark \ref{RemEquSolt} and Lemma \ref{DistriMN}.}

Next, we prove the second claim. Recalling \eqref{AlgoriVariaIneq} at the $k+1$-th iteration gives that
\begin{flalign}\label{ProofTheoreConbc}
\begin{gathered}
  \quad \langle {\psi  - {{\tilde \psi }_{k + 1}},\mathcal{W}({\psi _{k + 1}} - {{\tilde \psi }_{k + 1}}) + {{\tilde \zeta }_{k + 1}}} \rangle  \hfill \\
   \leq \mathcal{Y}(\psi ) - \mathcal{Y}({{\tilde \psi }_{k + 1}}) + \langle {\psi  - {{\tilde \psi }_{k + 1}},{\mathcal{K}_1}{{\tilde \psi }_{k + 1}}} \rangle  \hfill \\
  \quad + \langle {{\mathcal{K}_2}\psi  - {\mathcal{K}_2}{{\tilde \psi }_{k + 1}},{\mathcal{K}_3}{{\tilde \psi }_{k + 1}}} \rangle  \hfill \\
\end{gathered}
\end{flalign}
Assign ${\tilde \psi _{k + 1}}$ to $\psi $ in \eqref{AlgoriVariaIneq}, ${\tilde \psi _{k }}$ to $\psi$ in \eqref{ProofTheoreConbc}, and then sum the both inequalities to obtain
\begin{flalign}\label{ProofTheoreConcd}
\begin{gathered}
  \quad \langle {{{\tilde \psi }_k} \!-\! {{\tilde \psi }_{k + 1}},{\mathcal{K}_1}{{\tilde \psi }_{k + 1}} \!-\! {\mathcal{K}_1}{{\tilde \psi }_k}} \rangle  \hfill \\
   \quad + \langle {{\mathcal{K}_2}{{\tilde \psi }_k} \!-\! {\mathcal{K}_2}{{\tilde \psi }_{k + 1}},{\mathcal{K}_3}{{\tilde \psi }_{k + 1}} - {\mathcal{K}_3}{{\tilde \psi }_k}} \rangle  \hfill \\
   \geq \langle {{{\tilde \psi }_k} \!-\! {{\tilde \psi }_{k + 1}},\mathcal{W} {( {{\psi _{k + 1}} \!-\! {{\tilde \psi }_{k + 1}}} ) \!-\! ( {{\psi _k} - {{\tilde \psi }_k}} )} ) \!+\! {{\tilde \zeta }_{k + 1}} \!-\! {{\tilde \zeta }_k}} \rangle  \hfill \\
   = \frac{1}{2}\| {( {{\psi _{k + 1}} - {{\tilde \psi }_{k + 1}}} ) - ( {{\psi _k} - {{\tilde \psi }_k}} )} \|_{{\mathcal{W}^{\rm T}} + \mathcal{W}}^2 \hfill \\
  \quad + \langle {{\psi _k} - {\psi _{k + 1}},\mathcal{W}( {( {{\psi _{k + 1}} - {{\tilde \psi }_{k + 1}}} ) - ( {{\psi _k} - {{\tilde \psi }_k}} )} )} \rangle  \hfill \\
  \quad + \langle {{{\tilde \psi }_k} - {{\tilde \psi }_{k + 1}},{{\tilde \zeta }_{k + 1}} - {{\tilde \zeta }_k}} \rangle  \hfill \\
   = \frac{1}{2}\| {( {{\psi _{k + 1}} - {{\tilde \psi }_{k + 1}}} ) - ( {{\psi _k} - {{\tilde \psi }_k}} )} \|_{{\mathcal{W}^{\rm T}} + \mathcal{W}}^2 \hfill \\
  \quad - \langle {\mathcal{M}( {{{\tilde \psi }_k} - {\psi _k}} ),\mathcal{R}\mathcal{M}( {( {{\psi _{k + 1}} - {{\tilde \psi }_{k + 1}}} ) - ( {{\psi _k} - {{\tilde \psi }_k}} )} )} \rangle  \hfill \\
  \quad + \langle {{{\tilde \psi }_k} - {{\tilde \psi }_{k + 1}},{{\tilde \zeta }_{k + 1}} - {{\tilde \zeta }_k}} \rangle  \hfill \\
\end{gathered}
\end{flalign}
where the second equality follows from $\mathcal{R} = \mathcal{W}{\mathcal{M}^{ - 1}}$ and \eqref{AlgoriOne}. Note that $\langle {{{\tilde \psi }_k} - {{\tilde \psi }_{k + 1}},{\mathcal{K}_1}{{\tilde \psi }_{k + 1}} - {\mathcal{K}_1}{{\tilde \psi }_k}} \rangle  = 0$, $\langle {{\mathcal{K}_2}{{\tilde \psi }_k} - {\mathcal{K}_2}{{\tilde \psi }_{k + 1}},{\mathcal{K}_3}{{\tilde \psi }_{k + 1}} - {\mathcal{K}_3}{{\tilde \psi }_k}} \rangle  = 0$, and $\langle {\mathcal{M}({\psi _k} - {{\tilde \psi }_k}),\mathcal{R}\mathcal{M}(({\psi _{k + 1}} - {{\tilde \psi }_{k + 1}}) - ({\psi _k} - {{\tilde \psi }_k}))} \rangle  = \frac{1}{2}||\mathcal{M}({\psi _{k + 1}} - {{\tilde \psi }_{k + 1}})||_\mathcal{R}^2 - \frac{1}{2}||\mathcal{M}({\psi _k} - {{\tilde \psi }_k})||_\mathcal{R}^2 - \frac{1}{2}||\mathcal{M}({\psi _{k + 1}} - {{\tilde \psi }_{k + 1}}) - \mathcal{M}({\psi _k} - {{\tilde \psi }_k})||_\mathcal{R}^2$, which are substituted into \eqref{ProofTheoreConcd} to deduce that
\begin{flalign}\label{ProofTheoreConde}
\begin{gathered}
  \quad \| {\mathcal{M}( {{\psi _{k + 1}} - {{\tilde \psi }_{k + 1}}} )} \|_\mathcal{R}^2 \hfill \\
   \leq \| {\mathcal{M}( {{\psi _k} - {{\tilde \psi }_k}} )} \|_\mathcal{R}^2 + 2\langle {{{\tilde \psi }_k} - {{\tilde \psi }_{k + 1}},{{\tilde \zeta }_k} - {{\tilde \zeta }_{k + 1}}} \rangle  \hfill \\
  \quad - \| {( {{\psi _k} - {{\tilde \psi }_k}} ) - ( {{\psi _{k + 1}} - {{\tilde \psi }_{k + 1}}} )} \|_{\mathcal{\tilde W}}^2 \hfill \\
\end{gathered}
\end{flalign}
Since ${\varepsilon _k} \leq {\rho _k}$ and ${\{ {{{\tilde \psi }_k}} \}_{k \geq 0}}$ is bounded, there exists a $\xi >0$ such that $||\langle {{{\tilde \psi }_k} - {{\tilde \psi }_{k + 1}},{{\tilde \zeta }_k} - {{\tilde \zeta }_{k + 1}}} \rangle || \leq \xi {({\rho _{\max }})^k}$. Thus, \eqref{ProofTheoreConde} is derived as $||\mathcal{M}({\psi _{k + 1}} - {{\tilde \psi }_{k + 1}})||_\mathcal{R}^2 \leq ||\mathcal{M}({\psi _k} - {{\tilde \psi }_k})||_\mathcal{R}^2 + \xi {({\rho _{\max }})^k} = ||\mathcal{M}({\psi _k} - {{\tilde \psi }_k})||_\mathcal{R}^2 + {({\rho _{\max }})^k}\xi /(1 - {\rho _{\max }}) - {({\rho _{\max }})^{k + 1}}\xi /(1 - {\rho _{\max }})$, i.e., $||\mathcal{M}({\psi _{k + 1}} - {\tilde \psi _{k + 1}})||_\mathcal{R}^2 + \xi {({\rho _{\max }})^{k + 1}}/(1 - {\rho _{\max }}) \leq ||\mathcal{M}({\psi _k} - {\tilde \psi _k})||_\mathcal{R}^2 + \xi {({\rho _{\max }})^k}/(1 - {\rho _{\max }})$. Using \eqref{AlgoriOne} again results in $||{\psi _{k + 1}} - {\psi _{k + 2}}||_\mathcal{R}^2 + \xi {({\rho _{\max }})^{k + 1}}/(1 - {\rho _{\max }}) \leq ||{\psi _k} - {\psi _{k + 1}}||_\mathcal{R}^2 + \xi {({\rho _{\max }})^k}/(1 - {\rho _{\max }})$. Hence, the desired result is deduced via \cite[Proposition 1]{Shi2015Extra}.

Finally, {we use the above results to derive the first-order optimality residual. By \eqref{ProofAlgoriVa}, \eqref{ProofAlgoriVb}, \eqref{ProofAlgoriVc}, \eqref{ProofAlgoriVe} and \eqref{ProofAlgoriVf}, we have $0 \in \partial {g^ \star }({{\tilde w}_k}) + {\Gamma ^{ - 1}}({{\tilde w}_k} - ({w_k} + \Gamma R{x_k}))$, $0 \in \partial \mathcal{I}_{\tilde D}^ \star ({{\tilde \lambda }_k}) + {\Xi ^{ - 1}}({{\tilde \lambda }_k} - ({\lambda _k} + \Xi \tilde \Pi {x_k}))$, $0 \in \partial \mathcal{I}_{\bar D}^ \star ({{\tilde z}_k}) + {\Phi ^{ - 1}}({{\tilde z}_k} - ({z_k} + \Phi \bar \Pi {s_k}))$, $0 \in \partial {\mathcal{I}_\Theta }({{\tilde u}_k}) + {\Sigma ^{ - 1}}({{\tilde u}_k} - ({u_k} + \Sigma h({{\tilde x}_k})))$, and $0 \in \partial {\mathcal{I}_{{\Omega _{\text{d}}}}}({{\tilde s}_k}) + {\Upsilon ^{ - 1}}({{\tilde s}_k} - ({s_k} - \Upsilon {{\bar \Pi }^{\text{T}}}(2{{\tilde z}_k} - {z_k}) + \Upsilon \mathcal{C}({{\tilde x}_k})))$, which, together with \eqref{ProofAlgoriza} implies that $\mathcal{W}( {{\psi _k} - {{\tilde \psi }_k}} ) + {{\tilde \zeta }_k} \in \mathcal{T}{{\tilde \psi }_k}$. Recalling $\mathcal{W}$ in Lemma \ref{LemInetqua}, and $\mathcal{R}$ in Lemma \ref{LemFeryarInqua}, it follows from \eqref{AlgoriOne} that $\mathcal{W}( {{{\tilde \psi }_k} - {\psi _k}} ) = \mathcal{W}{\mathcal{M}^{ - 1}}( {{\psi _{k + 1}} - {\psi _k}} ) = \mathcal{R}( {{\psi _{k + 1}} - {\psi _k}} )$. Then, there exists $\tau >0$ such that
\begin{flalign*}
\begin{gathered}
  \quad {\text{di}}{{\text{st}}^2}( {0,\mathcal{T}{{\tilde \psi }_k}} ) \hfill \\
   \le {\| {\mathcal{R}( {{\psi _{k + 1}} - {\psi _k}} ) + {{\tilde \zeta }_k}} \|^2} \hfill \\
   \le \tau \| {{\psi _{k + 1}} - {\psi _k}} \|_\mathcal{R}^2 + 2\| {\mathcal{R}( {{\psi _{k + 1}} - {\psi _k}} )} \|{\varepsilon _k}+ {\varepsilon _k^2} \hfill \\
\end{gathered}
\end{flalign*}
where $||{{\tilde \zeta }_k}|| = ||{\zeta _k}|| \leq {\varepsilon _k}$ is used. Since $\| {{\psi _{k + 1}} - {\psi _k}} \|_\mathcal{R}^2 = o(1/k)$ and ${\varepsilon _{i,k}} = O(\rho _i^k)$, one then has ${\text{dis}}{{\text{t}}^2}( {0,\mathcal{T}{{\tilde \psi }_k}} ) = o(1/k) + O({({\rho _{\max }})^k})$.}
\end{proof}

\bibliographystyle{ieeetr}

\bibliography{paper_reference}

\end{document}